\documentclass[10pt]{article}

\usepackage{amsmath}

\usepackage{array}

\usepackage{appendix}

\usepackage{tocloft}                   

\usepackage{graphicx}

\usepackage{amsfonts}

\usepackage{amssymb}

\usepackage{mathrsfs}

\usepackage{yfonts}

\usepackage{euscript}

\usepackage{centernot}                 

\usepackage{ifsym}                     

\usepackage{upgreek}

\usepackage{mathtools}

\usepackage{color}

\usepackage{slantsc}
\usepackage{calligra}

\usepackage{bbold}          

\usepackage[T1]{fontenc}

\usepackage{epsf}

\usepackage{latexsym}

\usepackage{tipa}

\usepackage{makeidx}

\makeindex

\def\b{\begin{equation}}

\def\e{\begin{equation}}

\def\be{\begin{equation}}              

\def\ee{\end{equation}}

\def\beq{\begin{equation}}

\def\eeq{\end{equation}}

\def\bea{\begin{eqnarray}}

\def\eea{\end{eqnarray}}

\def\m{\mbox{ }}

\def\mma {\m , \m \m }

\def\!{\hspace{-1.6667em}}

\def\Proof{{\n{\u{Proof}}}\m}

\def\n{\noindent}

\def\u{\underline}

\def\w{\widetilde}

\def\s{\stackrel}

\def\mB{\mbox{B}}  

\def\mC{\mbox{C}}                        

\def\mD{\mbox{D}}                        

\def\mF{\mbox{F}}

\def\mG{\mbox{G}}

\def\mM{\mbox{M}}                        

\def\mO{\mbox{O}}

\def\mP{\mbox{P}}

\def\mS{\mbox{S}}                        

\def\mT{\mbox{T}} 

\def\mU{\mbox{U}}                        

\def\mW{\mbox{W}}

\def\mX{\mbox{X}}

\def\mY{\mbox{Y}}

\def\md{\mbox{d}}

\def\mo{\mbox{o}}

\def\mp{\mbox{p}}

\def\mv{\mbox{v}}

\def\upi{\mbox{$\pi$}}                              

\def\bv{\mbox{\bf v}}

\def\fO{\mbox{\sffamily O}}

\def\fP{\mbox{\sffamily P}}

\def\fQ{\mbox{\sffamily Q}}

\def\fT{\mbox{\sffamily T}}

\def\fX{\mbox{\sffamily X}}

\def\sa{\mbox{\scriptsize a}}

\def\scc{\mbox{\scriptsize c}}

\def\sd{\mbox{\scriptsize d}}

\def\se{\mbox{\scriptsize e}}

\def\si{\mbox{\scriptsize i}}

\def\sll{\mbox{\scriptsize l}}  

\def\sm{\mbox{\scriptsize m}}

\def\sn{\mbox{\scriptsize n}} 

\def\so{\mbox{\scriptsize o}} 

\def\sp{\mbox{\scriptsize p}}

\def\sr{\mbox{\scriptsize r}}

\def\st{\mbox{\scriptsize t}}

\def\sv{\mbox{\scriptsize v}}

\def\sx{\mbox{\scriptsize x}}

\def\sA{\mbox{\scriptsize A}}

\def\sC{\mbox{\scriptsize C}}

\def\sD{\mbox{\scriptsize D}}

\def\sG{\mbox{\scriptsize G}}

\def\sI{\mbox{\scriptsize I}}

\def\sP{\mbox{\scriptsize P}} 

\def\sQ{\mbox{\scriptsize Q}}

\def\sT{\mbox{\scriptsize T}}

\def\sU{\mbox{\scriptsize U}}

\def\sfA{\mbox{\sffamily{\scriptsize A}}}     

\def\sfP{\mbox{\sffamily{\scriptsize P}}}      

\def\sfQ{\mbox{\sffamily{\scriptsize Q}}}      

\def\sfT{\mbox{\sffamily{\scriptsize T}}}      

\def\sfX{\mbox{\sffamily{\scriptsize X}}}      

\def\tm{\mbox{\tiny m}}

\def\ttt{\mbox{\tiny t}}   

\def\tA{\mbox{\tiny A}}

\def\tQ{\mbox{\tiny Q}}

\def\tfQ{\mbox{\sffamily{\tiny Q}}}

\def\sumi2{\sum\mbox{}_{\mbox{}_{\mbox{\scriptsize $i$=1}}}^2}

\def\sumi3{\sum\mbox{}_{\mbox{}_{\mbox{\scriptsize $i$=1}}}^3}

\def\sumABcycles3{\sum\mbox{}_{\mbox{}_{\mbox{\scriptsize cycles $A,B$=1}}}^{3}}

\def\sumCDcycles3{\sum\mbox{}_{\mbox{}_{\mbox{\scriptsize cycles $C,D$=1}}}^{3}}

\def\sumj3{\sum\mbox{}_{\mbox{}_{\mbox{\scriptsize $j$=1}}}^3}

\def\sumk3{\sum\mbox{}_{\mbox{}_{\mbox{\scriptsize $k$=1}}}^3}

\def\prodiA1{\prod\mbox{}_{\mbox{}_{\mbox{\scriptsize $i$=1}}}^{A - 1}}

\def\bigtimes{\mbox{\Large $\times$}}

\def\pa{\partial}                                                   

\def\es{\m = \m}

\def\:={\m := \m}

\def\=:{\m =: \m}

\def\FrI{\mbox{$\mathfrak{I}$}}                                

\def\FrT{\mathfrak{T}}                                         
                                                       
\def\FrS{\mbox{\Large $\mathfrak{s}$}}                         
\def\sFrS{\mbox{\large$\mathfrak{s}$}}                         
   
\def\tFrS{\mbox{\footnotesize$\mathfrak{s}$}} 

\def\FrT{\mbox{\boldmath$\mathfrak{T}$}}                       
\def\FrG{\mathfrak{G}}                                         
\def\Hilb{\mbox{{\boldmath$\mathfrak{H}$}ilb}}                 

\def\bFrC{\mbox{\boldmath$\mathfrak{C}$}}                            
\def\bFrL{\mbox{\boldmath$\mathfrak{L}$}}                            

\def\Phase{\mbox{{\boldmath$\mathfrak{P}$}hase}}                     

\def\bFrR{\mbox{\boldmath$\mathfrak{R}$}}                            
\def\Rig-Phase{\bFrR\mbox{ig-}\Phase}                                
                                                                													   
\def\diam{\delta iam}                                                

\def\1mat{\u{\u{1}}}                                                 

\def\Positive-Modespace{\mbox{{\boldmath$\mathfrak{M}$}odespace$^+$}}

\def\POSITIVE-MODESPACE{\mbox{{\boldmath$\mathfrak{M}$}ODESPACE$^+$}}

\def\Leib{\bFrL\mbox{eib}}                                           

\def\Co{\bFrC\mbox{o}}                                               

\def\Top{\FrT\mo\mp}

\def\Kin-Hilb{\mbox{{\boldmath$\mathfrak{K}$}in-\Hilb}}                     

\def\Mid-Hilb{\mbox{{\boldmath$\mathfrak{M}$}id-\Hilb}}                     

\def\Dyn-Hilb{\mbox{{\boldmath$\mathfrak{D}$}yn-\Hilb}}                     

\def\5Star{\mbox{\Large$\star$}}              

\def\bigiota{\mbox{\Large $\iota$}}

\begin{document}

\begin{titlepage}

\begin{center}

\large{\bf THE SMALLEST SHAPE SPACES. II.} \normalsize

\large{\bf 4 Points in 1-$d$ Suffices to have a Complex Background-Independent Theory of Inhomogeneity} \normalsize 

\vspace{0.1in}

\normalsize

\vspace{0.1in}

{\large \bf Edward Anderson$^*$}

\vspace{.2in}

\end{center}

\begin{abstract}

\n The program of understanding Shape Theory layer by layer topologically and geometrically -- proposed in Part I -- 
is now addressed for 4 points in 1-$d$.  
Topological shape space graphs are far more complex here, 
whereas metric shape spaces are (pieces of) spheres which admit an intricate shape-theoretically significant tessellation.  
Metric shapes covers a far wider range of notions of inhomogeneity: 
collisions, symmetric states, mergers and uniform states are all distinctly realized in this model. 
We furthermore provide quantifiers for the extent to which various ways which configurations maximally and minimally realize these. 
Some of the uniform states additionally form cusps and higher catastrophes in the indistinguishable-particle and Leibniz shape spaces.  
We also provide shape-theoretically significant notions of centre for the indistinguishable-particle and Leibniz shape spaces.
4 points in 1-$d$ constitutes a useful and already highly nontrivial model of inhomogeneity and of uniformity -- both topics of cosmological interest --  
and also of background independence: of interest in the foundations of physics and in quantum gravity.  
We finally give the automorphism groups of the topological shape space graphs and the metric shape space (pieces of) manifolds,  
which is a crucial preliminary toward quantizing the indistinguishable-particle and Leibniz versions of the model.  

\end{abstract}

\n PACS: 04.20.Cv, Physics keywords: background independence, inhomogeneity, configuration space, Killing vectors, qualitative dynamics. 
 
\m

\n Mathematics keywords: shapes, spaces of shapes, Shape Geometry, Shape Statistics, Applied Topology, Applied Geometry, 
simple but new applications of Graph Theory to Shape Theory, simple examples of stratified manifolds.

\vspace{0.1in}
  
\n $^*$ Dr.E.Anderson.Maths.Physics@protonmail.com

\vspace{0.1in}

\section{Introduction}\label{Intro-II}

This treatise on the smallest shape spaces continues by considering (4, 1) similarity shapes: formed by 4 points-or-particles in 1-$d$.  
This is the critical model for which many notions -- jointly realized for Part I's (3, 1) model -- first admit distinct realizations. 
I.e.\ collisions, reflection-symmetric configurations, maximal diameter per unit moment of inertia configurations, mergers and various notions of uniformity.
In passing from (3, 1) to (4, 1) the number of qualitative types of shapes moreover goes up from 1 to 5 types to around 20 to 150 types.  
Due to this, the (4, 1) model is already sufficient to have each of the following. 

\m

\n 1) A complex theory of inhomogeneity, uniformity and structure. 
(4, 1) is thus already suitable to model inhomogeneous structure formation and dynamical departure from uniform states, 
which are both cosmologically interesting topics \cite{HH85, Penrose}. 

\m

\n 2) A complex theory of Background Independence 
\cite{A64, A67, BB82, I93, Giu06, FORD, APoT2, APoT3, BI, AMech, AConfig, ABook, Affine-Shape-1, Affine-Shape-2, Project-1, Project-2}, 
and thus (Part I and \cite{APoT2, AKendall, APoT3, ABook}) a realization of the Problem of Time 
\cite{DeWitt67, Battelle, Kuchar81, Kuchar91, Kuchar92, I93, Kuchar99, RovelliBook, KieferBook, APoT1, APoT2, APoT3, ABook}.  

\m 

\n 1) and 2) are foundational topics in General Relativistic and Quantum Gravitational Physics.  
For the (4, 1) model of 1) and 2), Part II's exposition of configuration space topology and geometry is a substantial and crucial first step, 
a  {\it sine qua non} for all of Dynamics, Probability and Statistics and, especially, Quantum Theory thereupon. 
The (4, 1) similarity shape space is a 2-sphere $\mathbb{S}^2$ \cite{Kendall} or some quotient thereof.  
In particular, this is the real projective space $\mathbb{RP}^2$ \cite{Kendall} for the mirror image identified shapes, 
an isosceles spherical triangle covering 1/24th of the sphere for the indistinguishable shapes \cite{AF, FileR}, 
and a scalene spherical triangle covering half of the previous for the mirror image identified {\sl and} indistinguishable particle case \cite{AF, FileR}:  
the {\sl Leibniz space} \cite{I} $\Leib_{\sFrS}(4, 1)$. 
This has comparable fundamental significance to $\Leib_{\sFrS}(3, 2)$, which is Kendall's much vaunted spherical blackboard 
\cite{Kendall84, Kendall89, Small, Kendall} from the Shape Statistics literature \cite{Bhatta, PE16}) for triangle similarity shapes.  
See also e.g.\ \cite{Roach, Watson, JM00} for other approaches to Shape Statistics on spheres.  

\end{titlepage}
 
\n So we study $\Leib_{\sFrS}(4, 1)$ in detail in Part II, and compare it and Kendall's $\Leib_{\sFrS}(3, 2)$ in Part III; a preliminary outline comparison is as follows.  
The (4, 1) shape sphere is tessellated by each of the aforementioned isosceles and scalene spherical triangles \cite{AF}; 
these are two of the more intricate tessellations of the sphere exposited in \cite{Magnus}.       
Some ways in which (4, 1) is more complicated than (3, 2) ensue: 
the (4, 1) tessellation is more complex, and (4, 1) has greater topological, uniformity and merger diversity of shapes.   
Conversely, (3, 2) requires a less trivial (`Hopf') realization \cite{Hopf, Dragt, +Tri, FileR, AConfig, ABook} 
for its shape sphere and also exhibits a number of nontrivial bundle phenomena \cite{III, A-Monopoles} which are not realized in 1-$d$ models.  

\m

\n See also Part IV \cite{IV} for how these two sources of complexity interact to produce 
a substantially greater amount of complexity in the (4, 2) model's theory of quadrilaterals \cite{Wolf-Quad, QuadI, QuadII}.  
In particular, Part II's diversity of notions of inhomogeneity underlies the substantial greater diversity that quadrilaterals, rather than triangles, afford.  
Understanding of quadrilaterals is moreover bounded by how well both (3, 2)'s triangles and (4, 1) models are understood, 
and the latter have hitherto received far less attention than the former.  
Due to the this, it is mostly Part II's advances which unlock {\sl many new} results about quadrilaterals in Part IV and subsequent papers \cite{IV, Affine-Shape-2, Forthcoming}   

\m

\n Study of (4, 1) as a small whole-universe model was initiated by the author and Franzen in \cite{AF} 
in the pure-shape case, and followed up by the author in the scaled case in \cite{Cones, ScaleQM, FileR}.  		 				
In Part II, we begin by considering (4, 1) topological shapes in Sec 2 and metric-level shapes in Sec 3, 
in each case with the corresponding shape spaces: graphs and (pieces of) manifolds respectively. 
These graphs are far larger and considerably more intricate than their (3, 1) counterparts.  
We show that the metric-level Leibniz space is a scalene spherical triangle. 
The double binary and ternary coincidences-or-collisions are two of its corners, 
whereas the interior and exterior binary coincidences-or-collisions are two of its edges.   
In Sec 4, we identify the remaining corner and edge of Leibniz space as the reflection-symmetric configurations and the centred binary coincidence-or-collisions respectively.  
We also identify the (most) uniform configuration and determine the configurations with maximal and minimal mass-weighted diameter per unit moment of inertia.    

\m

\n In Sec 5, we consider a notion of merger and a more general uniformity structure, each of which is formulable in terms of Lagrangian relative separations.
These mergers fold as boundary cusps and furthermore self-intersect in the form of a swallowtail catastrophe \cite{Arnol'd-Cat} in $\FrI\FrS(4, 1)$, 
                                                                            and of a  butterfly  catastrophe \cite{Arnol'd-Cat} in $\Leib_{\sFrS}(4, 1)$. 
We also identify two maximally uniform configurations -- the global such and the only isolated local such -- as centres for these shape spaces.  
We finally describe the qualitative types of (4, 1) shapes, count these out using Appendix B's graph-theoretic technique for all shape spaces, 
and show where these lie relative to each other in Leibniz space.  
In Sec 6 we show that there are further Jacobian notions of merger -- in which one particle is at the centre of mass of the other three --
and update the previous section's qualitative type descriptions, counts and locations in Leibniz space accordingly.
Secs 5 and 6 arrive at more than enough diversity to warrant support by Appendix A's quantifiers of clumping, uniformity and merger.  

\m

\n Sec 7 considers the automorphism groups for both the topological shape space graphs and the metric shape space manifolds, consisting of (similar) Killing vectors. 
$\Leib_{\sFrS}(4, 1)$(s) more irregular shape also leads to greater loss of Killing vectors, 
a crucial observation as regards quantizing the indistinguishable particle version of (4, 1)
The Conclusion (Sec 8) sums up the evidence justifying the (4, 1) similarity shapes as a first nontrivial theory of inhomogeneity and of uniformity. 
It also lists some prices to pay for working in Leibniz space and provides future research directions.

\vspace{10in}

\section{Topological level of structure}\label{Top-(4,1)}

\subsection{Topological shapes}\label{Top-Shapes}

{\bf Proposition 1} There are 6 topological types of (4, 1) shape, as per Fig \ref{(4, 1)-Top-Shapes}.
%
{            \begin{figure}[!ht]
\centering
\includegraphics[width=0.8\textwidth]{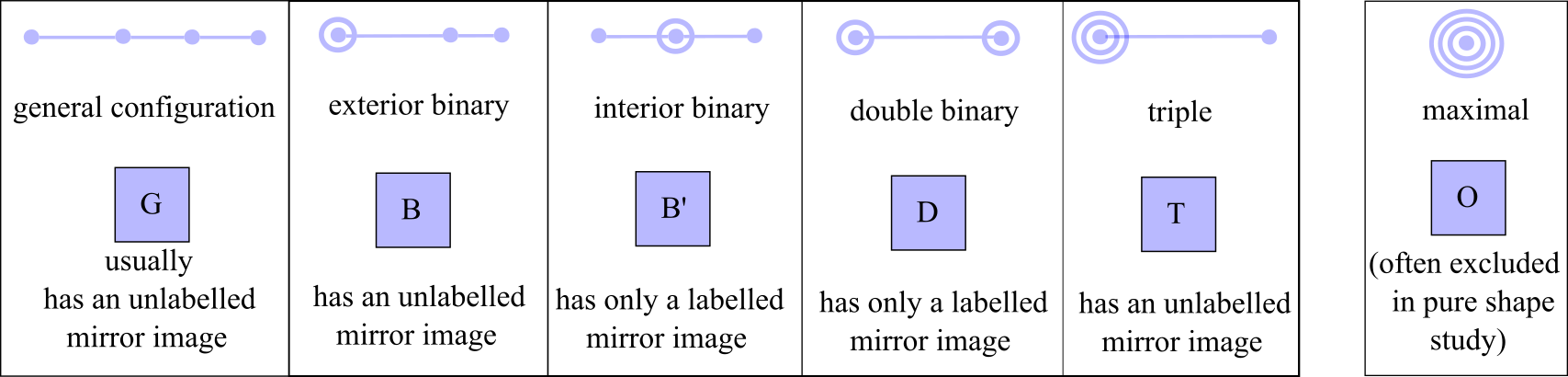}
\caption[Text der im Bilderverzeichnis auftaucht]{        \footnotesize{The 6 topological types of (4, 1) shape: 
generic configuration G, 
exterior and interior binary coincidences-or-collisions denoted by B and B$^{\prime}$ respectively, 
double binary coincidences-or-collisions D, 
ternary coincidences-or-collisions T, 
and the usually excluded maximal coincidence-or-collision O.} }
\label{(4, 1)-Top-Shapes} \end{figure}          }

\m

\n{\bf Remark 1} This contains the first indication that 1-$d$ mirror image identified topological shapes are not just partitions ($N \leq 3$ has just partitions). 
For $N = 4$ the                            $2 \, | \, 1 \, | \, 1$ partition is further fine-grained by the binary coincidences-or-collisions being split into 
the exterior     $\mB$'s                   $1 \, | \, 2 \, | \, 1$ 
and the interior $\mB^{\prime}$'s          $2 \, | \,1 \, | \, 1$. 
This indicates that the topological shapes in 1-$d$ have an additional notion of order in which partitions are realized. 

\m

\n{\bf Remark 2} If mirror images are distinct,      the $2 \, | \, 1 \, | \, 1$ partition 
                                    is distinct from  the $1 \, | \, 1 \, | \, 2$ as well.  
Under these modelling assumptions, however, (3, 1) already suffices to split the binary coincidences-or-collisions into $2 \, | \, 1$ and 
                                                                                                         and $1 \, | \, 2$ partitions.  

\m

\n{\bf Remark 3} For labelled and mirror image distinct (4, 1) shapes,  
\be
\#(\mG) = \left( \mbox{label permutations } \right) = 4 \, ! 
        = 24                                                  \mma \mbox{ and } 
\ee
\be 
\#(\mB) = \left( C(4, 2) \mbox{ choices of pair } \right)       \times 
          \left(  \mbox{ 2 orders for other particles } \right) \times 
		  \left( \mbox{ 2 mirror images } \right) 
        = 24                                                              \m . 
\ee
On the other hand, 
\be 
\#(\mB^{\prime}) = \left( C(4, 2) \mbox{ choices of pair} \right) \times \left(  \mbox{ 2 orders for other particles or 2 mirror images }  \right) 
                 = 12                                                                                                                                               
\ee
since now these two doubling effects are coincident, rather than cumulative, by the topologically symmetrical central positioning of the binary coincidence-or-collision.
Finally 
\be 
\#(\mD) = \left( \m C(4, 2) \mbox{ choices of pair } \right) 
        = 6                                                    \mma \mbox{ and } 
\ee
\be 
\#(\mT) = \left( \mbox{ 4 ways of leaving 1 particle out } \right) \times \left( \mbox{ 2 mirror images } \right) 
        = 8                                                                                                         \m .  
\ee

\subsection{Topological shape spaces}

\n{\bf Proposition 1} For mirror images held to be distinct and distinguishably labelled points, the topological shape space is
\be
\Top\mbox{-}\FrS(4, 1) = ( \mbox{ 74-vertex cube graph } )                                                                                                                \m ,
\ee
labelled as in Fig \ref{S(4, 1)-Top}.0)-1).     

\m

\n{\bf Derivation}. Continuity considerations show that these shapes fit together in the manner of Fig \ref{S(4, 1)-Top}.0), 
which closes up to form Fig \ref{S(4, 1)-Top}.1).  
See Fig {(4, 1)-Graphs}.a) for a full explicit graphic representation (as a planar graph).
%
{            \begin{figure}[!ht]
\centering
\includegraphics[width=1.0\textwidth]{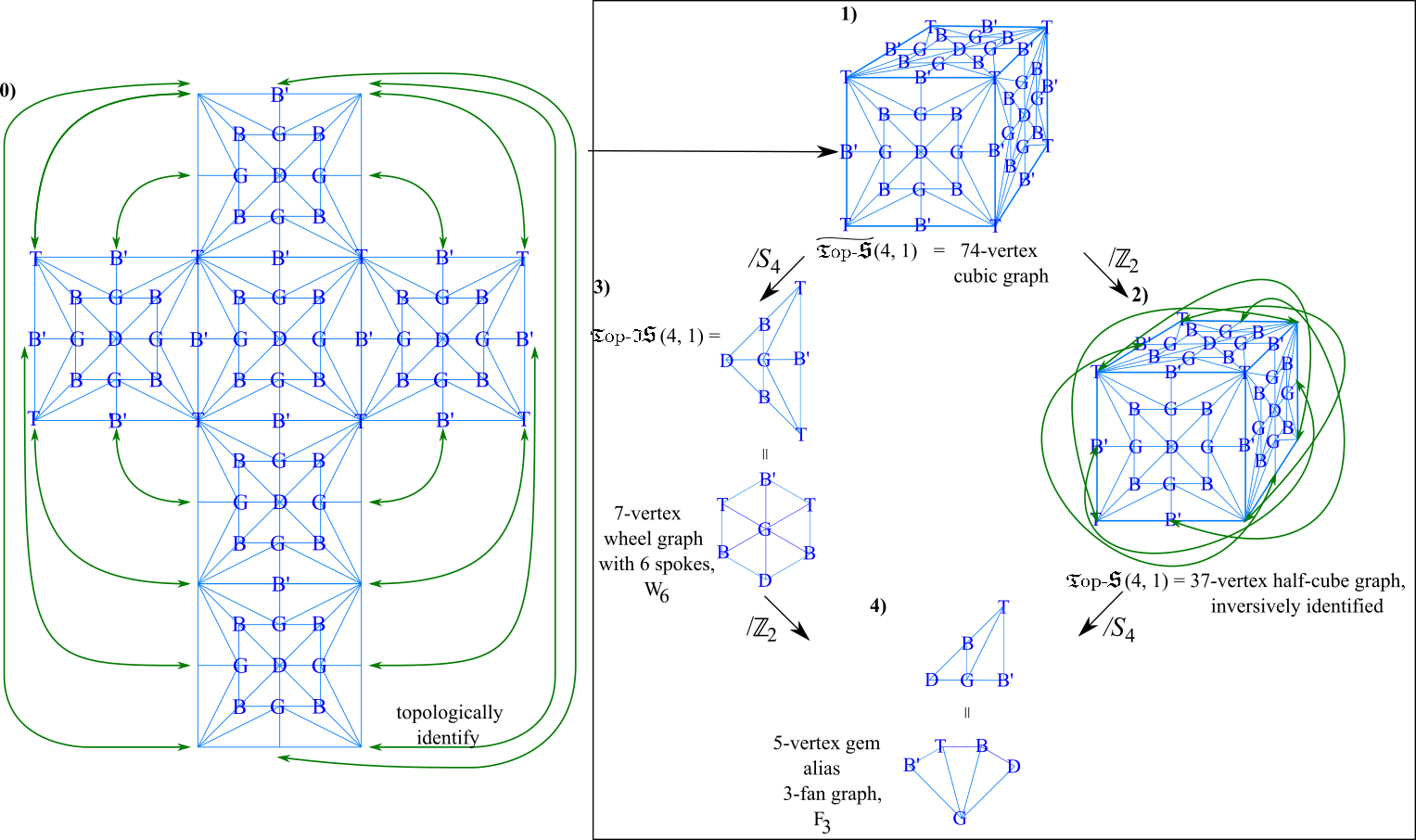}
\caption[Text der im Bilderverzeichnis auftaucht]{        \footnotesize{0) Continuity method for determining the topology of $\Top\mbox{-}\sFrS(4, 1)$. 
to be 1)'s cubic graph with its 74 vertices labeled as shown.     
2) $\Top\mbox{-}\w{\sFrS}(3, 1)$, on the other hand, is the inversively-identified half-cube graph with its 37 vertices labelled as shown.  
3) $\Top\mbox{-}\FrI\sFrS(3, 1)$ is the 7-vertex and thus 6-spoked wheel graph, $\mW_6$, reflection-symmetrically labelled with 5 distinct labels as shown.  
4) Finally $\Top\mbox{-}\Leib_{\tFrS}(3, 1)$ is the 5-vertex gem alias 3-fan graph $\mF_3$, with all labels distinct.} }
\label{S(4, 1)-Top} \end{figure}          }

\m

\n{\bf Proposition 2} If mirror images are now identified, the topological shape space is 
\be 
\Top\mbox{-}\w{\FrS}(4, 1)   \es  \frac{  \Top\mbox{-}\FrS(4, 1)  }{  \mathbb{Z}_2  } 
                             \es  ( \mbox{ 37-point topologically identified half-cube graph } )   \m 
\ee
as labelled in Fig \ref{S(4, 1)-Top}.2), and fully depicted as a graph in Fig \ref{(4, 1)-Graphs}.b).   

\m  

\n{\bf Proposition 3} If indistinguishable points are considered instead while retaining a sense of mirror image distinction, the corresponding shape space is 
\be
\Top\mbox{-}\FrI\FrS(4, 1) \es  \frac{\Top\mbox{-}\FrS(4, 1)}{\mathbb{S}_4} 
                           \es  \mW_6                                                      \m : 
\ee
the 6-spoked wheel graph labelled as per Fig \ref{S(4, 1)-Top}.3).

\m
 
\n{\bf Proposition 4} For indistinguishable points with mirror images identified, the topological-level Leibniz space is 
\be
\Top\mbox{-}\Leib_{\sFrS}(4, 1) \es  \frac{\Top\mbox{-}\FrS(4, 1)}{S_4 \times \mathbb{Z}_2} 
                                \es  \mbox{gem}                                                      \m :
\label{Leib(4, 1)}
\ee
the gem alias 3-fan graph $\mF_3$ with labels all distinct as per Fig \ref{S(4, 1)-Top}.4).
%
{            \begin{figure}[!ht]
\centering
\includegraphics[width=1.0\textwidth]{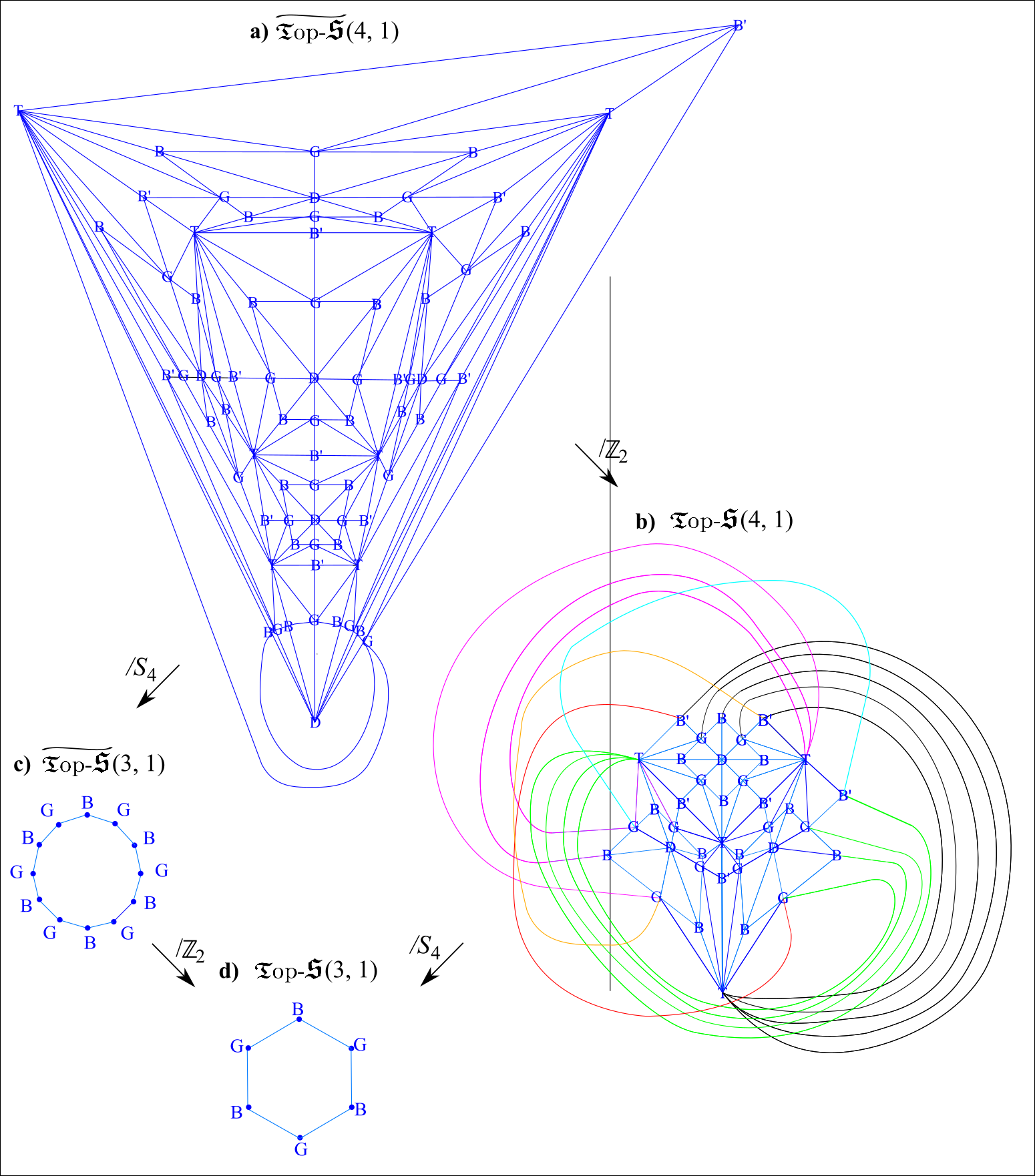}
\caption[Text der im Bilderverzeichnis auftaucht]{        \footnotesize{a) Planar graph representation of the 74-vertex cubic graph.  
b) An actual graph representation of the 37-vertex $\mathbb{RP^2}$-embedded] half-cube graph. 
c) and d) contrast these with their much simpler 3 point shape space counterparts.} }
\label{(4, 1)-Graphs} \end{figure}          }

\m

\n{\bf Remark 1} See Fig \ref{Graph-Count} for vertex, edge, face and vertex degree counts for these graphs, 
along with some simple topological and graph theoretic checks on these.  

\m

{            \begin{figure}[!ht]
\centering
\includegraphics[width=0.95\textwidth]{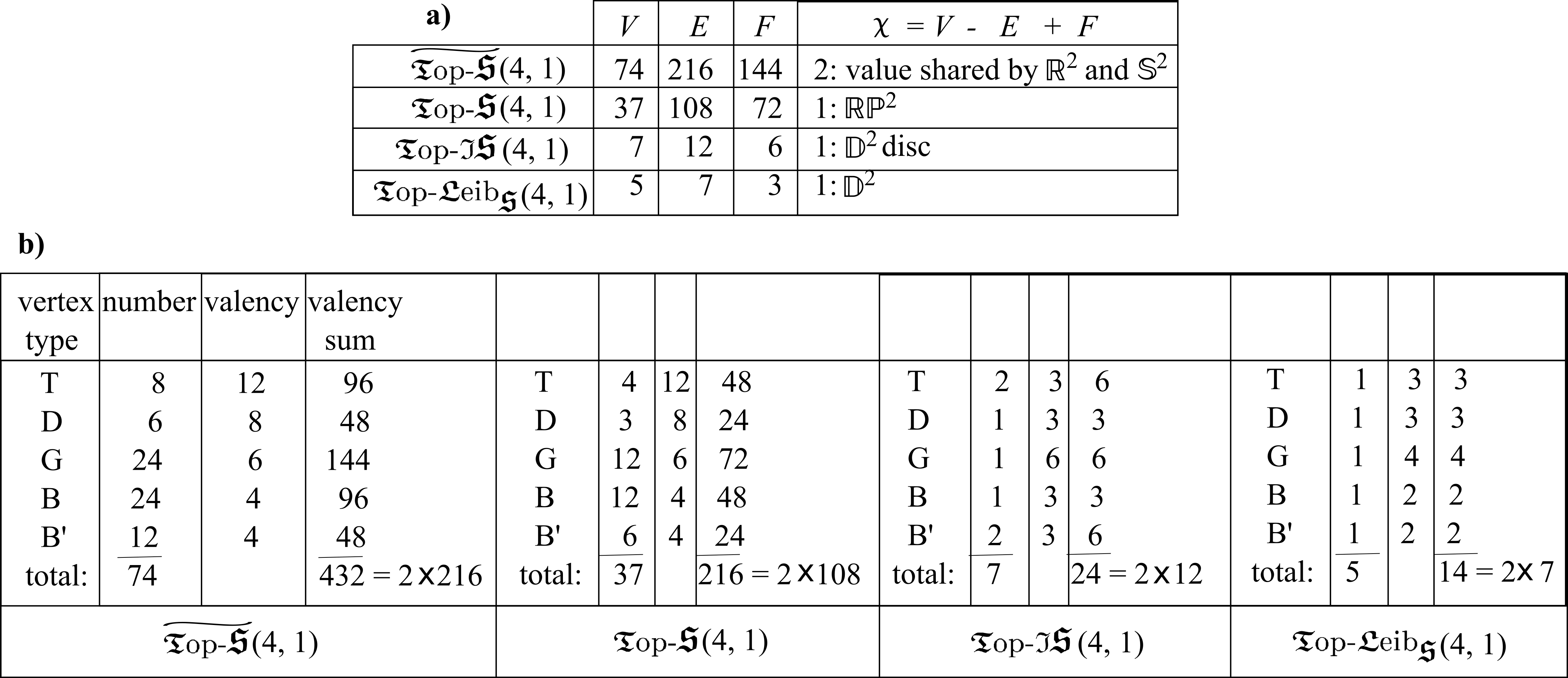}
\caption[Text der im Bilderverzeichnis auftaucht]{        \footnotesize{a) $F$, $E$, and $V$ counts, along with checking the Euler characteristic $\chi$.   
b) Vertex valency table for the topological shape space graphs for 4 points in 1-$d$, including how many of each vertex there are, 
and the basic graph-theoretic check of the first equation of Appendix I.A. } }
\label{Graph-Count} \end{figure}          }

\n{\bf Remark 2} By Remarks 1 and 2 of Sec \ref{Top-Shapes}, for (3, 1) only $\w{\FrS}$ and $\FrI\FrS$ are finer than partitions, 
whereas for (4, 1) all four of $\w{\FrS}$, $\FrS$, $\FrI\FrS$ and $\Leib_{\sFrS}$ are.
By this, Part I's notion of coincidence-or-collision diagram fails to carry enough information in all four cases for $N \geq 4$.  
In particular, $\Leib_{\sFrS}(4, 1)$ is the first Leibniz space which is not just equivalent to some space of partitions.  
This is significant since spaces of partitions constitute a simpler and more commonplace object of study.     

\vspace{11in}

\section{Metric level of structure}\label{Met-(4,1)}

\subsection{Metric shapes}
%
{            \begin{figure}[!ht]
\centering
\includegraphics[width=1.0\textwidth]{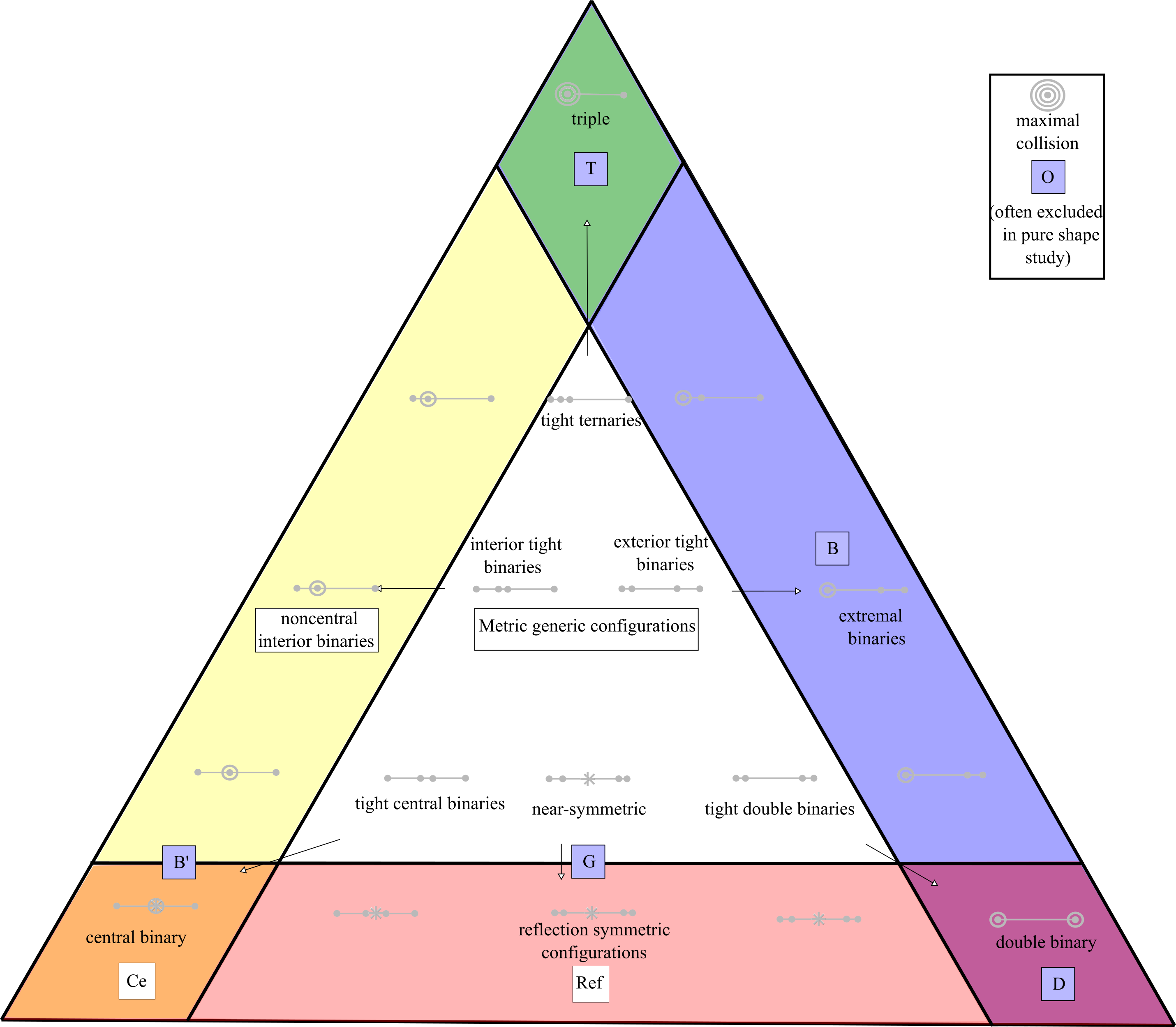}
\caption[Text der im Bilderverzeichnis auftaucht]{        \footnotesize{Some qualitatively different types of metric-level (4, 1) configurations. 
As compared to (3, 1), note the greater variety of 
types of coincidence-or-collision configurations, of notions of merger, 
and consequently of approximately clumped shapes, such as tight binaries, double binaries and ternaries).} }
\label{(4, 1)-Shapes} \end{figure}          }
%
See Fig \ref{(4, 1)-Shapes} for a first indication of (4, 1) supporting a substantial qualitative diversity of shapes.

\subsection{Metric shape spaces and their tessellations}

\n{\bf Proposition 1} The outcome of continuity assignment (Fig \ref{S(4, 1)-Met-Top}.0) for the case of distinguishable particles with mirror images distinct  
is a 2-sphere $\mathbb{S}^2$ decorated by the cube--octahaedron tessellation case of Fig \ref{S(4, 1)-Met-Top}.1) \cite{AF}. 

\m

\n{\bf Proposition 2} For distinguishable particles with mirror images identified, 
this yields instead 2-$d$ real projective space $\mathbb{RP}^2$ decorated with the half-cube--octahaedron tessellation of Fig \ref{S(4, 1)-Met-Top}.2) \cite{FileR}.  

\m

\n{\bf Proposition 3} For indistinguishable particles with mirror images distinct, we have the quarter-face of the cube, 
labelled as per Fig \ref{S(4, 1)-Met-Top}.3). 

\m

\n{\bf Proposition 4} The Leibniz space is the eighth-face of the cube, labelled as per Fig \ref{S(4, 1)-Met-Top}.4).
%
{            \begin{figure}[!ht]
\centering
\includegraphics[width=0.9\textwidth]{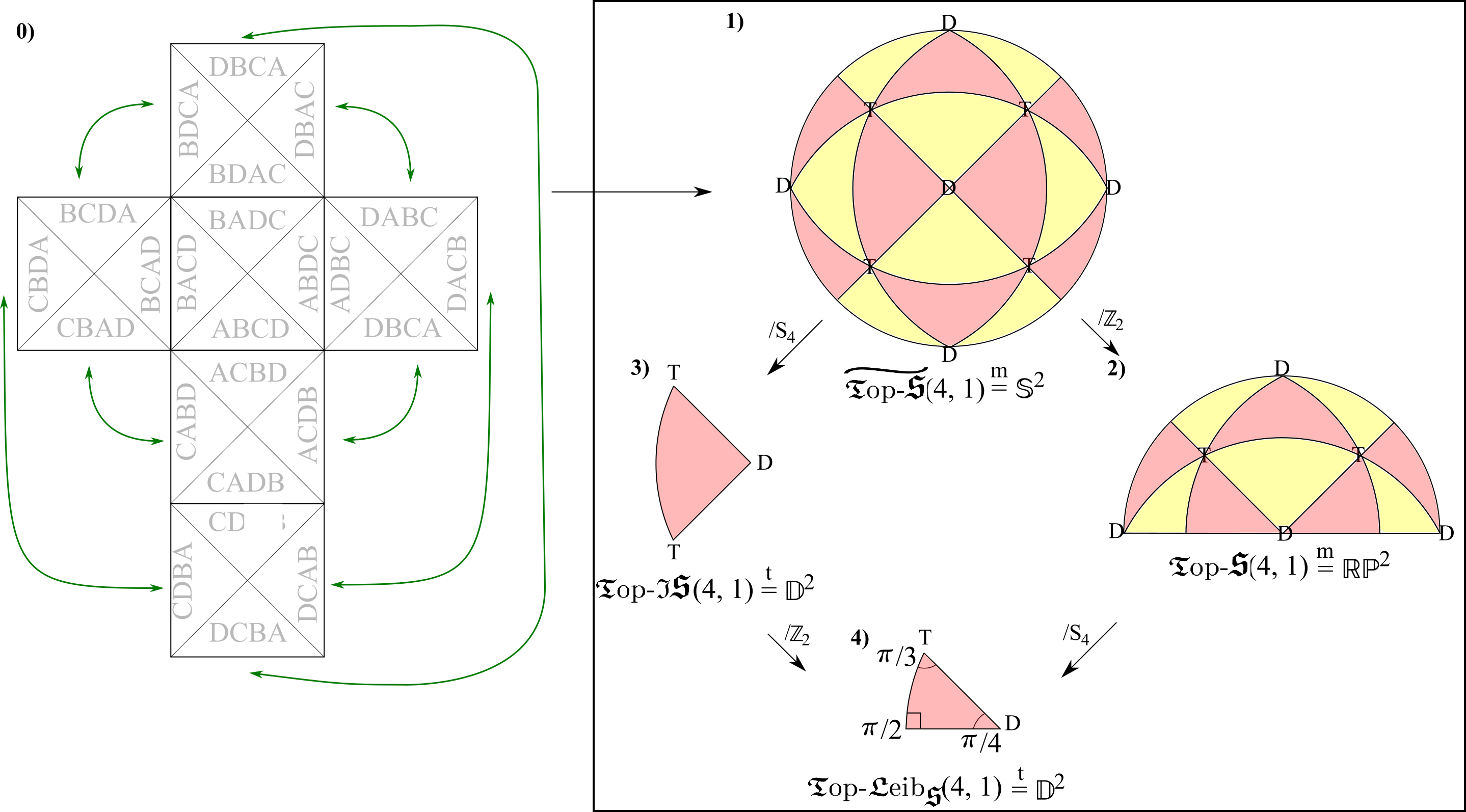}
\caption[Text der im Bilderverzeichnis auftaucht]{        \footnotesize{Topological-level features of metric shape spaces for (4, 1). 
The last two are topologically 2-discs $\mathbb{D}^2$. 
Here $\s{\tm}{=}$ denotes `equals as a manifold' and $\s{\ttt}{=}$ `equals at the topological level'.  
Note that arcs between T's are interior binaries $\mB^{\prime}$, whereas arcs between D and T configurations are exterior binaries.   
As regards the angles in the Leibniz space spherical triangle, its vertices have has valencies 8, 6, and 4, all equally split since we have a tessellation, 
so the angles are $2 \, \pi/8 = \pi/4$, $2 \, \pi/6 = \pi/3$ and $2 \, \pi/4 = \pi/2$.  
The excess angle relative to flat space is $\pi/2 + \pi/3 + \pi/4 - \pi = \pi/12$, so the total excess angle over the sphere is 
(24 tiles) $\times$ (excess angle $\pi/12$ per tile) = $2 \, \pi$, thus passing a basic spherical-geometric test.} }
\label{S(4, 1)-Met-Top} \end{figure}          }

\m

\n{\bf Remark 1} This cube--octahaedron pattern's regularity renders it an example of the following. 

\m 

\n{\bf Definition 1} A {\it tessellation} (or {\it tiling}) is a partition of a space into a number of equal-shaped {\it tile} regions.

\m

\n{\bf Remark 2} Consult \cite{Magnus} for details of the classification of all tessellations of the sphere, 
among which Fig \ref{S(4, 1)-Met-Top}.1)'s cube--octahaedron tessellation plays a prominent role.
This tessellation also appeared previously in Mitchell, Littlejohn and Aquilanti's work \cite{MLA99} on Molecular Physics in a somewhat different context.

\m

\n{\bf Remark 3} Faces, edges and vertices therein being physically significant in the current context,we are a fortiori dealing with {\sl labelled} tessellations.  

\m

\n{\bf Remark 4} In Shape Theory, such tessellations provide a useful `interpretational back-cloth', which facilitates the study of all of  
dynamical trajectories \cite{AF, +Tri, FileR}, 
probability distributions \cite{Kendall89} and 
quantum wavefunctions \cite{AF, +Tri, FileR, QuadII, ABook, QLS, Quantum-Triangles}.   
This method is originally due to Kendall in the context of Shape Statistics \cite{Kendall89}, 
involving the (3, 2) triangles' shape sphere and its corresponding Leibniz space which he termed `spherical blackboard'. 
These were given in Part I's motivational Figures 1 and 2, and further study of them is the main focus of Part III and \cite{A-Pillow, Max-Angle-Flow, A-Perimeter}.  

\m

\n{\bf Remark 5} In Shape Theory, the most prominent tiles one tessellates with have further significance as Leibniz space $\Leib_{\sFrS}$ and its 
mirror images distance double $\FrI\FrS$. 
Fig \ref{S(4, 1)-Met-Top}.a) tessellates with 24 $\FrI\FrS(4, 1)$ tiles, which are isosceles spherical triangle tiles as per Fig \ref{S(4, 1)-Met-Top}.c).  
On the other hand, Fig \ref{S(4, 1)-Met-Met}.a) tessellates with 48 $\Leib_{\sFrS}(4, 1)$ tiles, 
which are scalene spherical triangle tiles as per Fig \ref{S(4, 1)-Met-Top}.d)

\subsection{Jacobi H-coordinates in spherical polar form}\label{Jacobi-H}

The current treatise makes good use of the following spherical polar reformulation of Sec I.3.3's relative Jacobi H-coordinates
\beq
\rho_1 = \rho \, \mbox{sin} \, \theta \, \mbox{cos} \, \phi   \m ,
\label{(4, 1)-rho-1}
\eeq 
\beq
\rho_2 = \rho \, \mbox{sin} \, \theta \, \mbox{sin} \, \phi   \m ,
\label{(4, 1)-rho-2}
\eeq 
\beq
\rho_3 = \rho \, \mbox{cos} \, \theta                         \m .   
\label{(4, 1)-rho-3}
\eeq
These invert to 
\beq
\phi \es \mbox{arctan}
\left(
\frac{\rho_2}{\rho_1}
\right)                                                       \m ,
\eeq
\beq
\theta \es \mbox{arctan}
\left(
\frac{\sqrt{\rho_1\mbox{}^2 + \rho_2\mbox{}^2}}{\rho_3}
\right)                                                       \m ,
\eeq
\beq
\rho = \sqrt{\rho_1\mbox{}^2 + \rho_2\mbox{}^2 + \rho_3\mbox{}^2}                                                     \m .  
\eeq
\n{\bf Remark 1} For now, setting $\rho = constant$, one has coordinates on the shape sphere $\FrS(4, 1)$.
$\theta$ and $\phi$ are then geometrically standard spherical coordinates, with standard coordinate ranges $\theta \in (0, \pi)$, $\phi \in [0, 2 \, \pi)$ in the $\FrS(4, 1)$ case.
 
\m 
 
\n{\bf Remark 2} These have moreover now acquired shape-theoretic significance as functions of ratios, as follows.  

\m

\n 1) $\phi$ is a function of the ratio of sizes of the two binaries picked out by the underlying Jacobi H-clustering.
This is a measure of {\it contents inhomogeneity} \cite{AF}; this refers to viewing the (4, 1) model as containing two binaries. 
Then if these two binaries are equal in size one has contents homogeneity, whereas contents inhomogeneity quantifies the amount of departure from such equality.  

\m

\n 2) On the other hand, $\theta$ is the ratio of the two clusters together to the relative separation between them.  
This is a measure of {\it fractional size occupied by the contents}: what proportion of the system is occupied by the clusters.  
[The cosmological analogue of this is {\sl fractional volume}.]

\m

\n{\bf Application 1 of Jacobi H-coordinates} In terms of the coordinates introduced above, $\w{\FrS}(4, 1)$ corresponds to $\theta \in (0, \pi/2]$.
On the other hand, 
characterizing $\FrI\FrS(4, 1)$ and $\Leib_{\sFrS}(4, 1)$ has two boundary pieces of constant polar angle and one of constant azimuthal angle about a new axis.  
We return to this matter in more detail in Sec \ref{Leib-I-Coords}.

\subsection{Sub-shape-space structure}

\n{\bf Application 1 of Shape-Theoretic Aufbau Principle} (Sec I.1). 
A number of $\FrS(3, 1)$'s can be spotted within $\FrS(4, 1)$.  
For instance the DTTDTT sequences along great circles in Fig \ref{S(4, 1)-Met-Top}.1).
This identification clearly corresponds to picking out 3-subcluster in the (4, 1) model and treating it as a (3, 1) model.  

\m

\n{\bf Application 2 of Jacobi H-coordinates} is determination of points and curves of special configurations within shape space.  
The defining relation for each special configuration in terms of $\rho_i$ is converted to either a fixing of $\theta$, $\phi$ coordinates for points 
or a relation between $\theta$ and $\phi$ for curves. 
For instance in the face in which the point labels are ordered 1234,  
\be
\mB \m \mbox{ is the curve } \m \phi = 0   \m .  
\ee 
\be
\mB^{\prime} \m \mbox{ is the curve } \m \mbox{sin}\,\phi +  \mbox{cos}\,\phi = 2 \, \mbox{cot} \,\theta   \m .  
\ee

\subsection{Detail of the special (4, 1) metric shapes encountered so far}
%
{            \begin{figure}[!ht]
\centering
\includegraphics[width=1.0\textwidth]{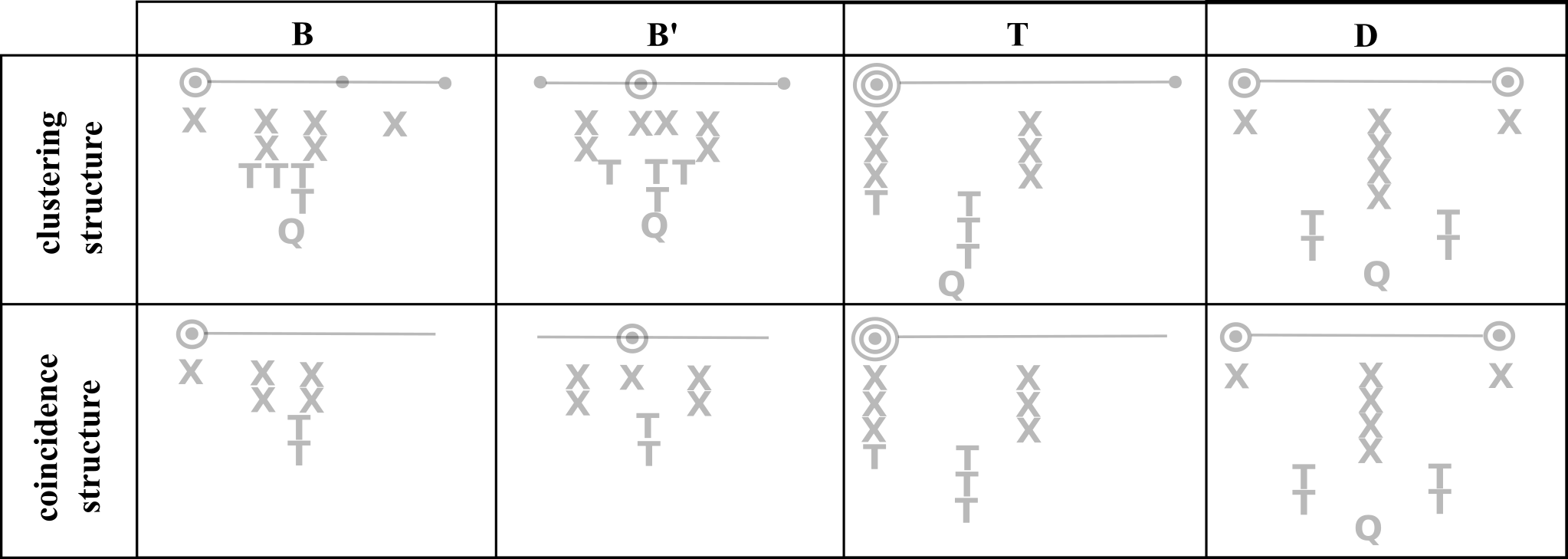}
\caption[Text der im Bilderverzeichnis auftaucht]{        \footnotesize{T, D, B and B$^{\prime}$ configurations' clustering  structure and 
                                                                                                                 coincidence-or-collision structure. 
[While further such figures of distinguished shapes include uniformity structure, this is trivial for each of the current figure's four cases.]} }
\label{T-D-B-B'} \end{figure}          }

\n This is given in Fig \ref{T-D-B-B'} using various of Part I's diagrammatic presentation concepts.

\section{On the nature of $\Leib_{\sFrS}(4, 1)$'s hitherto unlabelled edge and vertex}\label{Ref-Ce-Sec}
%
{            \begin{figure}[!ht]
\centering
\includegraphics[width=0.5\textwidth]{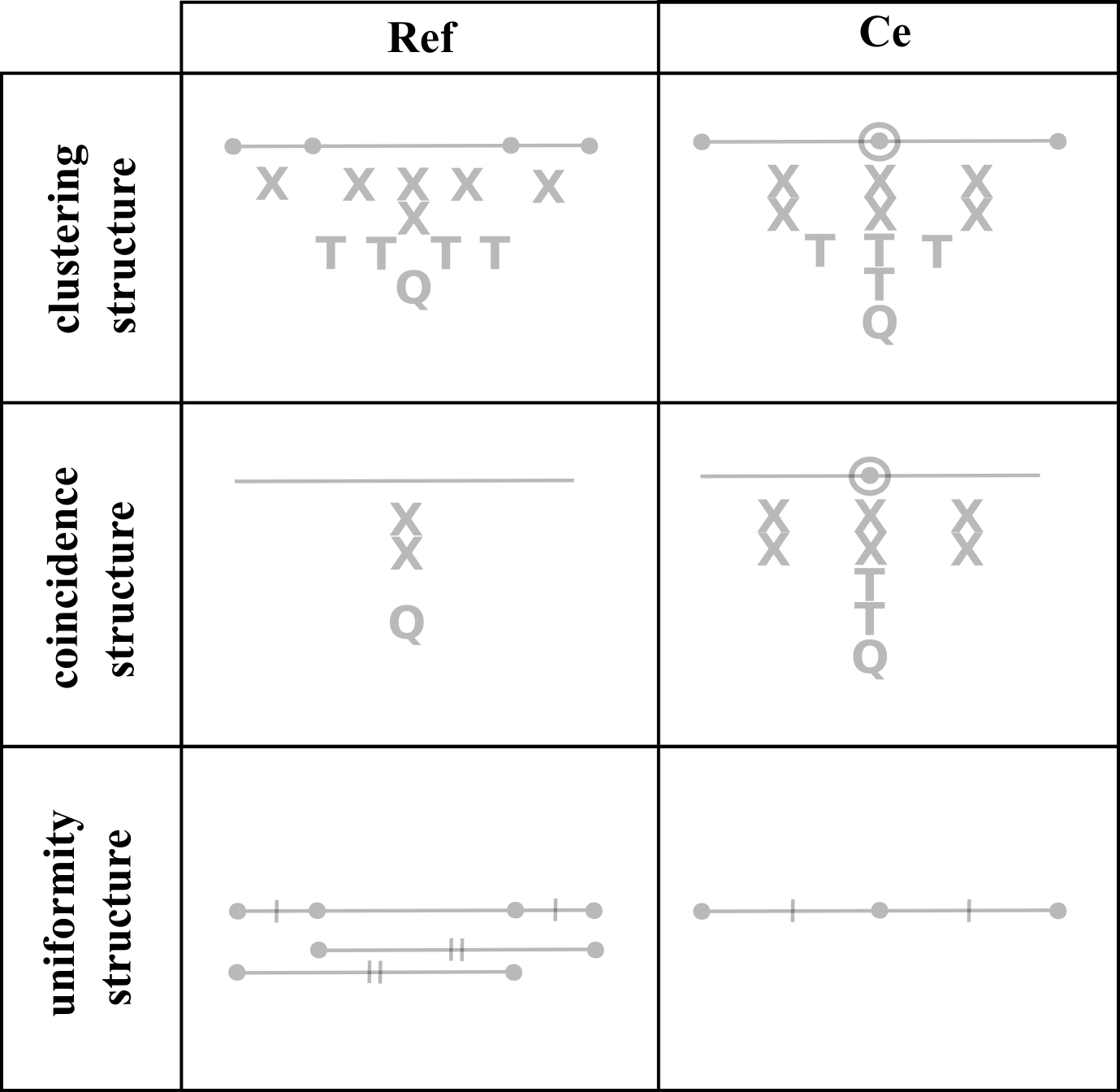}
\caption[Text der im Bilderverzeichnis auftaucht]{        \footnotesize{Ce and Ref configurations' clustering, coincidence-or-collision structure, and uniformity structures. 
The small perpendicular lines indicate equal-length pairs.} }
\label{Ref-Ce} \end{figure}          }

\n We next turn to $\Leib_{\sFrS}(4, 1)$ 's vertex and edge of an as-yet unidentified nature (these had no topological-level meaning).    

\m

\n{\bf Remark-and-Definition 1} The shapes lying on the other side of this edge in $\FrI\FrS(4, 1)$ are mirror images in space.
They are furthermore at mirror image positions in shape space with the unidentified edge playing the `line of reflection' role. 
This edge itself corresponds to shapes which are {\it reflection-symmetric} in space, by which we choose the notation Ref for this edge.  

\m

\n{\bf Proposition 1} Ref is an arc of a great circle in the shape space sphere.  

\m

\n{\bf Derivation} Each great circle in a sphere is the intersection of a sphere with a plane through its origin. 
The plane corresponding to the great circle that Ref is an arc of plays the role of plane of reflection in the above construction. $\Box$

\m

\n{\bf Remark 1} In contrast, (3, 1) had just one reflection-symmetric shape -- also denoted Ref -- i.e.\ a point rather than an arc in the corresponding Leibniz space. 

\m

\n{\bf Characterization 1} Reflection-symmetric (4, 1) shapes admit the further useful and memorable characterization that the binaries to either end of the shape are of equal size. 
In other words, these are {\it contents-homogeneous} shapes.  
This refers moreover exactly to the contents that the corresponding Jacobi H-coordinates focus on.

\m

\n{\bf Remark 2} The Ref curve has the double-binary D at one end.
D clearly enjoys (column 4 of Fig \ref{T-D-B-B'})the Ref property and its merger consequences (Fig \ref{Ref-Ce}.a), 
alongside a number of further coincidence-or-collision and merger properties.  

\m

\n In $\FrI\FrS(4, 1)$, D is most naturally thought of as the obvious confluence of two B's.

\m

\n On the other hand, in $\Leib_{\sFrS}(4, 1)$, D has another characterization: as the confluence of just one B with Ref's symmetry property. 
This reflection symmetry property then has the consequence of forcing a doubling of the number of B's present, from 1 to 2.  

\m 

\n{\bf Remark-and-Definition 2} It is then natural to ask which shape lies at Ref's other endpoint.  
This shape also lies on $\mB^{\prime}$, which uniquely fixes it to have the form given in column 2 of Fig \ref{Ref-Ce}.  
We call this the {\it centred binary coincidence-or-collision}, and denote it by Ce.  
It is the particular interior binary coincidence-or-collision $\mB^{\prime}$ whose binary coincidence-or-collision $ab$ is additionally at CoM($cd$).

\m

\n{\bf Remark 3} See Column 2 of Fig \ref{Ref-Ce} for Ce's further clustering structure.     
This is again in excess of that of the two $\Leib_{\sFrS}(4, 1)$ edges which meet there, $\mB^{\prime}$ and Ref.  

\m

\n Contrast also with T's clustering structure in Column 3 of Fig \ref{T-D-B-B'}), 
which is also in excess of that of the two $\Leib_{\sFrS}(4, 1)$ edges meeting there: now $\mB^{\prime}$ and $\mB$.

\subsection{Tessellation of the shape sphere by Leibniz space tiles}
%
{            \begin{figure}[!ht]
\centering
\includegraphics[width=0.6\textwidth]{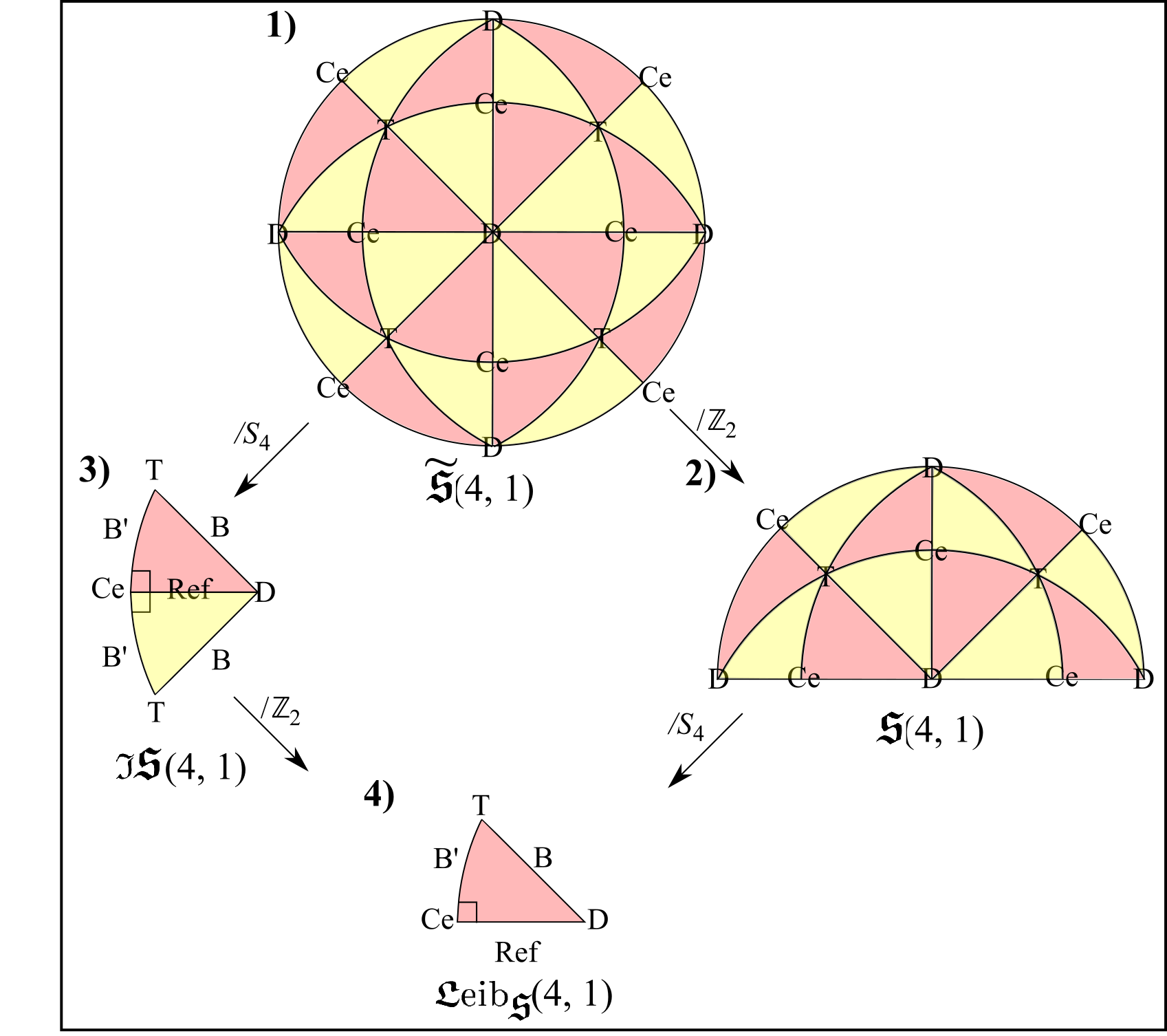}
\caption[Text der im Bilderverzeichnis auftaucht]{        \footnotesize{Qualitatively different types of metric configuration for (4, 1).
This exhibits a greater variety of types of coincidence-or-collision configuration, of notions of merger, 
and consequently of near-coincidence-or-collision clumping (tight binaries and ternaries).} }
\label{S(4, 1)-Met-Met} \end{figure}          }

\n Having obtained the Ref arcs and Ce points from metric-level considerations, we understand the entirety of the boundaries and corners of Leibniz space. 
We celebrate by presenting the aforementioned shape space tessellations by Leibniz space tiles in Fig \ref{S(4, 1)-Met-Met}.

\subsection{Extrinsic and relational space considerations}

\n{\bf Application 3 of Jacobi H-coordinates} 
The $\mathbb{\rho}_i$, $i = 1$, to $3$  furthermore form a Cartesian axis system for an ambient $\mathbb{R}^3$ that the $\FrS(4, 1)$ sphere sits in.

\m 

\n{\bf Remark 1} The North pole $\theta = 0$ is at one of the D's. 
The corresponding principal axis is then a {\it face diagonal} of the cube: 
a line from the centre of one face through the centre of the cube to the opposite face's centre.  
In the current shape-theoretic context, the South pole coincides with a D shape as well (Fig \ref{D--D-Axis-System}.0).  
This axis corresponds setting $\rho_3 = 0$: this D-axis is perpendicular to the plane of zero crossbar (Fig \ref{D--D-Axis-System}.2).

\m

\n Setting $\rho_1 = 0$ and $\rho_2 = 0$ instead gives in each case an axis with a Ce at either end.
Thus, since all three axes are perpendicular, we have overall a Ce--Ce--D axis system. 
These other planes' significance are zero left post and zero right post (Fig  \ref{D--D-Axis-System}.3-4), which are cluster-specific 
subcases of binary coincidences-or-collisions.

\m

\n{\bf Remark 2} One can moreover rotate the coordinate system by $\pi/4$ about the principal axis,  -- 
\be
\phi \longrightarrow \phi - \frac{\pi}{4} \es  (\mbox{ redefined } \phi \m ) \m , 
\ee
so as to to take advantage of the D--D lines between opposite pairs of faces all being mutually perpendicular (Fig  \ref{D--D-Axis-System}.1) 
and hence eligible as a Cartesian axis system D--D--D.  
%
{            \begin{figure}[!ht]
\centering
\includegraphics[width=0.8\textwidth]{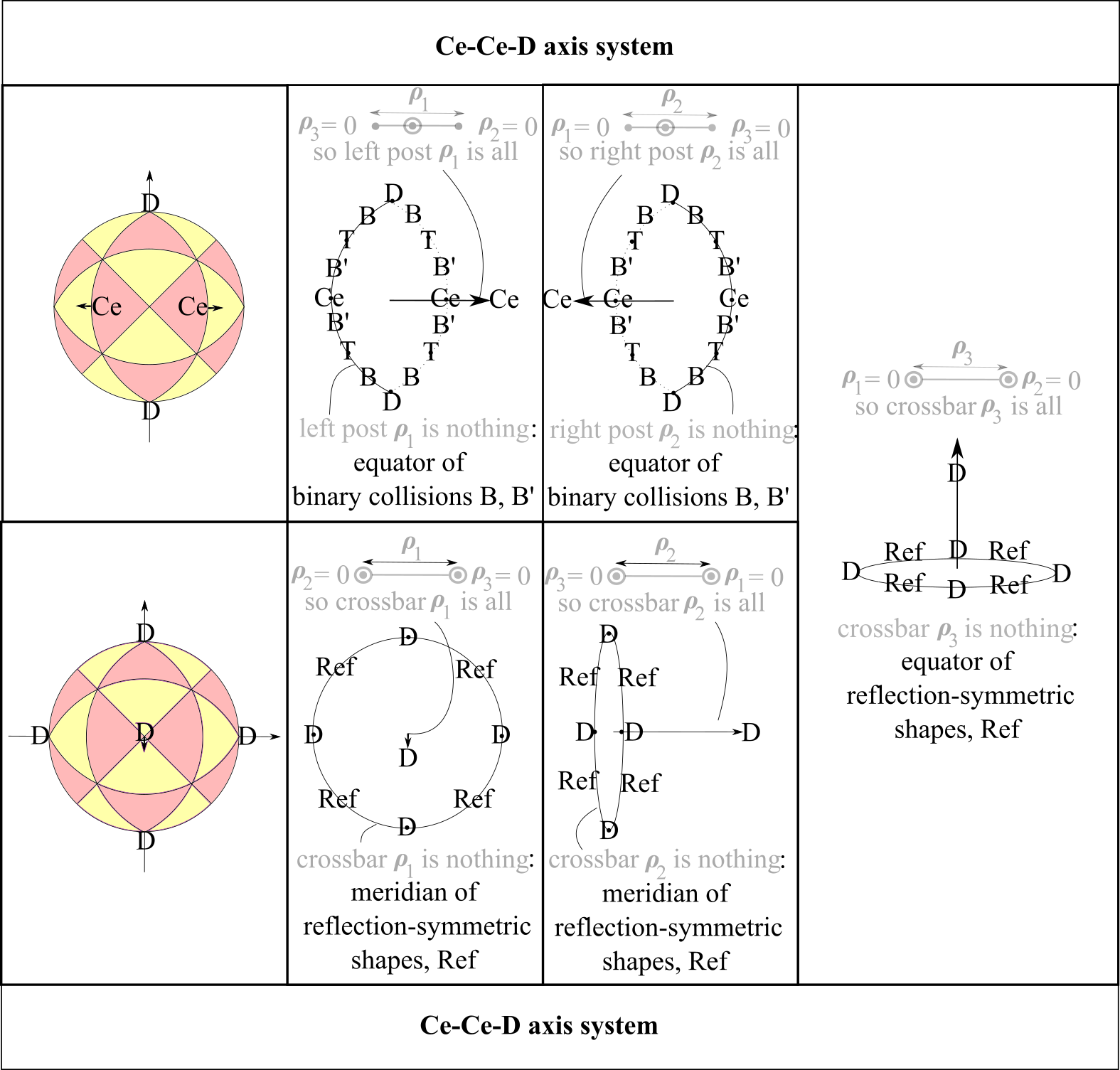}
\caption[Text der im Bilderverzeichnis auftaucht]{        \footnotesize{D--D axis system and interpretation of plane perpendicular to each axis.} }
\label{D--D-Axis-System} \end{figure}          }

\m

\n{\bf Remark 3} This ambient $\mathbb{R}^3$ is moreover ${\cal R}(4, 1)$, 
since the cone over the shape space sphere $\FrS(4, 1)$ returns $\mathbb{R}^3$ both topologically and metrically.  

\m

\n{\bf Proposition 1} The corresponding topological notions of scaled shapes are furthermore 
the cones over the corresponding topological shape space in Fig \ref{S(4, 1)-Top}.  
I depict the cones over $\mW_6$, gem, $\FrI\FrS(4, 1)$ and $\Leib_{\sFrS}(4, 1)$ in Fig \ref{R(4, 1)}, 
along with identifying which graphs the first of these two are.   
The cones over the 74-point cubic graph, 
                   37-point inversively identified cubic graph, 
				   cubic tessellation of the sphere 
			   and half-cubic tessellation of $\mathbb{RP}^2$ are also straightforward to envisage.     			   
%
{            \begin{figure}[!ht]
\centering
\includegraphics[width=1.0\textwidth]{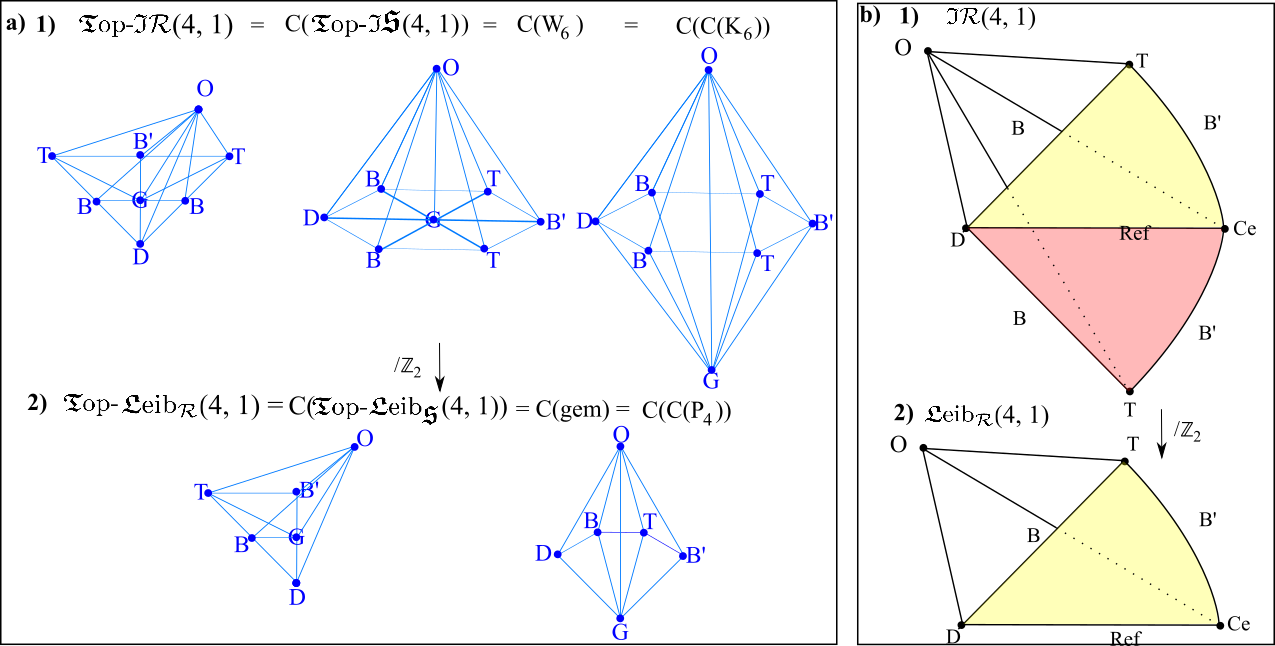}
\caption[Text der im Bilderverzeichnis auftaucht]{        \footnotesize{Metric and topological versions of $\FrI\FrS(4, 1)$ and $\Leib_{\tFrS}(4, 1)$.} }
\label{R(4, 1)} \end{figure}          }

\m

\m

\m

\subsection{Coordinatization of $\FrI\FrS(4, 1)$ and $\Leib_{\sFrS}(4, 1)$}\label{Leib-I-Coords}
%
We can now also give the polar and off-axis azimuthal coordinates for the sides of $\FrI\FrS(4, 1)$ and $\Leib_{\sFrS}(4, 1)$ in Fig \ref{Displaced-Coord}.  
%
{            \begin{figure}[!ht]
\centering
\includegraphics[width=0.5\textwidth]{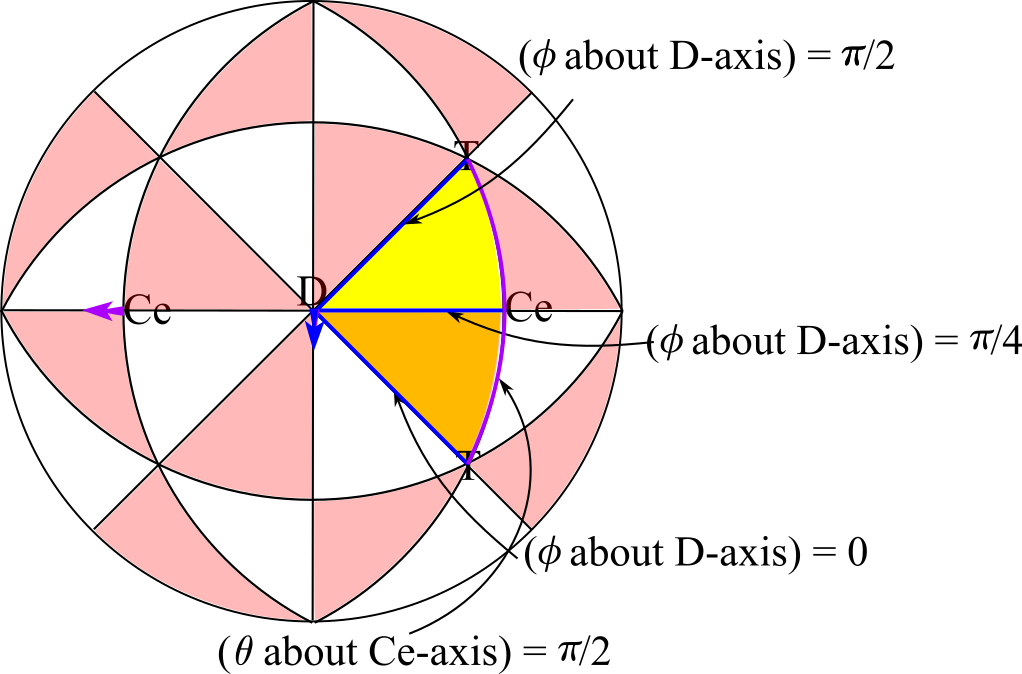}
\caption[Text der im Bilderverzeichnis auftaucht]{        \footnotesize{Coordinate description of the boundaries of $\FrI\FrS(4, 1)$ and $\Leib_{\sFrS}(4, 1)$.} }
\label{Displaced-Coord} \end{figure}          }

\subsection{(4, 1) coincidence-or-collision space}
%
{            \begin{figure}[!ht]
\centering
\includegraphics[width=0.75\textwidth]{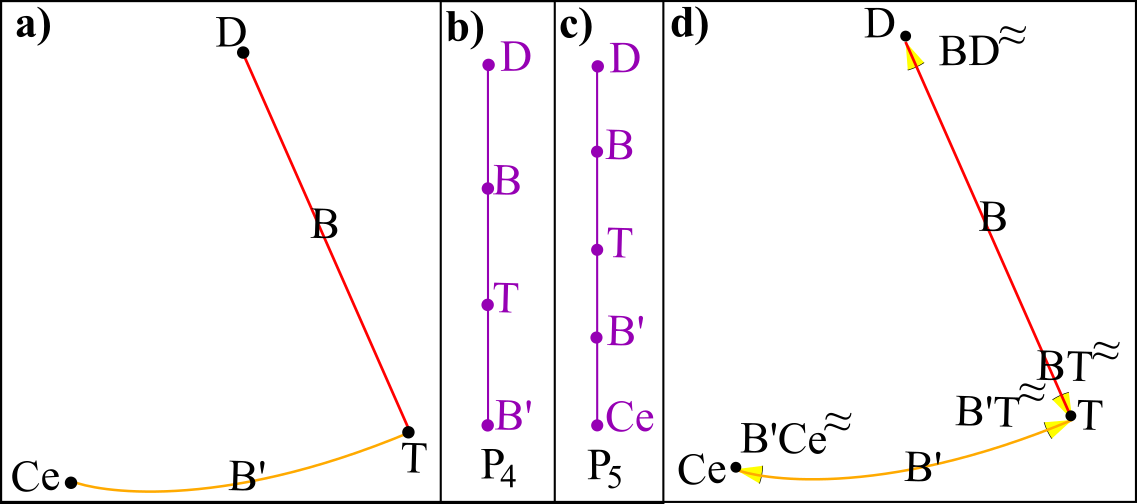}
\caption[Text der im Bilderverzeichnis auftaucht]{        \footnotesize{Coincidence-or-collision structure $\Co(\Leib_{\sFrS}(4, 1))$  
a) at the metric-level, 
b) as a topological-level adjacency graph, 
c) the metric-level struture's own topological adjacency graph, and 
d) the metric-level version including approximate notions.} }
\label{Collision-Set} \end{figure}          }

\n{\bf Proposition 1} The metric-level coincidence-or-collision structure $\Co(\Leib_{\sFrS}(4, 1))$ is as given in Fig \ref{Collision-Set}.a).   

\m

\n{\bf Proposition 2} At the topological level, the corresponding adjacency graph is the 4-path graph $P_4$ 
                                                                                                                labelled as per Fig \ref{Collision-Set}.b). 
This already featuring in a previous section represents that at the topological level cooincidences-or-collisions are all, so $\Co(\FrG\FrS) = \Top-\FrG\FrS$.  																												

\m

\n{\bf Proposition 3} The topological adjacency graph for the metric level of structure of the collision structure for (4, 1) shapes 
is instead the 5-path graph $P_5$ labelled as per Fig \ref{Collision-Set}.c). 

\m

\n These last two graphs two differ because Ce has no additional topological significance at the primary space level.  

\m

\n{\bf Proposition 4} There are 4 types of approximate coincidences-or-collisions at the metric level. 
These are as indicated in yellow on Fig \ref{Collision-Set}.d). 
Thus there is a 
\be
\mbox{grand total of 5 + 4 = 9 qualitative types of coincidence-or-collision at the metric level } .
\ee  
\n{\bf Remark 1} The nomenclature for the approximate types is along the following lines. 

\m

\n 1) BD$^{\approx}$ means `a B which is approximately a D', i.e.\ a binary coincidence-or-collision and a separate tight binary.  

\m

\n 2) BT$^{\approx}$ means `a B which is approximately a T', i.e.\ a tight ternary with includes now an explicitly exterior binary coincidence-or-collision.  

\m

\n 3) B$^{\prime}$T$^{\approx}$ means `a B$^{\prime}$ which is approximately a T', 
i.e.\ a tight ternary with includes now an explicitly interior binary coincidence-or-collision.  

\m

\n 4) Finally B$^{\prime}$Ce$^{\approx}$ means `a B$^{\prime}$ which is approximately a Ce', i.e. an almost-centred interior binary coincidence-or-collision.  

\m

\n This nomenclature can and will be built up for many further kinds of approximate shapes.  

\subsection{The maximally uniform shape}

\n{\bf Definition 1} The {\it (maximally) uniform shape} for (4, 1) is the one with equally-spaced adjacent points, 
\be
r_{12}   =   r_{23} 
         =   r_{34} 
	   \neq  0  
\label{r-adj}
\ee
in the 1234-ordered case.
We denote this by $\mU(4, 1)$, or simply by U when no confusion arises.  

\m

\n{\bf Remark 1} This is an opposite extreme to maximal clumpiness, which for (4, 1) is T (or O if allowed).

\m

\n{\bf Remark 2} Compare (\ref{r-adj}) with the separation coincidences of D, 
\be
r_{12}   =    0 
         =    r_{34}  \mma 
r_{13}   =    r_{23} 
         =    r_{14} 
	     =    r_{24} 
	   \neq   0       \m , 
\ee
and T: 
\be
r_{12}   =   r_{13} 
         =   r_{23}       \mma 
r_{14}   =   r_{24} 
         =   r_{34} 
	   \neq   0           \m . 
\ee
\n{\bf Remark 3} The uniform shape U enjoys the clustering coincidences indicated in column 2 of Fig \ref{U-CeA-UA}.a).
%
{            \begin{figure}[!ht]
\centering
\includegraphics[width=1.0\textwidth]{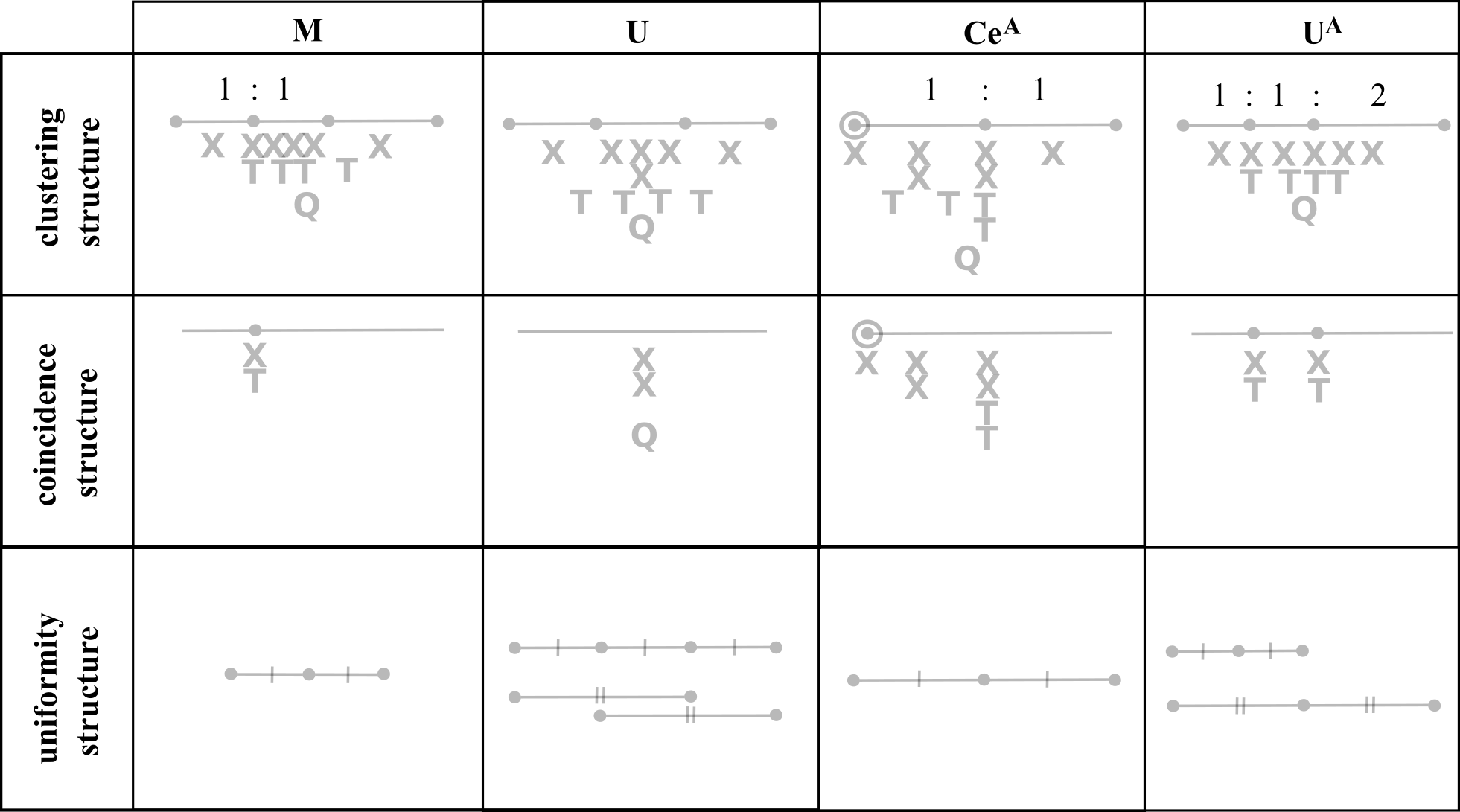}
\caption[Text der im Bilderverzeichnis auftaucht]{        \footnotesize{M, U, Ce$^{\tA}$ and U$^{\tA}$ configurations' clustering  structure, 
                                                                                                                       coincidence-or-collision structure 
																												    and uniformity structure.     } }
\label{U-CeA-UA} \end{figure}          }

\subsection{Model diameter per unit moment of inertia maximizing and minimizing shapes}\label{cal L-(4, 1)}
%
{            \begin{figure}[!ht]
\centering
\includegraphics[width=0.2\textwidth]{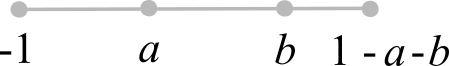}
\caption[Text der im Bilderverzeichnis auftaucht]{        \footnotesize{General (4, 1) configuration centred about $\fQ$ at 0;   
                                                                        we subsequently normalize this configuration using the total moment of inertia.     } }
\label{(4, 1)-General-Shape} \end{figure}          }

\n In Fig \ref{(4, 1)-General-Shape}'s parametrization, the total moment of inertia is 
\beq
\bigiota = 2\{1 + a^2 + b^2 - a - b - a b\}  \m , 
\eeq
Various possible numerators for $\diam$ (I.260) are then 
\beq
\{2 - a - b\}^2        \mma 
\{1 + b\}^2            \mma
\{a - b\}^2            \mma
\{1 - b - 2 \, a\}^2   \m .  
\eeq
\n While 
\be
{\cal D}_{\sm\sa\sx} = \mU 
                     = \mbox{Ref}   \m , 
\label{(3, 1)-Coincidence}
\ee 
for (3, 1) subsequent generalization is to the configuration with a minimal amount of mass on either side with the rest concentrated at $\fO$.  
This description indeed returns U(3, 1) for (3, 1), but returns Ce and not U(4, 1) for (4, 1).
Thus for (4, 1) the (3, 1) coincidence-or-collision (\ref{(3, 1)-Coincidence}) is replaced by 
\be
\diam_{\sm\sa\sx}  \subset \mbox{Ref} \supset \mU \m .  
\ee
So the $\mbox{Ref}$ arc contains all of these configurations, but each of $\diam_{\sm\sa\sx}$ and $\mU$ further single out a distinct point, 
with $\diam_{\sm\sa\sx}$ being at one end of the $\mbox{Ref}$ arc and $\mU$ an interior point of this arc.  
%
{            \begin{figure}[!ht]
\centering
\includegraphics[width=0.4\textwidth]{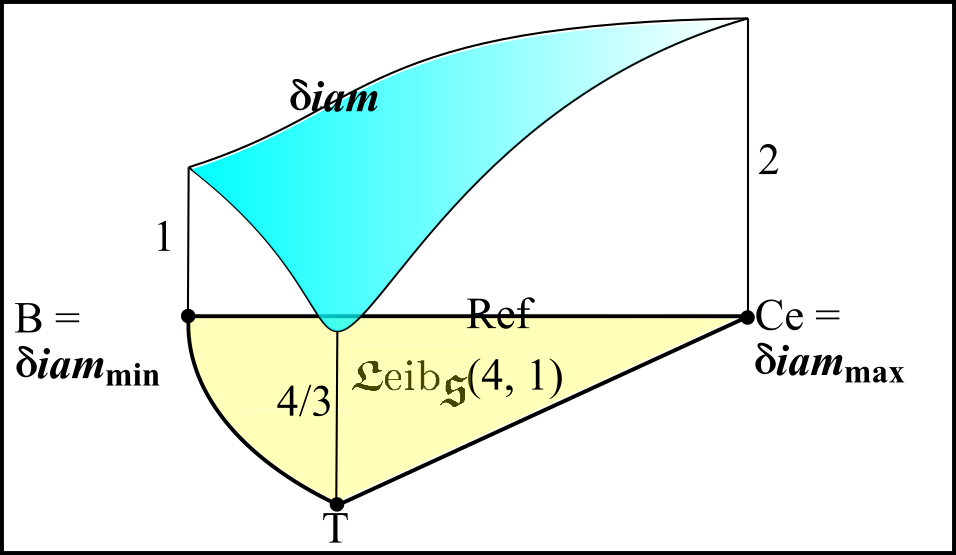}
\caption[Text der im Bilderverzeichnis auftaucht]{        \footnotesize{(4, 1)'s mass-weighted diameter squared per unit moment of inertia.} }
\label{Max-Min-(4, 1)} \end{figure}          }

\m

\n Furthermore, for (4, 1) 
\be 
\diam_{\sm\si\sn}  = \mD
\ee 
-- another corner of $\Leib_{\sFrS}(4, 1)$ -- with the final corner T also being minimal, but merely locally, as sketched in Fig \ref{Max-Min-(4, 1)}.
Thus the symmetric minimality -- corresponding to half the matter being at each end of the model universe: $\mD$ in (4, 1) -- is globally minimal. 
On the other hand, the asymmetric minimality -- corresponding to one particle at one end and all the others at the other end: $\mT$ in (4, 1) -- is local.

\subsection{Summary and qualitative type count of shapes encountered so far}\label{Plain-Q}

See Fig \ref{Leib(4, 1)-Plain} for a summary of the qualitative type names and their topological, geometrical and proximity relations in Leibniz space. 
%
{            \begin{figure}[!ht]
\centering
\includegraphics[width=0.87\textwidth]{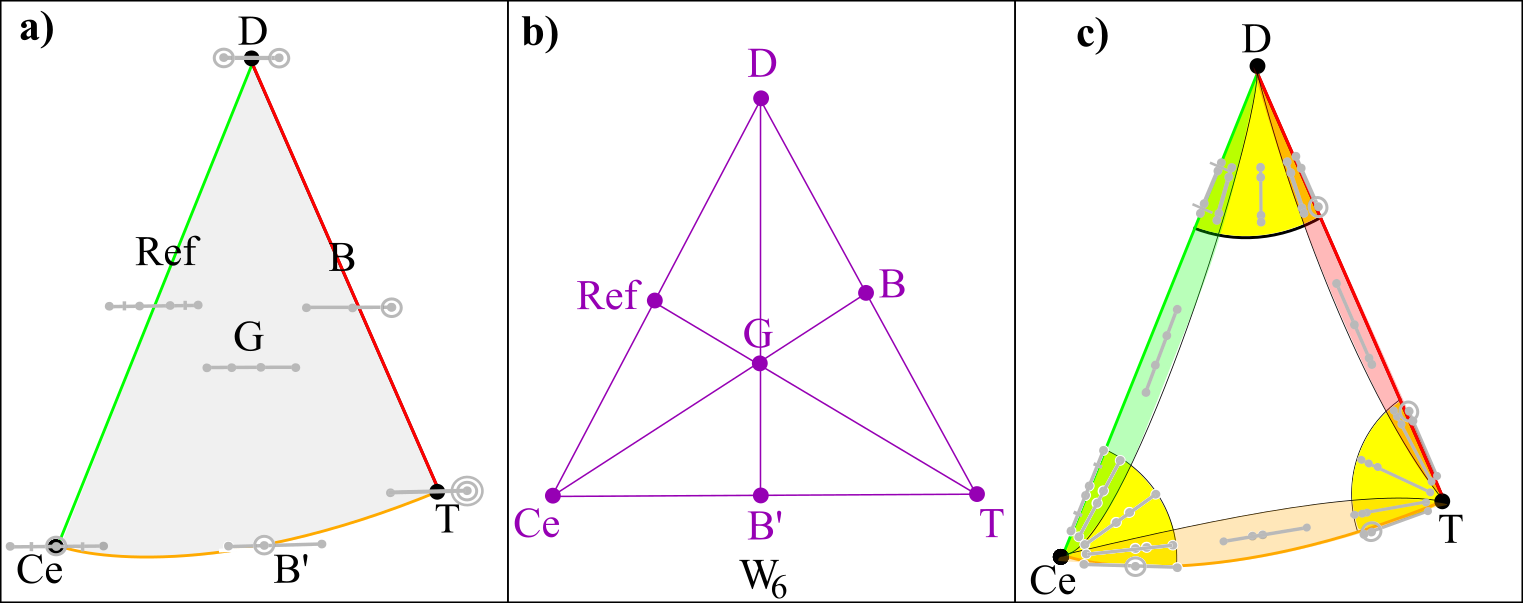}
\caption[Text der im Bilderverzeichnis auftaucht]{        \footnotesize{a) Incipient metric-level decor on $\Leib_{\tFrS}(4, 1)$, along with b) 
its adjacency graph and c) a count of its approximate qualitative types.} }
\label{Leib(4, 1)-Plain} \end{figure}          }

\m

\n{\bf Definition 1} The number of {\it (exact) qualitative types} in a shape space of (top manifold) dimension 2 is 
\be
Q(\mG) := F + E + V   \m .  
\label{Q}
\ee 
\n{\bf Remark 1} In Part I, the $F$ contribution was not yet supported.  
We now identify $Q$ as the plain -- rather than alternating-sign -- sum counterpart of the Euler characteristic $\chi(\mG)$, 
an identification which continues to make sense in arbitrary dimension.  

\m

\n{\bf Remark 2} In the case of a Leibniz space, every face, vertex and edge contributes a more strongly distinct qualitative 
type due to Leibniz space affording no symmetry redundancy. 
For the other shape spaces, some of the qualitative types are mirror images of each other and/or point relabellings of each other.  

\m

\n{\bf Remark 3} See Appendix B for definitions of the other four definitions of qualitative types used in the current treatise, 
alongside the computational propositions and corollaries used to build this treatise' tables of numbers of qualitative types, 
of which Fig \ref{(4, 1)-Q-Top} is the first.  
%
{            \begin{figure}[!ht]
\centering
\includegraphics[width=0.75\textwidth]{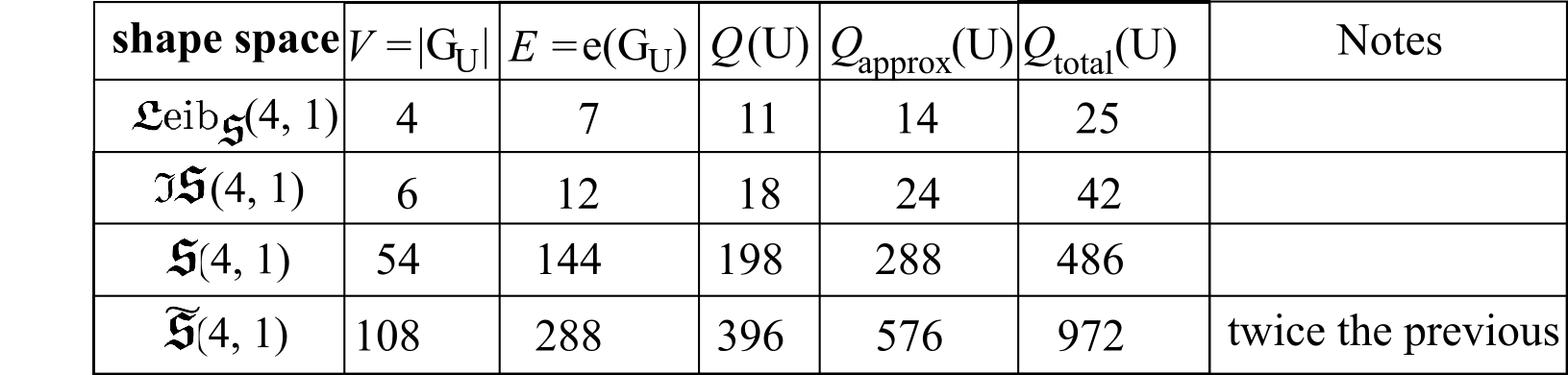}
\caption[Text der im Bilderverzeichnis auftaucht]{        \footnotesize{Number of qualitative types at the topological level. } }
\label{(4, 1)-Q-Top} \end{figure}          }

\section{Further notions of Lagrangian uniformity}

\subsection{A first example}\label{M-Arc}

\n In \cite{FileR}, a piece of this uniformity curve emanating from Ce was pointed out.  

\m

\n{\bf Proposition 1} In the shape space face whose shapes are ordered 1234 along the line;  
\be
\mU          \m \mbox{ is the curve } \m \mbox{sin}\,\phi + 3 \, \mbox{cos}\,\phi = 2 \, \mbox{cot} \, \theta   \m ,
\ee
\n{\bf Remark 1} These shapes contain the (3, 1) maximally uniform shape as a subconfiguration. 

\m

\n{\bf Proposition 2} This U curve intersects with the opposite face of $\Leib_{\sFrS}(4, 1)$ at the shape depicted in column 3 of Fig \ref{U-CeA-UA}.  

\m

\n{\bf Definition 1} We term this shape the {\it asymmetric central shape}, $\mbox{Ce}^{\sA}$.  
Like for Ce, centrality refers to this configuration being in a 1 : 1 ratio, 
just now with the binary on one edge rather than itself being central, whence the further qualification of asymmetry.

\subsection{Geodesic extension gives catastrophe curves of mergers}\label{Extension}

\n {\bf Remark 1} The preceding subsection's curve of uniformities is moreover a geodesic, 
so we can work out its continuation after $\mbox{Ce}^{\sfA}$ in $\FrS(4, 1)$ (Fig \ref{Swallow-Butter}.1).   
It folds up in the form of a boundary reflection in each of $\FrI\FrS(4, 1)$ and $\Leib_{\sFrS}(4, 1)$, and we can continue to follow each of these these curves.   

\m

\n{\bf Proposition 1} The first new result thus obtained is that the reflected geodesic next strikes a boundary at the (maximally) uniform state U(4, 1)  

\m

\n{\bf Proposition 2} The second new result is that the third arc of the uniformity geodesic U in $\Leib_{\sFrS}(4, 1)$ intersects the first arc. 
This reflects that it is possible for a configuration to concurrently be a uniformity of this kind twice over, 
as per column 4 of Fig \ref{U-CeA-UA}'s 1 : 1 : 2 configuration.  

\m 

\n{\bf Definition 1} We term this configuration the asymmetric uniform state, $\mU^{\sfA}$.
This name alludes to this shape being a local maximum in uniformity: 
as many elements of (Lagrangian) uniformity as are possible away from the arc of reflection symmetric states, Ref.

\m

\n{\bf Proposition 3} The third new result is that this U arc ends at the T configuration.

\m

\n{\bf Proposition 4} $\FrI\FrS(4, 1)$ has twice as far to go between striking corners.

\m

\n The fifth and sixth new results are the overall shape formed by each of these two sequences of folded geodesic arcs.  

\m

\n{\bf Proposition 5} For $\FrI\FrS(4, 1)$, the left and right uniformity great circle arcs present fold up 
in the form of the butterfly catastrophe \cite{Arnol'd-Cat}  (Fig \ref{Swallow-Butter}.2).  

\m

\n{\bf Proposition 6} For $\Leib_{\sFrS}(4, 1)$, on the other hand, the single (Identity of Indiscernibles!) 
uniformity great circle arc folds up twice to give the swallowtail catastrophe \cite{Arnol'd-Cat} (Fig \ref{Swallow-Butter}.3).  

\m

\n{\bf Remark 1} These complications can be taken to indicate there being geometrical and dynamical prices to pay in working with (scale and) shape space quotients.  
%
{            \begin{figure}[!ht]
\centering
\includegraphics[width=0.45\textwidth]{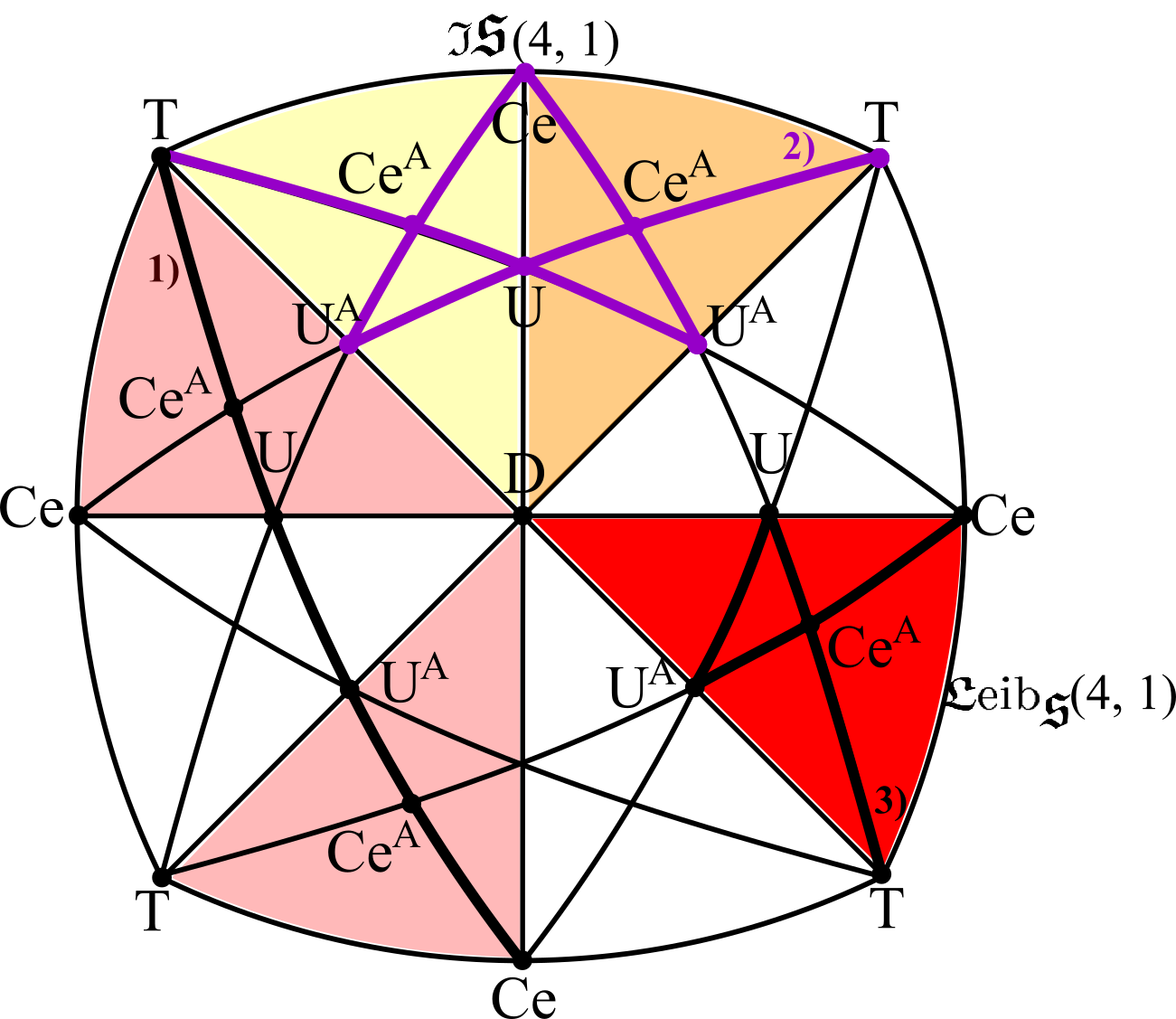}
\caption[Text der im Bilderverzeichnis auftaucht]{        \footnotesize{Metric and topological versions of $\FrI\sFrS(4, 1)$ and $\Leib_{\tFrS}(4, 1)$.
The uniformity curve U is represented as follows. 
In the $\sFrS(4, 1)$ sphere, it is a great circle.   
1) depicts a cube face's worth of this as a thick black curve.  
2) In $\FrI\sFrS(N, 1)$, the corresponding geodesic bounces off the edge at $\sU^{\tA}$ and then again at Ce, 
where it joins to another copy of the geodesic bouncing off the mirror image $\sU^{\tA}$ point to end at the mirror image T point. 
This is depicted as the thick purple curve, and takes the overall form of a butterfly catastrophe.  
3) In $\Leib_{\tFrS}(4, 1)$, the corresponding geodesic bounces off the edges at U and $\sU^{\tA}$, 
forming the swallowtail catastrophe as marked by the other, self-intersecting, thick black curve.} }
\label{Swallow-Butter} \end{figure}          }

\m

\n{\bf Remark 2} We next explain what `continuing an arc' entails.  

\m 

\n Case 1) $\Leib_{\sFrS}(4, 1)$ and $\FrI\FrS(4, 1)$'s boundaries exhibit {\sl reflective boundary conditions}. 
Let us note in passing that such were hypothesized by DeWitt in the case of GR superspace \cite{DeWitt70}.  
Moreover, with reflective boundary conditions, if there is a `head on' i.e.\ $\pi/2$ incidence with a boundary, a curve retraces itself backward. 

\m

\n Case 2) The other way a curve can start or end is if it strikes a corner; in this case, within the reduced configuration space, 
the curve retraces itself backward as well. 

\m

\n For the current subsection, case 2 suffices.  
The periodic nature of great circle geodesics is preserved under passage to $\FrI\FrS(4, 1)$ and $\Leib_{\sFrS}(4, 1)$.  
On the other hand, Case 1) ocurrs in Sec \ref{MQ} and in Part III.

\m

\n{\bf Remark 3} Fig \ref{S(4, 1)-Metric-b} puts together the $\Leib_{\sFrS}(4, 1)$ decor uncovered so far.
%
{            \begin{figure}[!ht]
\centering
\includegraphics[width=0.35\textwidth]{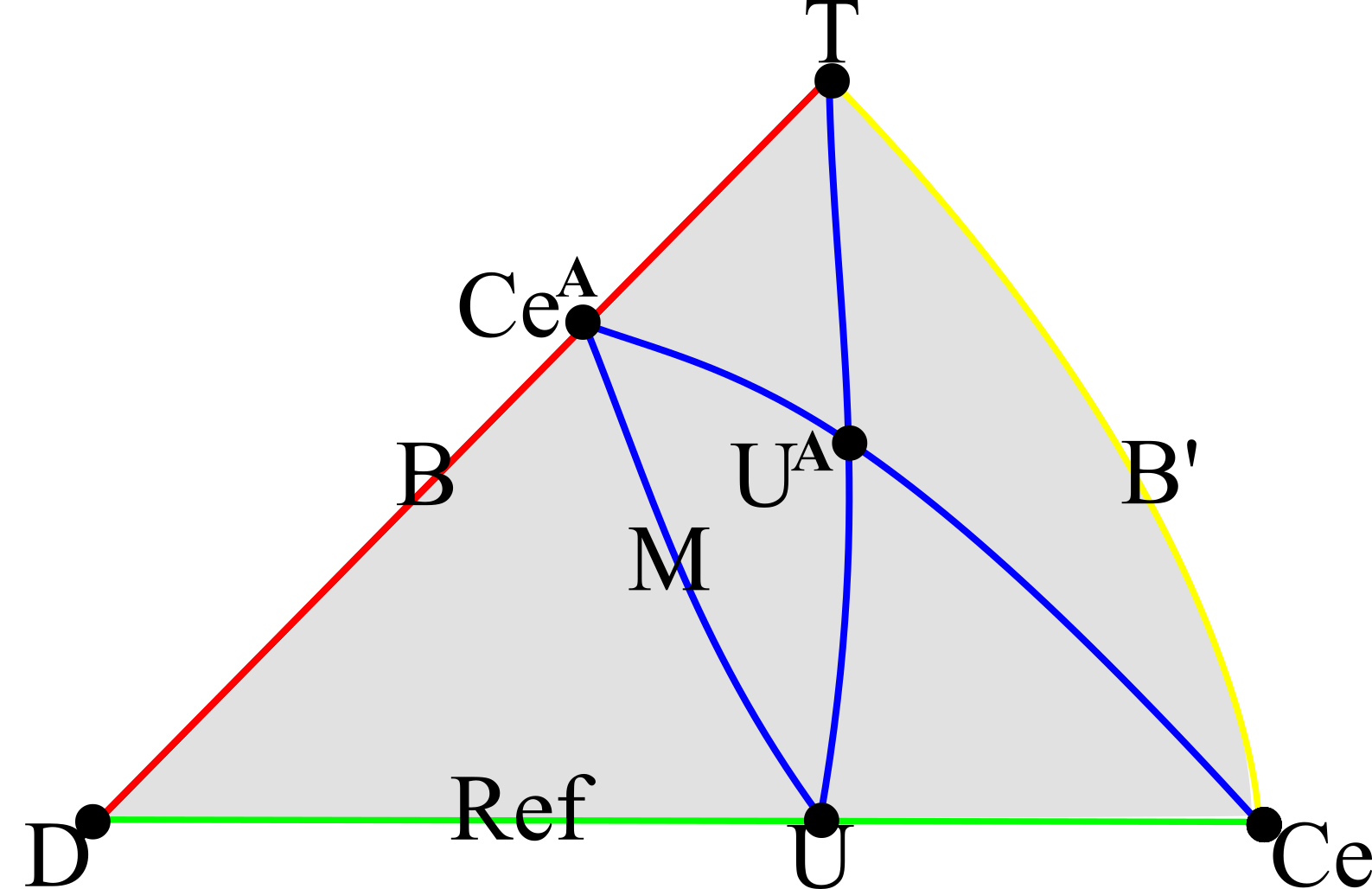}
\caption[Text der im Bilderverzeichnis auftaucht]{        \footnotesize{Metric-level extension of decor on $\Leib_{\tFrS}(4, 1)$ 
due to uniformity arc U 
and arc of reflection Ref.
} }
\label{S(4, 1)-Metric-b} \end{figure}          }

\subsection{Edge, vertex and algebraic methods for probing $\FrI\FrS(4, 1)$ and $\Leib_{\sFrS}(4, 1)$}

\n Techniques available for exploring the merger structure of shape spaces include the following.  

\m

\n 1) Shape space arc method.

\m

\n 2) Shape space point method. 

\m

\n 3) Algebraic method formulated at the level of relative space.  

\m

\n{\bf Remark 1} One can think of 1) and 2) as edge and vertex methods, which exhibit complementarity.  
Sec \ref{M-Arc}'s features were first detected by a vertex method: 
noting that Ce has clustering features in excess of those of the arcs hitherto known to intersect there: Ref and $\mB^{\prime}$.
This excess is partly accounted for by an arc U passing through Ce. 
Following this arc through to its endpoint as described in Sec \ref{Extension} is an example of edge method.  
In turn, this arc intersects other already-known arcs, pointing to new vertices to investigate (Sec \ref{Extension}).  
This gives a fair illustration of how these two methods complement in an alternating manner. 
A vertex's excess is moreover relative to the {\sl known} arcs at that stage in the calculation, so it needs to be updated whenever another edge is found to emanate from there.  
This method lacks precision, however, or at least requires careful interpretation due to excess being an unsigned concept, 
by which overcrowding's negative excess could mask positive excess due to hitherto unidentified arcs.  

\m

\n{\bf Remark 2} The algebraic method, on the other hand, finds all merger structures in one step and independently of each other. 
This independence is in contrast with the edge and vertex method's cumulative dependence on knowledge. 
`In one step and independently' is to be interpreted as still requiring solution of multiple equalities, 
but in a manner that none of these equalities affects the outcome of any of the others (see Appendix A for more).    

\m 

\n{\bf Remark 3} The algebraic linear merger equations in question simply involve setting the position of any subsystem CoM 
(including particles themselves as 1-particle CoMs) equal to any other.
Their solution is a network of arcs -- 1 degree of freedom solutions -- intersecting at vertices: 0 degrees of freedom solutions. 
This is with reference to (4, 1) shapes; for ($N \geq 5$, 1) higher simplices occur as well. 
In this way, the algebraic method has an immediate and obvious generalization to larger particle numbers, 
whereas the edge and vertex method spirals out of control by increasing $N$ bringing in further types of simplex.
The solving process does require restricting the solutions to values within the shape-theoretically meaningful range of parameters. 
From the point of view of basic Optimization, this is also a standard procedure: specification of a region that admissible solutions must belong to.

\m

\n{\bf Remark 4} All in all, the advantages of the algebraic method are as follows. 

\m 

\n A) It solves for all mergers in one step. 

\m

\n B) It cannot be tricked by overcrowding masking excess, 

\m

\n C) It can be formulated as a simple and entirely straightforward algorithm, so it is programmable. 

\m

\n D) It remains just as straightforward to conceive of and program if the particle number is further increased.  

\m

\n{\bf Remark 5} (4, 1) calculations -- and (5, 1) concepts without calculations -- thus suffice to determine the primality of algebraic methods over vertex-and-edge methods.  

\m 

\n Moreover, with increasing $N$, it is simpler to solve along a known merger arc or manifold for its intersection points with hitherto unknown further merger arcs or manifolds.  
In this way, all three of the methods can be combined iteratively.

\subsection{Uniformity structure}\label{11-Sec}

\n{\bf Proposition 1} U and Ref form the (Lagrangian) uniformity structure, in accord with Fig \ref{Uniform-Structure}.a).  

\m

\n{\bf Derivation} Let the points-or-particles be at a general (unnormalized) position 0, 1, $a$ and $b$.  
The relative particle separations are then 
\be
1        \mma 
|a|      \mma 
|b|      \mma 
|a - 1|  \mma 
|b -1|   \m \mbox{ and } \m \m 
|b - a|  \m .
\ee
The (4, 1) Lagrangian uniformity equations are then the result of equating any two of these and disregarding solutions 
whose equal separations are purely coincidence-or-collision features.  
The solution of this (4, 1) Lagrangian uniformity is the folded U arc -- minus its T endpoint -- 
                                                alongside the Ref arc -- minus its D endpoint, as per Fig \ref{Uniform-Structure}.a).  
This gives the solutions with {\sl some element} of uniformity, i.e.\ equal nonzero separations not induced by coincidences-or-collisions. $\Box{ }$
%
{            \begin{figure}[!ht]
\centering
\includegraphics[width=1.0\textwidth]{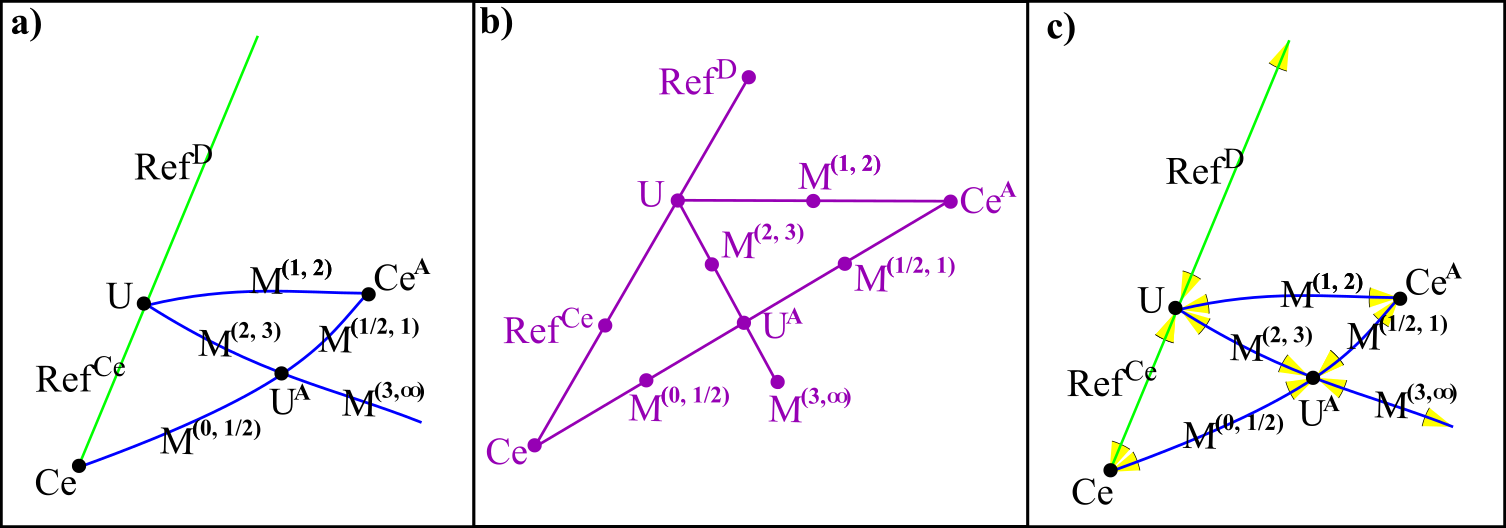}
\caption[Text der im Bilderverzeichnis auftaucht]{        \footnotesize{a) (Lagrangian) uniformity structure's differential-geometric structure is as labelled; 
see the next subsection for the maeaning of the suffixes in use.
This consists of 4 special points and 7 arcs, by which there are 11 exact qualitative types of (Lagrangian) uniformity.  
b) The topological adjacency graph of the uniform structure, with 11 vertices joined by 12 adjacency relation arcs. 
c) There are additionally 14 types of approximate notions of uniformity 
} }
\label{Uniform-Structure} \end{figure}          }

\m 

\n{\bf Remark 1} One can moreover quantify {\sl how much} uniformity each shape possesses by counting how many equal separations of each size there are, 
as per Appendices \ref{Uniformity-Measures}.  

\m

\n{\bf Proposition 2} The topological adjacency graph of the uniformity structure is as in Fig \ref{Uniform-Structure}.b).  

\m

\n{\bf Proposition 3} The number of qualitative types of uniform states for each of the four (4, 1) shape spaces is as per Fig \ref{(4, 1)-Q-Uni}.
%
{            \begin{figure}[!ht]
\centering
\includegraphics[width=0.7\textwidth]{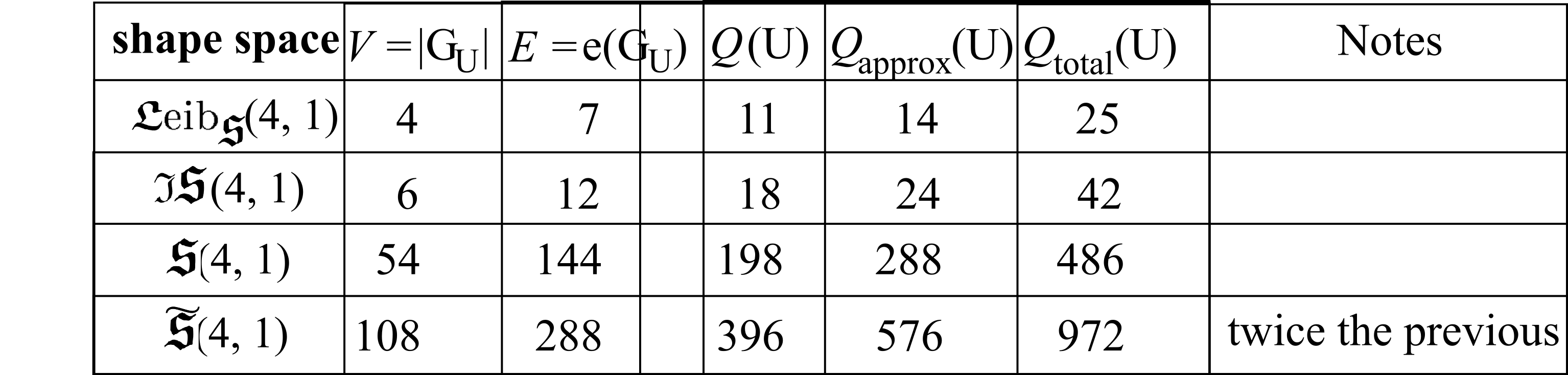}
\caption[Text der im Bilderverzeichnis auftaucht]{        \footnotesize{Number of qualitative types including the uniformity decor.}}
\label{(4, 1)-Q-Uni} \end{figure}          }

\m 

\n{\bf Remark 2} Approximate {\it near-uniform configurations} are a simple model of perturbatively small inhomogeneities, itself a popular topic in Cosmology.

\m

\n {\bf Remark 3} While a uniform system makes as much sense as a uniform subsystem, 
some types of inhomogeneous state are only meaningful in a subsystem context.
One example is void subsystems.  
Another is inhomogeoneous universes whose contents are nevertheless themselves highly homogeneous, 
such as a universe consisting of an inhomogeneous distribution of binaries which are similarly tight to each other.

\subsection{Qualitative types of Lagrangian shapes}\label{21-Sec}
%
{            \begin{figure}[!ht]
\centering
\includegraphics[width=1.0\textwidth]{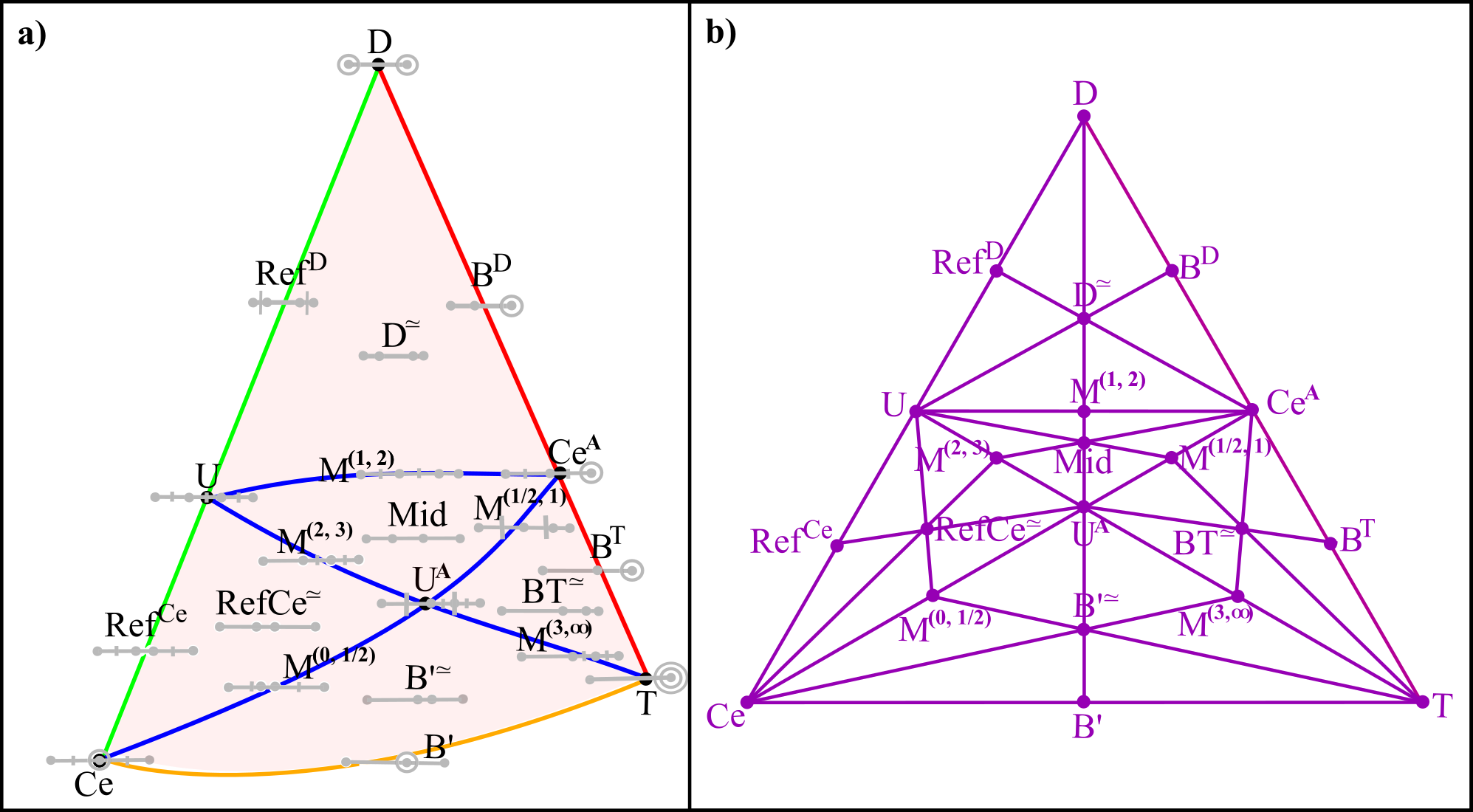}
\caption[Text der im Bilderverzeichnis auftaucht]{        \footnotesize{a) 21 qualitative types of Lagrangian shapes. 
b) Their topological adjacency graph, which has 45 edges and can readily be envisaged as a join of 5 $\mW_6$ -- 6-wheel -- subgraphs (one per face).} }
\label{21} \end{figure}          }
 
\n{\bf Remark 1} At the topological-and-metric Lagrangian level, 
Leibniz space is split up into the following qualitative types (inter-related as per Fig \ref{21}). 

\m

\n A) 6 special points: D, T, Ce, U, Ce$^{\sA}$ and U$^{\sA}$.

\m

\n B) 10 special arcs, as follows.  

\m

\n The B$^{\prime}$ arc of interior binary coincidences-or-collisions. 

\m

\n The B arc of exterior binary coincidences-or-collisions, 
now split by its 1 : 1 ratio shape Ce$^{\sA}$ into {\it ternary-concentrated} 
                                             and {\it peripheral} alias {\it double-binary-concentrated} subarcs, $B^{\sT}$ 
											                                                                  and $B^{\sD}$ respectively.    
\m
								 
\n The Ref arc of reflection-symmetric configurations, 
now split by its uniform   shape    U(4, 1)       into {\it central-binary-concentrated}  
                                            and  {\it peripheral} alias {\it double-binary-concentrated} subarcs, $\mbox{Ref}^{\sC\se}$  
											                                                                  and $\mbox{Ref}^{\sD}$ respectively.
																											  
\m

\n The arc of uniformities U, as split in five by its self-intersection point U$^{\sfA}$, 
                                                   the 1 : 1 binary shape Ce$^{\sA}$, 
												   the uniform shape U(4, 1), 
											   and its self-intersection point once again. 																											  

\m

\n A coarser naming is according to whether the remaining point is {\it interior} or {\it exterior} to the 1 : 1 ratio's three points. 
As ordered above, the first two arcs are interior U$^{\si}$, whereas the last three are exterior, U$^{\se}$.  
This topological idea can furthermore be metrically refined 
by ascribing to each U subarc the range of ratios that its remaining point's position makes with the central particle in the 1 : 1 ratio.
This gives the alternative `U-centric' vertex names U$^0$ for Ce, 
                                                    U$^1$ for Ce$^{\sfA}$, 
													U$^2$ for U(4, 1), 
													U$^{\infty}$ for T 
											    and U$^{\{3, 1/2\}}$ for the self-intersection U$^{\sfA}$.   
This also gives the more concise names U$^{(0, 1/2)}$, 
                                       U$^{(1/2, 1)}$, 
									   U$^{(1, 2)}$, 
									   U$^{(2, 3)}$ and 
									   U$^{(3, \infty)}$ for the five subarcs of U in the same order as above.

\m

\n C) The folded U-arc cuts up Leibniz space into five 2-$d$ regions.  
A suitable nomenclature for these involves giving priority to Leibniz space's corners followed by its (external) edges. 

\m

\n The top region in the figure is adjacent to just one corner, D, 
suggesting the name {\it double-binary-concentrated} shapes, denoted D$^{\approx}$, alias {\it peripheral} shapes.  

\m

\n The bottom region is adjacent to two corners but only one external edge, 
suggesting the name {\it interior-binary-concentrated} shapes, denoted B$^{\prime\approx}$.  

\m

\n The right-side region is adjacent to T, but `ternary-concentrated' is insufficiently descriptive as it also applies to one end of the B$^{\prime\approx}$. 
Thus we use the cluster hierarchical names {\it exterior-binary-concentrated within ternary-concentrated}, denoted BT$^{\approx}$ for this qualitative type, 
along with {\it interior-binary-concentrated within ternary-concentrated}, 
denoted by B$^{\prime}$T$^{\approx}$, for the {\sl lesser} qualitative type of being in the bottom region {\sl and} near T.  
The same reasoning gives yields {\it approximately-reflection-symmetric central-binary-concentrated}, denoted RefCe$^{\approx}$, 
for the qualitative type of the left-side region, alias interior binaries.

\m

\n The final region, which is adjacent to no corners or (more than a point's worth of) edges.
We call this region Mid, standing for the `middling' (4, 1) shapes. 

\m

\n{\bf Remark 3} This gives a total of $Q = 21$ exact qualitative types within Leibniz space with Lagrangian decor.
See table \ref{(4, 1)-Q-Lag} for many further counts of qualitative types for (4, 1) shape spaces.  
%
{            \begin{figure}[!ht]
\centering
\includegraphics[width=1.0\textwidth]{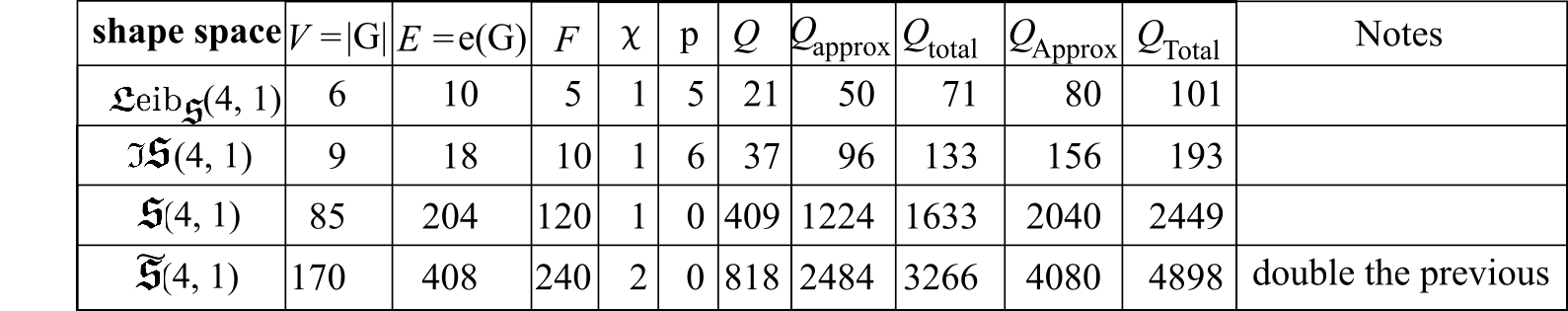}
\caption[Text der im Bilderverzeichnis auftaucht]{        \footnotesize{Number of qualitative types including the Lagrangian decor. } }
\label{(4, 1)-Q-Lag} \end{figure}          }

\m 

\n{\bf Remark 4} Each special point and special arc at the Lagrangian level is one or both of a coincidence-or-collision or a Lagrangian uniformity, 
with the sole overlaps being Ce and Ce$^{\sA}$. 
Thus Ce and Ce$^{\sA}$ acquire a further name as uniform states which are also coincidences-or-collisions.   
One way of denoting this is as B$^{\prime}$(1 : 1) and B(1 : 1), 
which makes clear that the difference between the two is in whether the binary coincidence-or-collision is interior or exterior.

\subsection{$\FrI\FrS(4, 1)$ and $\Leib_{\sFrS}(4, 1)$'s uniform state centres}

\n{\bf Remark 1} This subsection is in the context that, over the years, 
very many notions of centre have been ascribed to triangles, with at least some of these carrying over to spherical triangles. 
Thus we pose the question of whether the U(4, 1) and U$^{\sfA}$ `shape centres' of these shape space spherical triangle regions coincide 
with any previous definitions of spherical triangle centres.  

\m

\n{\bf Structure 1} The globally maximally uniform configuration U(4, 1)  
is a shape-theoretically meaningful notion of centre for the $\FrI\FrS(4, 1)$ isosceles spherical triangle.  
We denote this by $Z(\FrI\FrS(4, 1))$; like all notions of centre for isosceles triangles, it lies on the line of reflection symmetry (Fig \ref{(4, 1)-Centres}.a).   

\m

\n{\bf Structure 2} On the other hand, $\Leib_{\sFrS}(4, 1)$,is neither isosceles nor enjoys such a clear-cut shape-theoretic notion of centre.
$\mU^A$ -- the intersection of the folded U arc with itself -- is a shape-theoretic notion of centre for $\Leib_{\sFrS}(4, 1)$ (Fig \ref{(4, 1)-Centres}.b).

\m 

\n{\bf Remark 2} $\mU^{\sA}$ is moreover a weaker notion of centre because $\Leib_{\sFrS}$ is scalene, and its shape is also a weaker notion of uniformity: 
now a local rather than global maximum.   
This reflects another trade-off in maximizing reduction. 
Namely that while $\Leib_{\sFrS}(4, 1)$ has the advantage of representing all shapes precisely once, it is a scalene spherical triangle region, 
whereas the more redundant representation as $\FrI\FrS(4, 1)$ is geometrically simpler by being an isosceles spherical triangle region. 

\m

\n{\bf Remark 3} N.B. that both of these centres are particularly uniform states: the global maximum in uniformity and the local maximum away from the Ref line.  
Papers I, III and IV provide further evidence that shape space centres' co-realiation with uniform states is a common occurrence.  
%
{            \begin{figure}[!ht]
\centering
\includegraphics[width=0.7\textwidth]{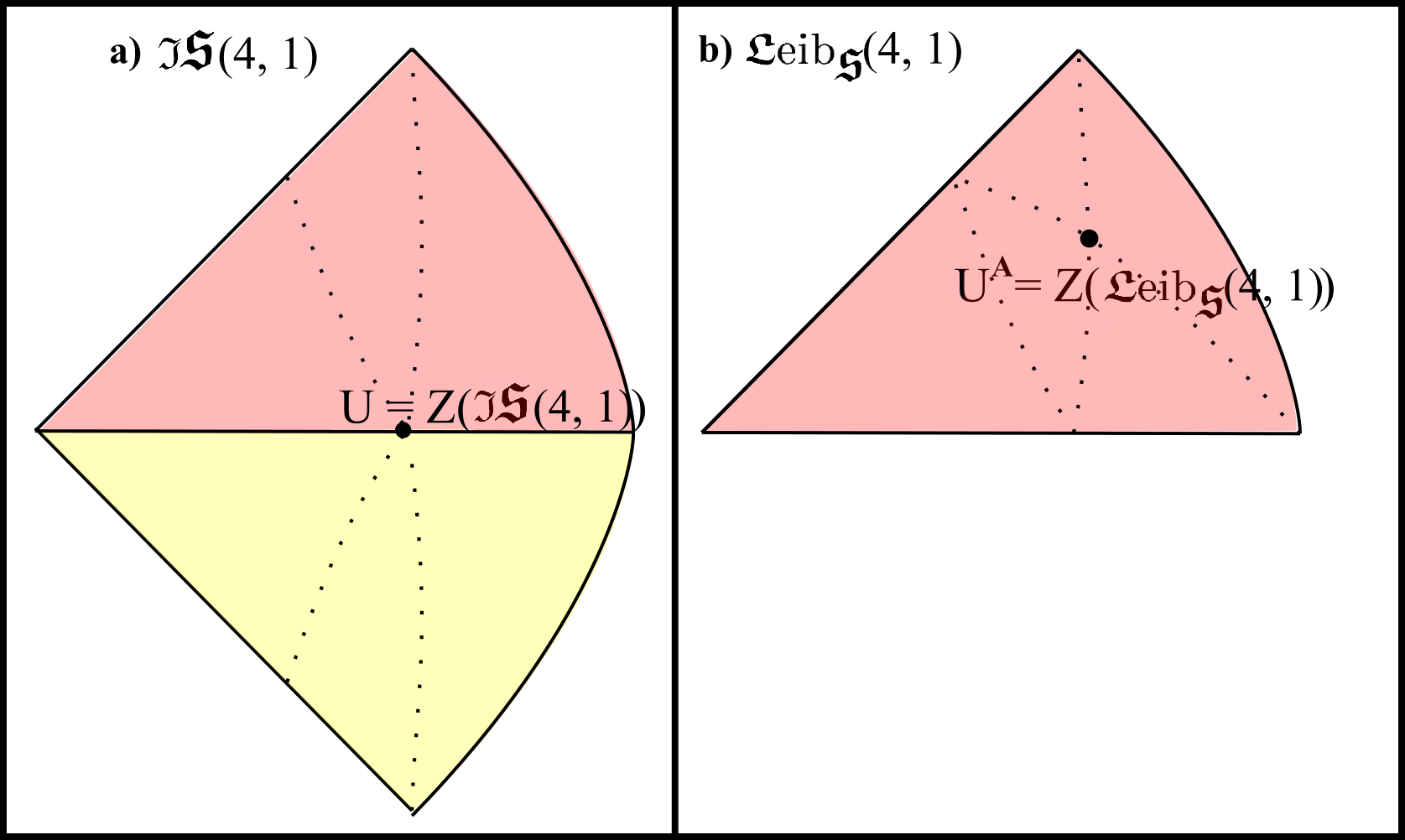}
\caption[Text der im Bilderverzeichnis auftaucht]{        \footnotesize{Shape-theoretically significant centres for $\FrI\sFrS(4, 1)$ and $\Leib_{\tFrS}(4, 1)$.} }
\label{(4, 1)-Centres} \end{figure}          }

\subsection{Approximate qualitative types of Lagrangian shapes}
%
{            \begin{figure}[!ht]
\centering
\includegraphics[width=0.85\textwidth]{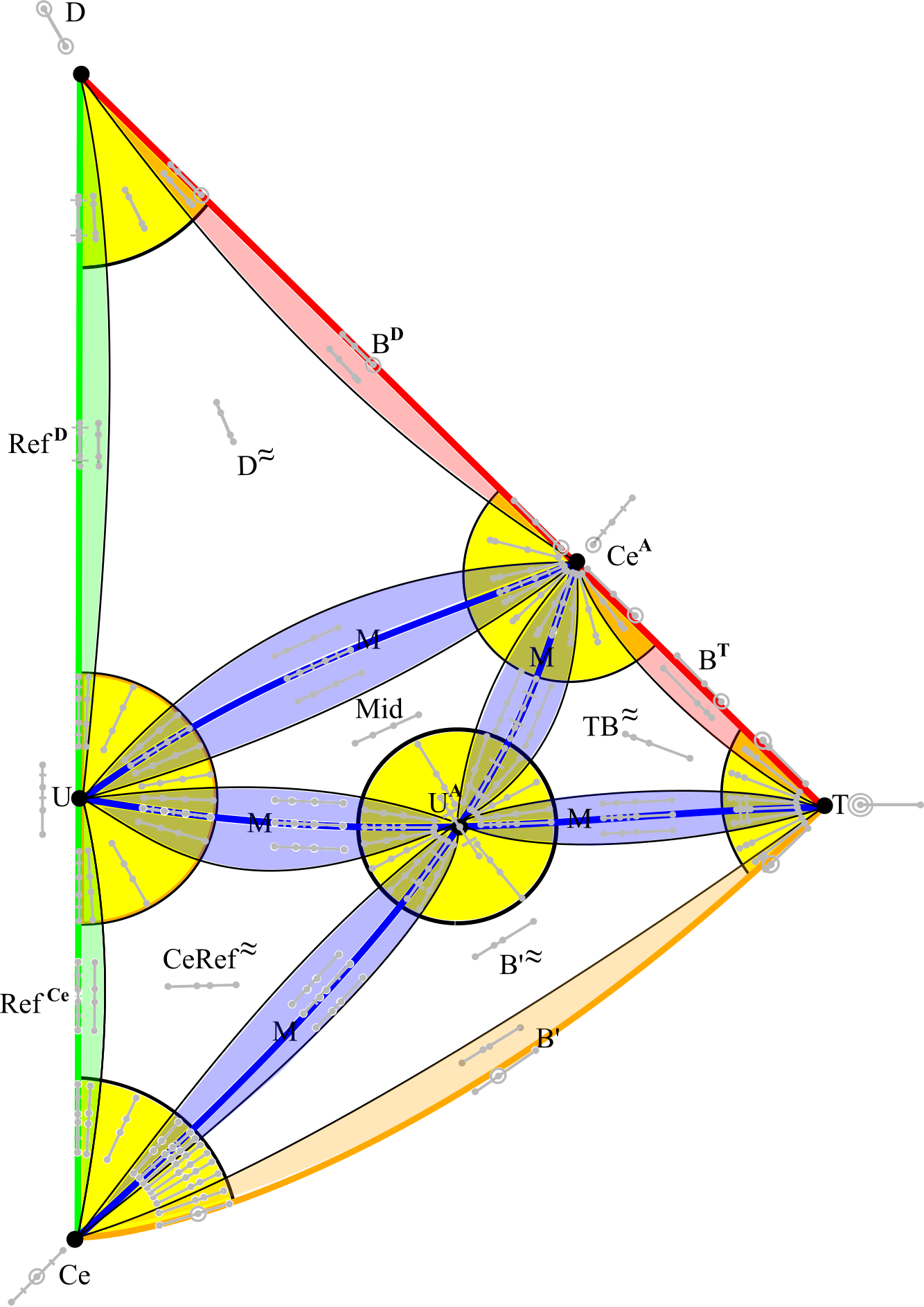}
\caption[Text der im Bilderverzeichnis auftaucht]{        \footnotesize{80 qualitative types of Lagrangian shapes, alongside 21 exact qualitative types.} }
\label{79} \end{figure}          }
%
\n We already counted these out as part of Fig \ref{(4, 1)-Q-Lag}.  
We end this Section by now plotting out an example of each of these approximate qualitative types in Fig \ref{79}.  
These can be named systematically by what stratum they are in and what special points and/or arcs they are near to.  
Some cases can also be given more memorable or picturesque names, such as the {\it pair of tight binaries separated by a void} (near D) 
                                                                  and the {\it tight ternary and void} (near T).

\section{Further Jacobian notions of merger}

\subsection{Merger reinterpretation of some of the shapes found so far}

\n We next accord merger interpretations to special shapes that we have already found; the subsequent subsection finds the remaining mergers.  

\m

\n {\bf Characterization 1} Basic calculations as summarized in Fig \ref{Ref-Ce} 
show that these reflection-symmetric shapes constituting the Ref curve additionally enjoy various merger features.
For now, we note in particular that all of these shapes have two binary centres of mass $\fX$ -- of the two outermost and the two innermost points -- 
coincide with each other and with $\fO = \fQ$.
I.e.\ we have the alias 
\be
\mbox{Ref} = \mM^{\sfX\sfX\sfQ}  \m . 
\ee
\n{\bf Remark 1} In the (3, 1) model, the only notion of merger M is the same as the only notion of reflection-symmetric shape and of uniform state U. 
The merger between the central particle B and CoM(AC) is the only normalizable way of having a $\fQ$ merger, whether with a $\fP$ or with a $\fX$. 
We denote this by 
\be
\mM^{\sfP\sfT} = \mM^{\sfP\sfX\sfT} = \mM^{\sfX\sfT} \m .  
\ee
This shape's characterization as a uniform shape is moreover Lagrangian: $r_{12} = r_{23}$.

\m

\n{\bf Remark 2} A third point of view, back in  $\FrI\FrS(4, 1)$ and for which we provide some tools in Appendix A, 
is that the amount of clustering information in D is in excess of that in two B configurations. 
D is thus a confluence of more than two significant curves.  
In this way, we can arrive at Ref being the third ingredient of the confluence 
even without ever considering $\Leib_{\sFrS}(4, 1)$ and its a priori uninterpreted edge.  

\m

\n{\bf Remark 3} Inter-relations between types of merger, and uniformity and Lagrangian characterizations of mergers, are also common of not ubiquitous in (4, 1).  
The (4, 1) model's uniformities considered so far are moreover M's for (3, 1) subsystems.

\m

\n{\bf Remark 4} Finally, we henceforth simplify notation from U(3, 1) to M and U(4, 1) to U.

\subsection{The further M$^{\sfQ}$ arc of mergers}\label{MQ}
%
{            \begin{figure}[!ht]
\centering
\includegraphics[width=0.72\textwidth]{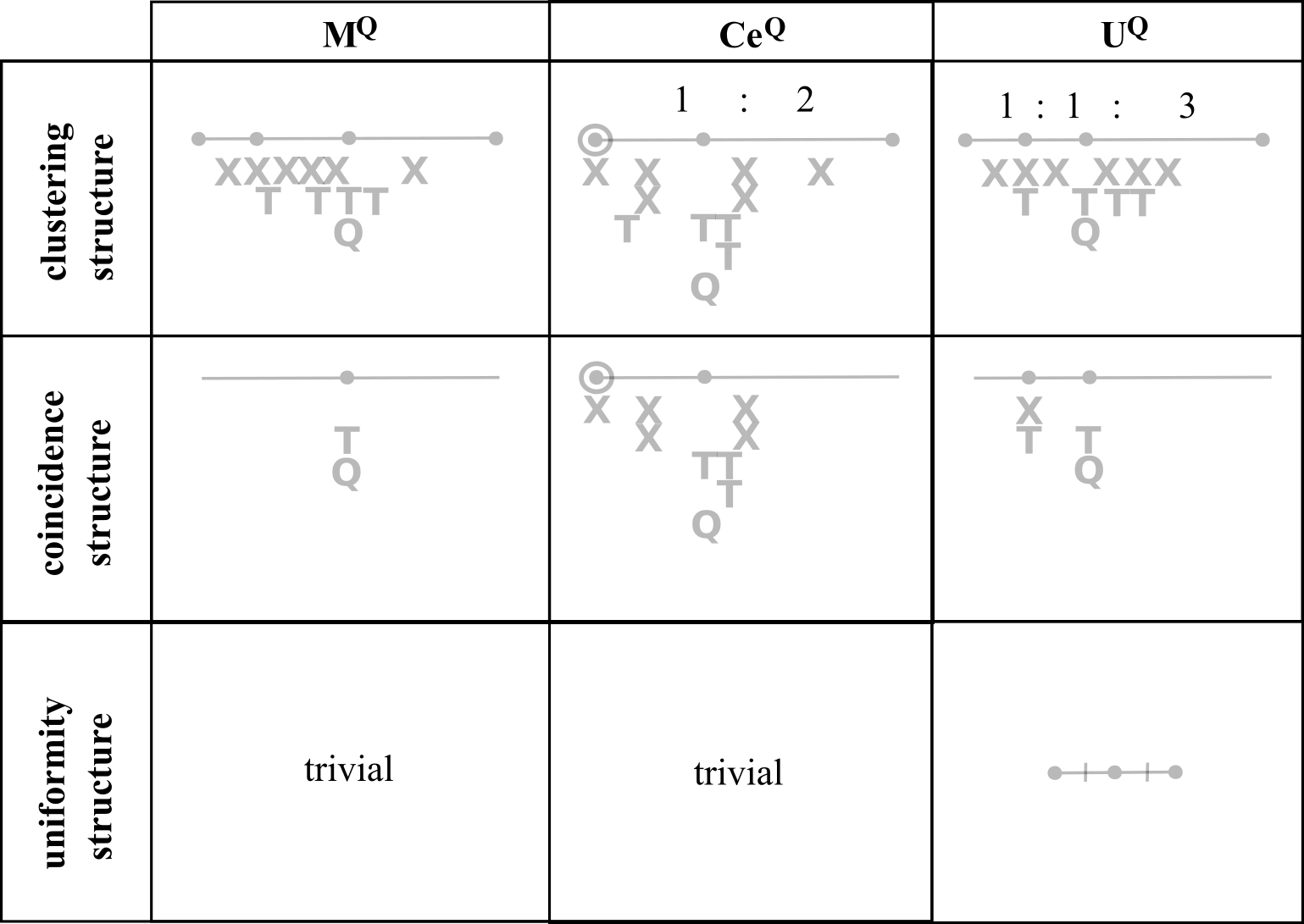}
\caption[Text der im Bilderverzeichnis auftaucht]{        \footnotesize{M$^{\tQ}$, B(1 : 2) and 1 : 1 : 3 configurations' clustering  structure, 
                                                                                                                    coincidence-or-collision structure, 
												                    											   and uniformity structure.} }
\label{CeQ-UQ-MQ} \end{figure}          }

\n The (4, 1) shapes have a further type of merger not accounted for in the previous section: M$^{\sfT\sfQ}$ = M$^{\sfQ}$. 
%

\m

\n Investigating this benefits from relative Jacobi K rather than H coordinates; see the next subsection.     

\m

\n{\bf Proposition 1} The M$^{\sfQ}$ arc's endpoints in $\Leib_{\sFrS}(4, 1)$ are Ce and a new 1 : 2 ratio binary coincidence-or-collision shape, B(1 : 2).  

\m

\n{\bf Proposition 2} The equation for the M$^{\sfQ}$ arc in $\Leib_{\sFrS}(4, 1)$ is 
\be
\mM^{\sT}    \m \mbox{ is the curve } \m \mbox{cos} \, \phi =  \mbox{cot} \, \theta                          \m . 
\ee
\n{\bf Proposition 3} The M$^{\sfQ}$ arc moreover intersects with the M arc at the 1 : 1 : 3 ratio shape.  

\m

\n{\bf Remark 1} See Fig \ref{CeQ-UQ-MQ} for details of these shapes-in-space, 
                     Fig \ref{S(4, 1)-Metric-c} for where the arc of these shapes lies in shape space, 
			     and Fig \ref{(4, 1)-Q-Jac} for qualitative type counts including the M$^{\sfQ}$ decor.  

\m

\n{\bf Remark 2)} 1 : 1 : 3 is a second more heterogeneous notion of shape-theoretically significant centre, now the confluence of M and M$^{\sfQ}$.  
This is a part-Jacobian notion of shape space centre, in contrast to the previous two notions of shape space centre being purely Lagrangian.
It also has no more Lagrangian uniformity elements than any other general point on the M arc.  
%
{            \begin{figure}[!ht]
\centering
\includegraphics[width=1\textwidth]{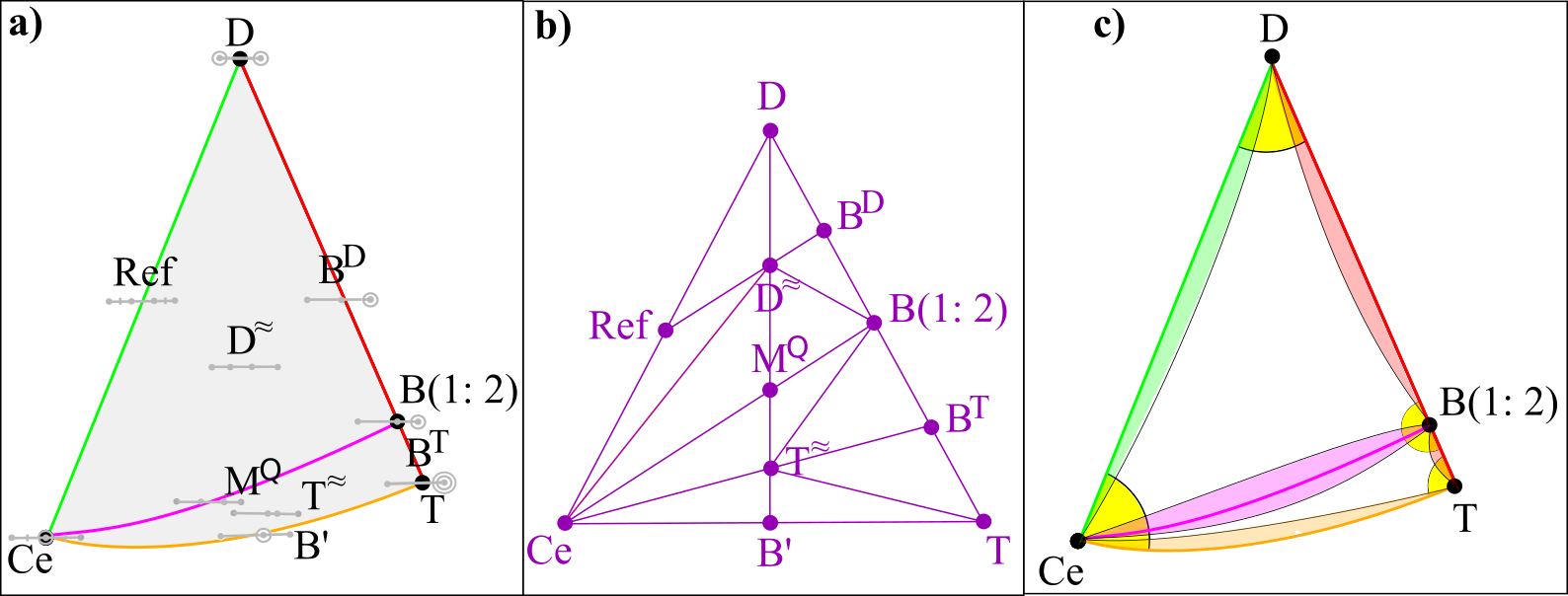}
\caption[Text der im Bilderverzeichnis auftaucht]{        \footnotesize{a) Metric-level decor on $\Leib_{\tFrS}(4, 1)$ due to merger M$^{\tfQ}$, in magenta, with 
b) the corresponding adjacency graph and c) count of its approximate qualitative types.} }
\label{S(4, 1)-Metric-c} \end{figure}          }

\n{\bf Remark 3} The number of exact qualitative types is
\be
Q :=  V + E + F = 4 + 5 + 2 
   =  11                     \m .
\ee
\n{\bf Remark 4} Also, the Euler characteristic is 
\be
\chi :=  V - E + F = 4 - 5 + 1 
      =  1                      \m ,
\ee 
again consistent with the disc topology.  

\m

\n{\bf Remark 5} The adjacency graph \ref{S(4, 1)-Metric-c}.b) has $V(\mbox{Adj}) = Q = 11$, 
and is moreover 2 $\mW_{6}$ wheel subgraphs with 2 common edges, so $E(\mbox{Adj}) = 12 \times 2 - 2 = 22$.

\m

\n{\bf Remark 6} The number of approximate qualitative Lagrangian types is 
\be 
Q_{\sa\sp\sp\sr\so\sx} = 34 \m .
\ee
\n{\bf Remark 7} This gives a grand total of 
\be
Q_{\st\so\st\sa\sll} := Q + Q_{\sa\sp\sp\sr\so\sx} 
                      = 11 + 34 
					  = 45                         \m \mbox{ Hopf qualitative types } \m , 
\ee
as exhibited in Fig \ref{S(4, 1)-Metric-c}.c).

\subsection{Supporting consideration of Jacobi K-coordinates and their axis system}

\n{\bf Remark 1} In Jacobi K-coordinates, the North pole $\theta = 0$ is at one of the T's.   
The principal axis is then a (vertex) diagonal of the cube, meaning a line from a vertex through the centre of the cube to the antipodal vertex.
In the current shape-theoretic context, this is labelled T at either end.  
This axis results from setting $\rho_3 = 0$: this T-axis is perpendicular to the plane of triple mergers corresponding to the K-cluster in question. 
I.e. the fourth particle $q_4$ is at $\mbox{CoM}(123)$.  
This corresponds to mergers of form M$^{\sfX\sfQ}$.

\m

\n{\bf Remark 2} Now however no further such diagonals can be concomitantly chosen as Cartesian axes.
This is because T-axes -- between opposite corners of the cube -- are not perpendicular.
We can however choose a perpendicular edge midpoint diagonal. 
These give the same orthonormal systems as the 3-body problem Jacobi vectors -- eq (I.33) -- which are now respectively a T-axis, a Ce-axis and a 1 : 1 : 3 axis.

\subsection{How each of the 8 special points encodes the arcs which intersect there.}
%
\n See Fig \ref{No-Excess} for how each special point's clustering hierarchy coincidences is underpinned by the collection of special arcs through that point. 
%
{            \begin{figure}[!ht]
\centering
\includegraphics[width=1.0\textwidth]{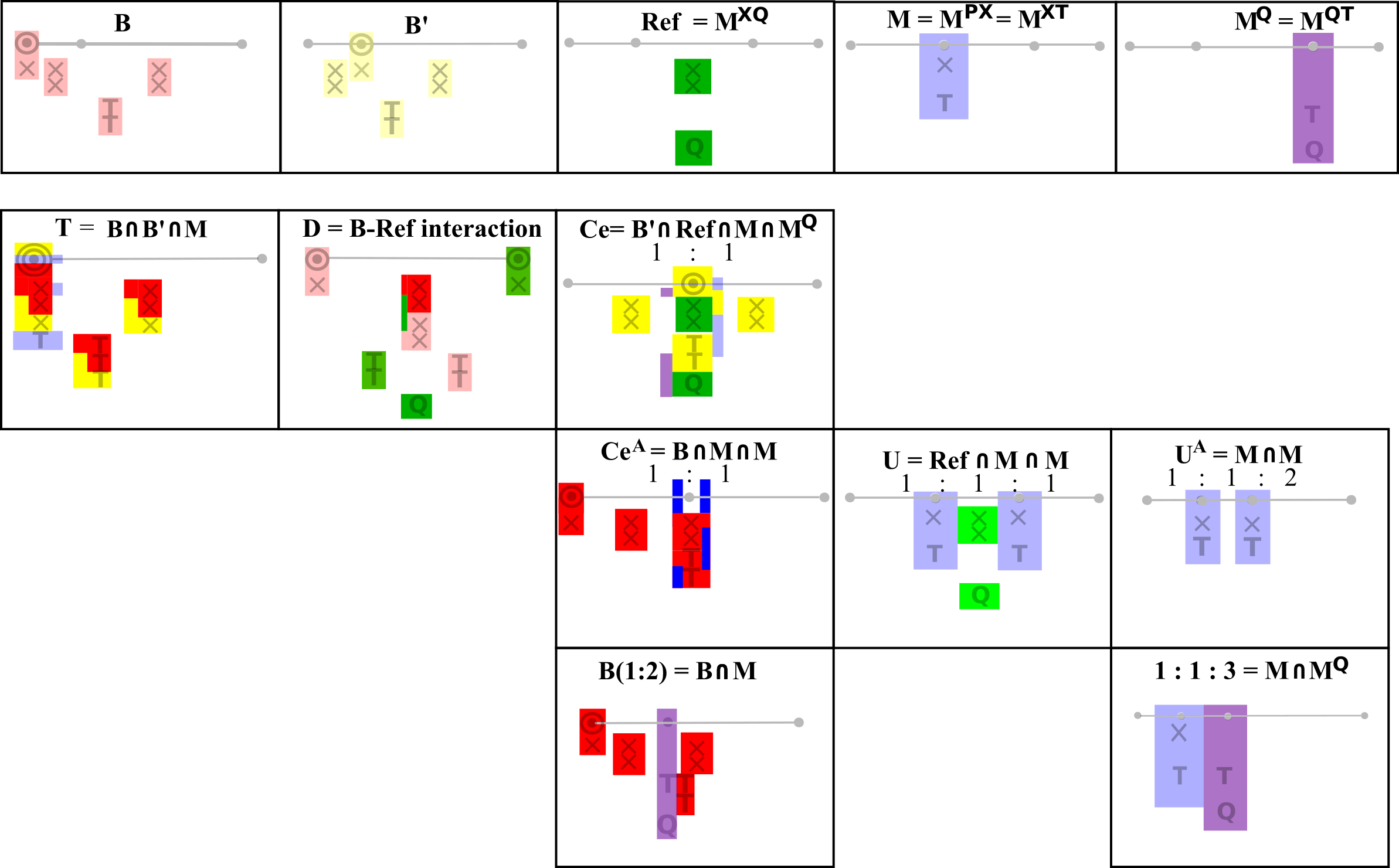}
\caption[Text der im Bilderverzeichnis auftaucht]{        \footnotesize{Clustering hierarchy coincidences for the five special types of arc 
and the eight special points.} }
\label{No-Excess} \end{figure}          }

\subsection{Jacobian versus Lagrangian notions of uniformity}
%
%
%
\n{\bf Remark 1} The Lagrangian-uniform states are a strict subset of the mergers for (4, 1): compare Figs \ref{Uniform-Structure} and \ref{29}.    
Aside from the T and D multiple coincidences-or-collisions, the arc of mergers M$^{\sfQ}$ is not included in the Lagrangian uniform structure.  
On the other hand, all of these points are included in some of the notions of Jacobian-uniform structure, 
corresponding to equating further (or all) differences of relative Jacobi (rather than relative Lagrange) separations.  

\m

\n{\bf Remark 2} All in all, in 1-$d$ uniform shapes are in general a strict subset of merged shapes, 
if only because our intuitions about uniformity are Lagrangian whereas the notion of merged shapes is Jacobian.

\subsection{Qualitative types of Jacobian shapes}
%
{            \begin{figure}[!ht]
\centering
\includegraphics[width=1.0\textwidth]{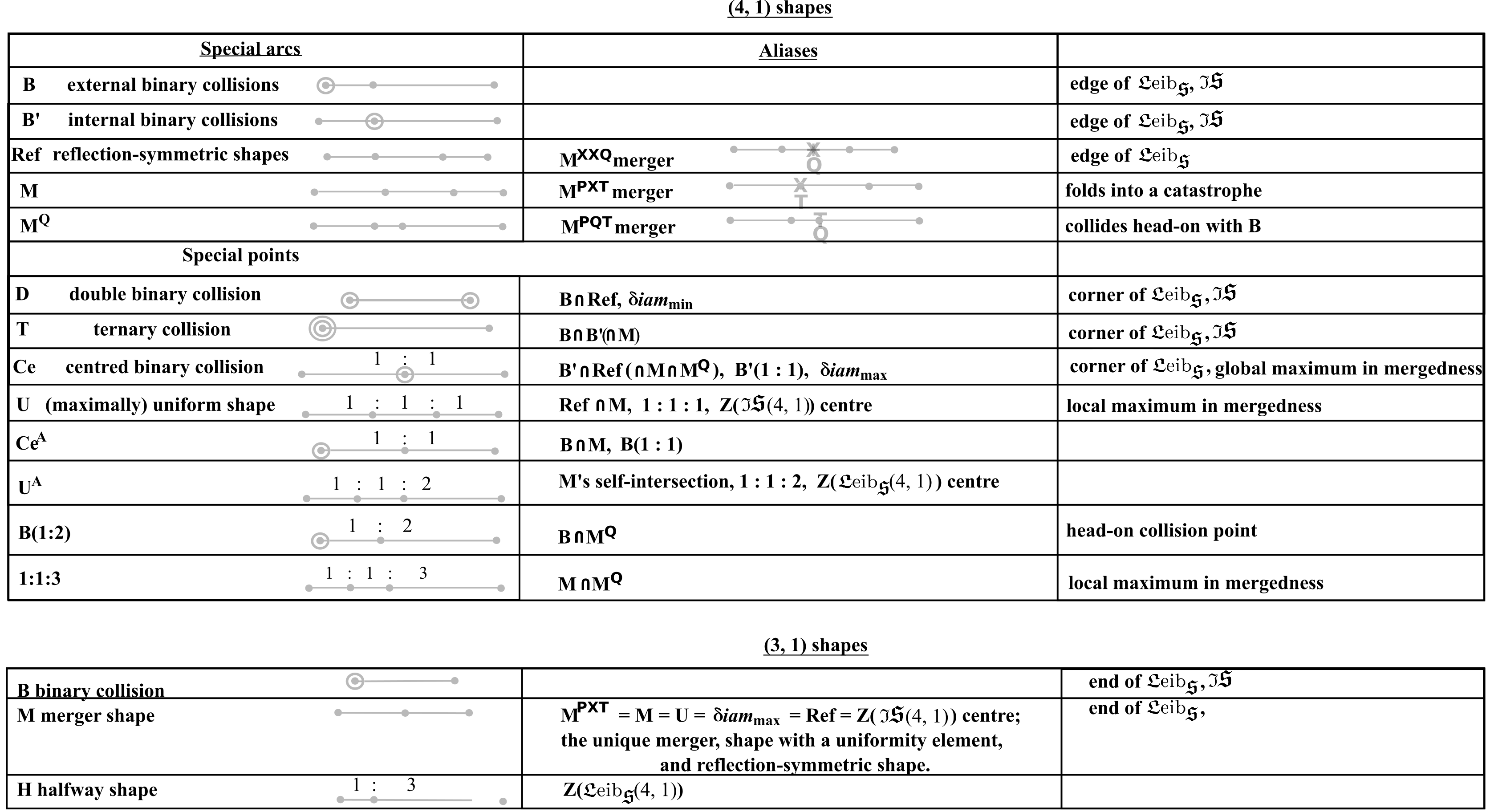}
\caption[Text der im Bilderverzeichnis auftaucht]{        \footnotesize{End summary of the special points and arcs.} }
\label{End-Summary-Arcs-and-Points} \end{figure}          }
%
{            \begin{figure}[!ht]
\centering
\includegraphics[width=0.85\textwidth]{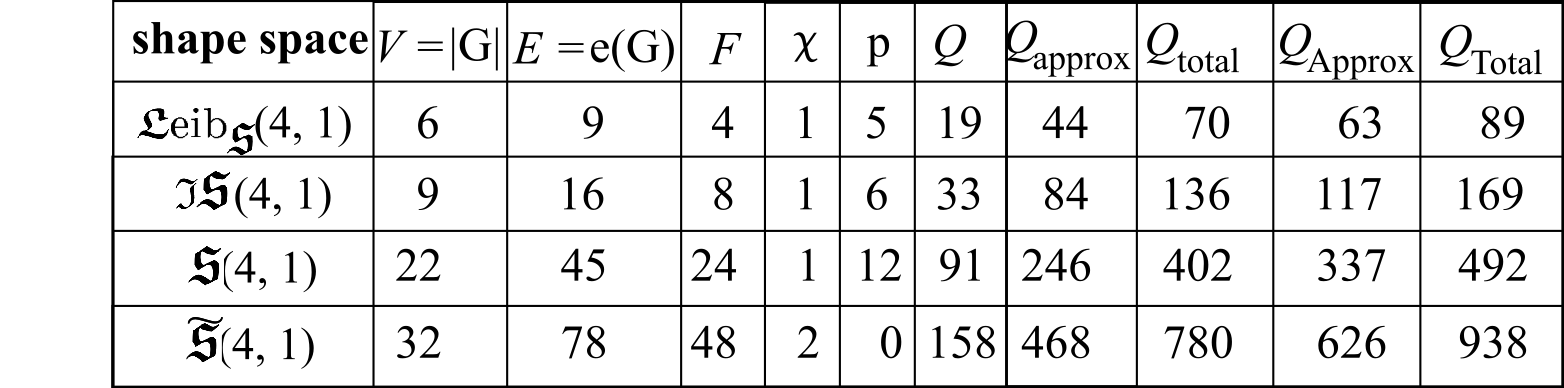}
\caption[Text der im Bilderverzeichnis auftaucht]{        \footnotesize{Number of qualitative types including the full Jacobian decor.}}
\label{(4, 1)-Q-Jac} \end{figure}          }

\n{\bf Remark 1} We have already seen that in Leibniz space including the full Jacobian decor, 
there are $V = 8$ special points T, D, Ce, U, U$^{\sA}$, 1 : 1 : 3, B(1 : 1) and B(1 : 2).   

\m

\n{\bf Remark 2} These are joined by $E = 14$ special arcs: B$^{\prime}$, B subdivided into three pieces, 
Ref and M$^{\sQ}$ each subdivided into two pieces, and M subdivided into six pieces. 
These are all as in Sec \ref{21-Sec}, except that 

\m

\n i) M$^{(3, \infty)}$ has been split into M$^{(3, 4)}$ and M$^{(4, \infty)}$ by the point 1 : 1 : 3 = M$^4$.

\m

\n ii) B$^{\sT}$ is also split into two arcs, say B$^{\sT 2}$ and B$^{\sm\si\sd}$.

\mbox{} 

\n iii) Finally the new M$^{\sfQ}$ arc is split by the 1 : 1 :3 point into M$^{\sfQ}$B and M$^{\sfQ}$Ce subarcs. 

\m

\n{\bf Remark 3} These arcs cut Leibniz space up into $F = 7$ regions; these are as before, except that 

\m

\n a) the B$^{\prime\approx}$ region is split by the M$^{\sfQ}$ arc into regions B$^{\prime 2\approx}$ still bordering B and the more interior mid$^{\sfQ}$.  

\m

\n b) The BT$^{\approx}$ region is split by the M$^{\sfQ}$ arc into regions BT$^{2\approx}$ still bordering T and the more interior B$^{\sm\si\sd\approx}$.  

\m

\n{\bf Remark 4} Thus there is a total of 29 exact qualitative types in Leibniz space with full Jacobi decor. 
See Fig \ref{End-Summary-Arcs-and-Points} for an end-summary of special arcs and special points among the (4, 1) shapes.  
See Fig \ref{(4, 1)-Q-Jac} for further counts of qualitative types in the four (4, 1) shape spaces.  

\m

\n{\bf Proposition 1} The qualitative types of shape moreover occur in shape space in the pattern of Fig \ref{29}. 
%
{            \begin{figure}[!ht]
\centering
\includegraphics[width=1.0\textwidth]{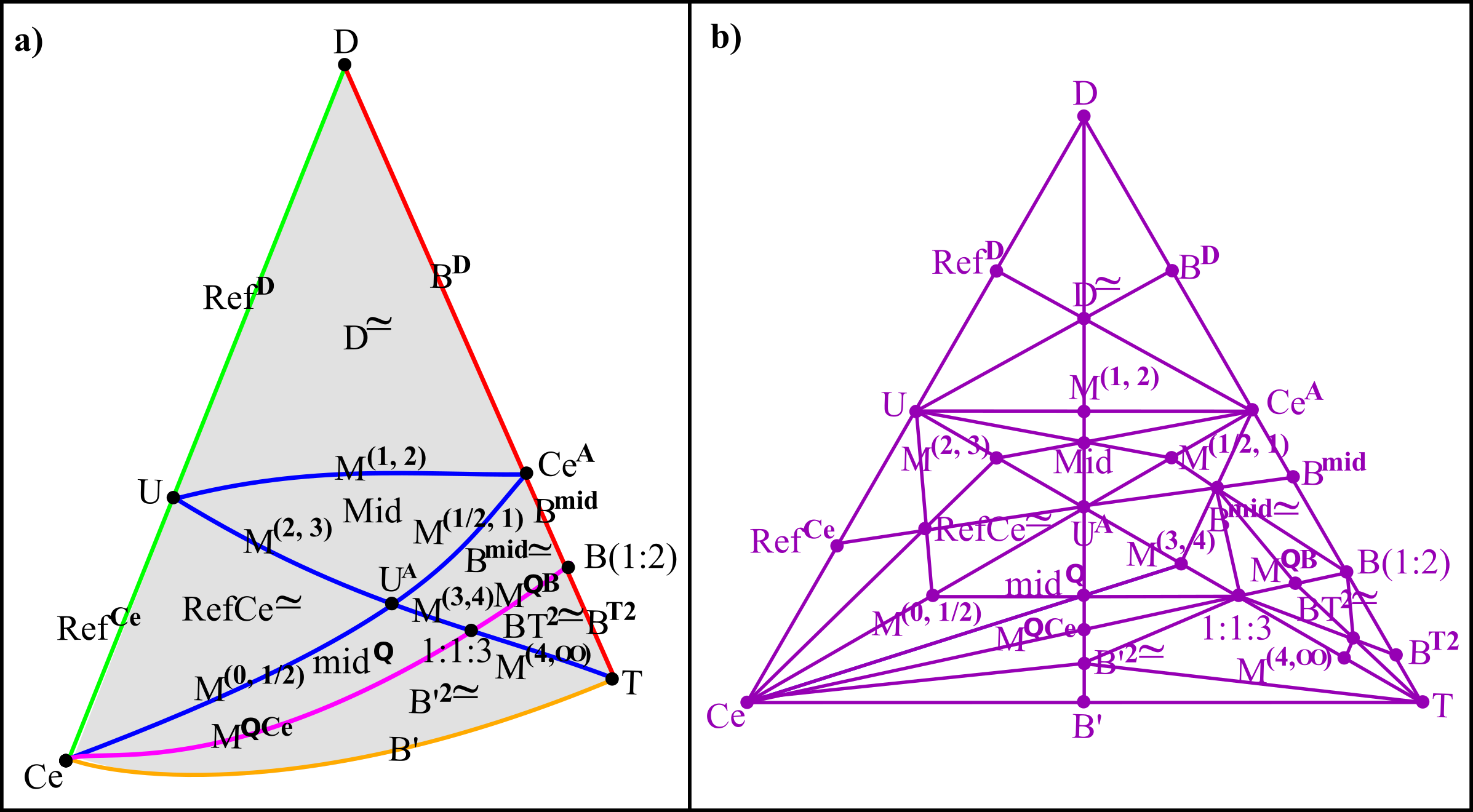}
\caption[Text der im Bilderverzeichnis auftaucht]{        \footnotesize{a) Conceptual names for the 29 qualitative types of shapes in (4, 1) Leibniz space with full Jacobi decor.
b) The corresponding adjacency graph, which is readily envisaged as a join of 6 $\mW_6$ -- 6-wheel -- and one $\mW_8$ -- 8-wheel -- subgraphs, 
corresponding to the seven faces.
c) For counting out the approximate qualitative types.}}
\label{29} \end{figure}          }

\subsection{Qualitative types of (4, 1) approximate Jacobi shapes}
%
{            \begin{figure}[!ht]
\centering
\includegraphics[width=0.85\textwidth]{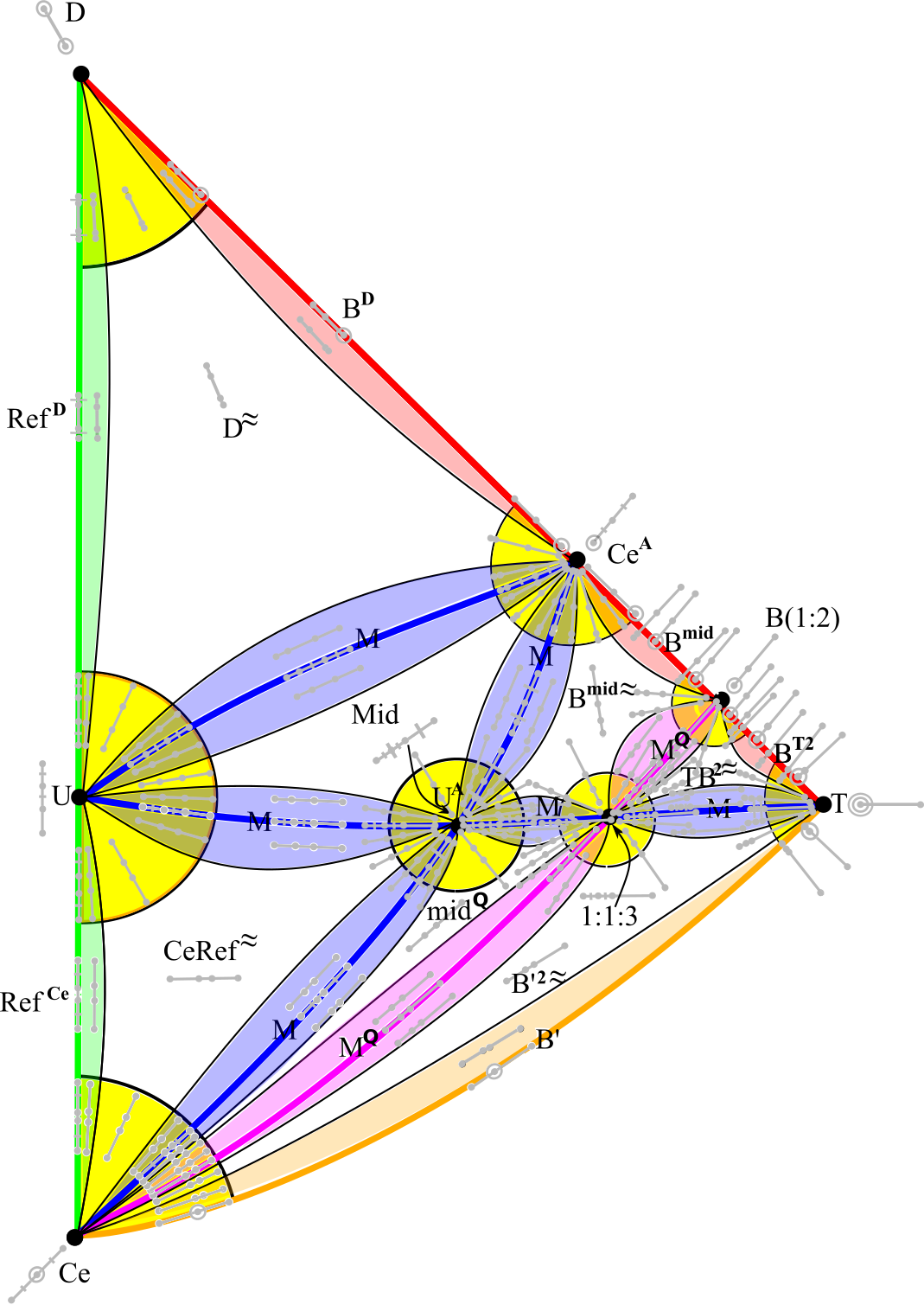}
\caption[Text der im Bilderverzeichnis auftaucht]{        \footnotesize{Grand total of 145 qualitative types of (4, 1) shapes.}}
\label{94} \end{figure}          }

\n We already counted these out as part of Fig \ref{(4, 1)-Q-Jac}.  
We end this Section by extending our plot of approximate qualitative types from the Lagrangian  level to the Jacobian level in Fig \ref{94}.

\section{Configuration space automorphism groups}\label{Killing-(4,1)}

\subsection{Graph automorphisms}

\n{\bf Proposition 1}
\be
Aut(\Top\mbox{--}\FrS(4, 1)) = Aut(\mbox{cube graph as labelled}) 
                             = \mbox{cube-octahaedron group O} 
					         = S_4 \times \mathbb{Z}_2                            \m , 
\ee
\be
Aut(\Top-\w{\FrS}(4, 1))    = Aut(\mbox{identified half-cube graph as labelled}) 
						    = S_4                                                 \m , 
\ee
\be
Aut(\Top-\FrI\FrS(4, 1))    = Aut(W_{6} \mbox{ as labelled}) 
                            = \mathbb{Z}_2                                        \m , 
\ee
and 
\be
Aut(\Top\mbox{--}\Leib_{\sFrS}(3, 1)) = Aut(\mbox{gem as labelled}) 
                            = id                                                  \m . 
\ee
\n{\bf Derivation} The first two of these follow from permutation of the T's fixing everything in the second case, and everything bar orientation in the first case.   
By the argument of Appendix I.B,  the above automorphism groups remain unchanged under $\FrS \longrightarrow {\cal R}$.

\subsection{Dilational momenta}

\n{\bf Definition 1} The (4, 1) shape theory has three dilational momenta 
\be
\pi_i \es  \frac{\pa}{\pa \rho_i}  \m .  
\ee
\n{\bf Definition 2} In the scale--shape version, one has additionally a {\it configuration space radial momentum}  
\be 
\pi_{\rho} \es  \frac{\pa}{\pa \rho}  \m .  
\ee 
Let us define also the {\it absolute dilational momentum}, 
\beq
\pi_{\rho} \es   \rho \, \frac{\pa}{\pa \rho} 
           \es   \rho \, \pi_{\rho}            \m .
\eeq

\subsection{Killing vectors and isometry groups}

\n{\bf Structure 1} The (4, 1) shape spaces are pieces of $\mathbb{S}^2$, which has an $SO(3)$ of Killing vectors  
\beq
\sD_i := \epsilon_{ijk} \nu_j \frac{\pa}{\pa \nu_k}  \m , 
\label{Di}
\eeq
among which one as usual takes a notably simpler form in polar coordinates: 
\beq
\sD_3   \es  \frac{\pa}{\pa\varphi}  \m .
\eeq
\n{\bf Remark 1} While there are no angles in 1-$d$, 
the mathematics familiar from angular momentum in 3-$d$ recurs under the new interpretation of {\it relative dilational momenta} \cite{AF}. 

\m

\n{\bf Remark 2} Let us also use $\sD_3$ as an example of what dilational momentum is.  
Expanding (\ref{Di}) 
\beq
\sD_3  =   \nu_1\upi_2 - \nu_2\upi_1  
      \es  \nu_2\frac{\nu_1}{\nu_2} - \sD_1\frac{\nu_2}{\nu_1}  \m ,   
\label{WillBeA3}
\eeq
so we have a weighted relative dilational quantity. 
                   In Jacobi H-coordinates, this corresponds to a particular exchange of dilational momentum between the left and right post clusters.  
On the other hand, in Jacobi K-coordinates, the exchange is between the blade-face and the axe-handle (defined as per Fig I.8).  

\m

\n{\bf Proposition 1} All of these Killing vectors remain globally valid for 
\beq
\w{\FrS}(4, 1)   \es   \frac{\FrS(4, 1)}{\mathbb{Z}_{2\,\si\sn\sv}} 
                 \es   \mathbb{RP}^2                                 \m . 
\eeq
Thus 
\beq
Isom(\FrS(4, 1)) = Isom(\w{\FrS}(4, 1)) 
                 = SO(3)                                       \m .   
\eeq
On the other hand, for $\FrI\FrS(4, 1)$ and $\Leib_{\sFrS}(4, 1)$, we need a $\sD_3^{\prime}$ about the pole whose equator includes the region's T--Ce edge.
Thus
\beq 
Isom(\FrI\FrS(4, 1)) =  Isom(\Leib_{\sFrS}(4, 1)) 
                     =  SO(2) 
					 =  U(1) 
					 =  \mathbb{S}^1              \m .   
\eeq

\m

\n{\bf Structure 2} The (4, 1) scale and shape spaces are pieces of $\mathbb{R}^3$, which has ab inito three translational Killing vectors,  
\beq
\pi_i \es  \frac{\pa}{\pa\rho_i}  \m ,
\eeq
and an SO(3) of Killing vectors, now most naturally expressed as 
\beq
\sD_i \es  \epsilon_{ijk} \rho_j \frac{\pa}{\pa \rho_k}  \m .  
\label{Di-rho}
\eeq
\n{\bf Proposition 2} In $\w{\cal R}(4, 1)$, the identification of the half-plane (Fig S(4, 1)-Met-Top.2) globally breaks all three translations but not the rotations.
The same occurs for $\FrI{\cal R}(4, 1)$ in Fig \ref{R(4, 1)}.3), and for $\w{\cal R}(4, 1)$ in Fig \ref{R(4, 1)}.4). 
Thus 
\beq
Isom(\w{\cal R}(4, 1)) =   Isom(\FrI{\cal R}(4, 1)) 
                       =   Isom(\Leib_{\cal R}(4, 1)) 
					   =   SO(3) 
					   =   U(1) 
					   =   \mathbb{S}^1                \m .
\eeq
%
{\begin{figure}[ht]
\centering
\includegraphics[width=0.7\textwidth]{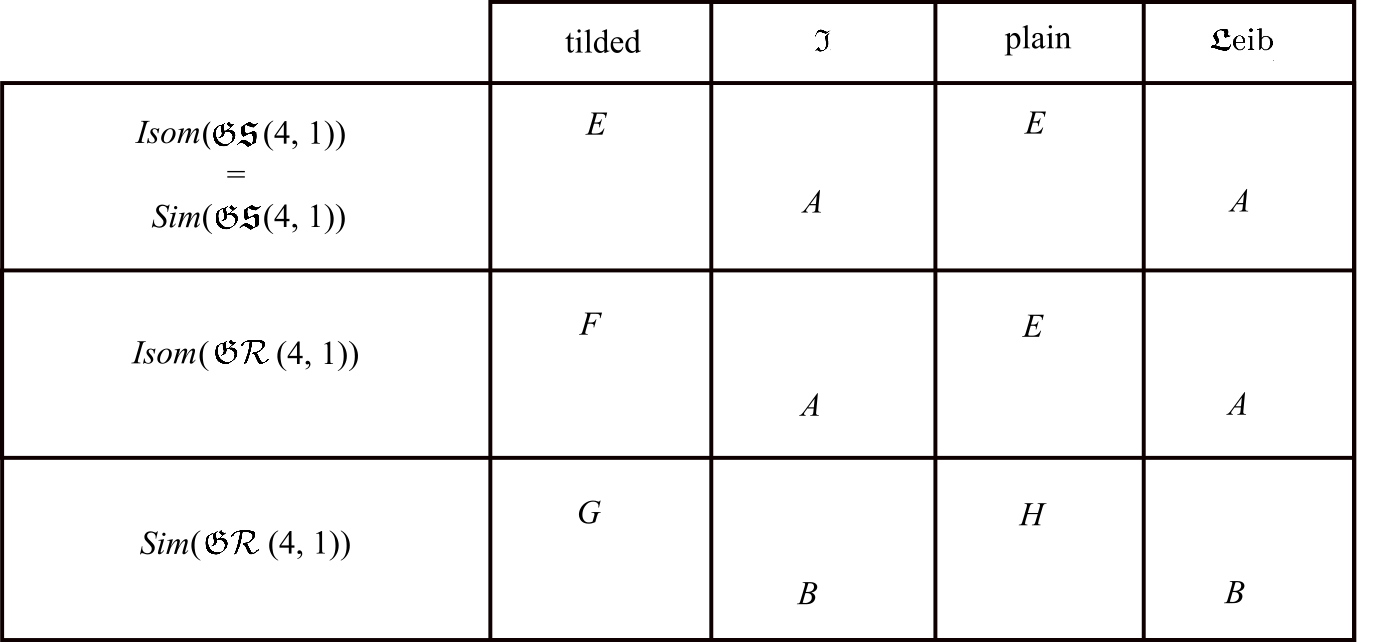}
\caption[Text der im Bilderverzeichnis auftaucht]{\footnotesize{Table of pattern of isometry and similarity groups for (scale and) shape spaces for 4 points in 1-$d$.}} 
\label{12-2}\end{figure} } 
 
\n{\bf Structure 3} See also summary figure \ref{12-2}, which now uses, in addition to Fig I.45's notation, 
\be
E                 :=         Rot(3) 
                   =         SO(3) 
            \m \s{m}{=} \m  \mathbb{RP}^3                                                    \m ,
\ee
\be
F              :=       Eucl(3) 
                =       Tr(3)         \rtimes  Rot(3) 
				=       \mathbb{R}^3  \rtimes  SO(3) 
        \m \s{m}{=} \m  \mathbb{R}^3  \rtimes  \mathbb{RP}^3                              \m ,
\ee
\be
G              :=        Sim(3) 
                =        Tr(d)  \rtimes \{ Rot(3) \times Dil \}
                =       \mathbb{R}^3  \rtimes \{ SO(3) \times \mathbb{R}_+ \}				
        \m \s{m}{=} \m  \mathbb{R}^3  \rtimes  \{ \mathbb{RP}^3 \times \mathbb{R}_+ \}
                =       \mathbb{R}^3  \rtimes  \mC(\mathbb{RP}^3)                         \m ,
\ee
\be
H              :=        Rot(d) \times Dil  
                =        Tr(3)         \times   \mathbb{R}_+  
        \m \s{m}{=} \m  \mathbb{RP}^3  \rtimes  \mathbb{R}_+ 
                =       \mC( \mathbb{RP}^3)                                               \m .  
\ee
\n{\bf Remark 3} Finally note that the constant $\phi$-curves in each of H and K coordinates are furthermore privileged as Killing vector preserving curves.
Such `adapted' submanifolds privileged by Killing vector preservation will play a substantial role in the theory of (4, 2) shapes \cite{IV, Affine-Shape-2, Forthcoming}.

\section{Conclusion}\label{Conclusion-II}

\subsection{(4, 1) as the smallest nontrivial theory of inhomogeneity and uniformity}

We have demonstrated in Part II that there are many more qualitative types of shape in (4, 1) than in (3, 1).
For (3, 1), many notions of shape can be defined, but most of these end up being co-realized by the very limited range of special shapes available: 
binary coincidence-or-collision B, merger M and half-way shape H.  
For (4, 1), there are enough special shape points and arcs that most notions of shape -- most uniform shape, largest and smallest shape per unit moment of inertia, 
most symmetric shape, coincidence-or-collision structure, uniformity structure, merger structure, symmetry structure in each of space, 
space of separations and space of Jacobi vectors -- are realized by different subsets of shape space. 
This further justifies Part I's introduction of such diversity of notions of shape, alongside Part II's Appendix A of quantifiers for many of these notions 
so as to distinguish between manifestations of different strength of these shape properties, and indeed Part II's subtitle.  
In this way, (4, 1) succeeds as a first nontrivial theory of both exact and approximate notions of each of inhomogeneity and uniformity, 
which are all matters of interest in Classical and Quantum Cosmology, Foundations of Physics and Background-Independent Quantum Gravity.   
Points A) to F) below provide more specific justification.  

\m

\n A) Already at the topological level where the coincidence--collision structure is the only distinguishable feature, 
(4, 1) suffices to get away from (3, 1)'s topological Leibniz space reducing to a space of partitions.
The (4, 1) Leibniz space is further fine-grained by discerning between exterior and interior binary coincidences-or-collisions, B and B$^{\prime}$ respectively.  
Furthermore, the distinguishable particle topological shape spaces have 74 and 37 vertices for (4, 1) to just 12 and 6 for (3, 1), 
with the (4, 1) ones additionally forming a complicated graph structure (Fig 3) to the (3, 1) ones just being cyclic graphs. 
This is partly underpinned by the corresponding (4, 1) metric shape spaces being $\mathbb{S}^2$ and $\mathbb{RP}^2$ as topological manifolds 
as compared to the (3, 1) shape spaces being just circles $\mathbb{S}^1$.
It is partly underpinned by the (4, 1) relabelling symmetry group $S_4$ being rather larger than (3, 1)'s $S_3$, 
and furthermore by the former having a rather more interesting action on $\mathbb{S}^2$, giving the cube-octahaedron tessellation.  
The 24 tiles involved here moreover have the interpretation of $\FrI\FrS(4, 1)$: the indistinguishable-particle configuration space. 

\m

\n B) As for (3, 1), topologically meaningful coincidences-or-collisions do not suffice to understand the periphery of Leibniz space, 
since some of the shapes in question have solely metric distinctive properties. 
For (3, 1), the $\Leib_{\sFrS}$ interval's missing end-point shape turned out to be all of the most uniform, merger, reflection symmetric and maximum sized shape, M. 
For (4, 1), the $\Leib_{\sFrS}$ scalene spherical triangle has 2 edges -- B and B$^{\prime}$ -- and 2 vertices -- ternary T and double-binary D coincidences-or-collisions -- 
provided by the coincidence-or-collision structure;  
its remaining edge and vertex turn out to be, respectively, a whole arc of reflection symmetric shapes Ref, and the central binary coincidence-or-collision Ce. 
This is clearly the intersection of Ref and the interior binaries B$^{\prime}$.  
On the other hand, $\FrI\FrS$ has a fully understood periphery at the topological level for each of (3, 1) and (4, 1).
For (3, 1), this is two B points, whereas for (4, 1) it is one D and two T points, joined by one B$^{\prime}$ and two B lines.  
This represents a first price to pay for taking Leibniz's Identity of Indiscernibles to its logical conclusion for the (4, 1) similarity model: 
$\FrI\FrS$'s periphery already admits an adequate topological characterization whereas $\Leib_{\sFrS}$ requires a topological-and-metric level characterization.   
At the metric level, moroever, $\FrI\FrS(4, 1)$ is an isosceles spherical triangle. 
This greater symmetry relative to $\Leib_{\sFrS}(4, 1)$ represents a second price to pay for taking Leibniz's Identity of Indiscernibles to its 
logical conclusion for the (4, 1) similarity model. 

\m

\n C) While Ref is still a type of merger for (4, 1), there turn out to be two other types of merger, M and M$^{\sfQ}$. 
In Leibniz space, $\mM^{\sfQ}$ emanates from the corner Ce and ends with a head-on collision 
%
%
with the edge B [at the 1: 2 ratio exterior binary coincidence-or-collision shape, B(1 : 2)]. 
On the other hand, M also emanates from Ce and collides with B  but now not head-on, so the M-arc folds as a cusp at this intersection point 
(the 1: 1 ratio exterior binary coincidence-or-collision shape)
It next collides with the edge Ref at U -- the most uniform configuration -- folding as another cusp -- before ending at the corner T. 
This twice-folded M trajectory moreover contains a self-intersection at the 1 : 1 : 2 ratio shape, by which it forms a swallowtail catastrophe. 
The $\FrI\FrS$ version of the M trajectory has three folds [at Ce and its two B(1: 2) shapes] and three self-intersections [at U and its two 1 : 1 : 2 shapes], 
by which it forms the more complex butterfly catastrophe. 
This represents a third price to pay in taking Leibniz's Identity of Indiscernibles to its logical conclusion for the (4, 1) similarity model:  
the removal of redundancy can turn straightforward geodesics into cusped and self-intersecting trajectories and moreover higher-order catastrophes appear.  
Finally M and M$^{\sfQ}$ also intersect, at the 1 : 1 : 3 ratio shape, by which the shape space contains 8 special points (summarized within Fig 30).  

\m

\n D) The various shape quantifiers' maxima and minima are then quite well spread out over these 8 special points. 
In particular, the three corners of $\Leib_{\sFrS}$ have the characterization that T is the largest coincidence, 
whereas Ce and D are the largest and smallest shape per unit moment of inertia. 
The global maximum of uniformity is indeed at U, but the 1 : 1 : 2 ratio shape has a local maximum of uniformity, 
by which we also name is $\mU^{\sA}$ for (maximally) uniform asymmetric shape.  
U and U$^{\sA}$ serve moreover as shape-theoretic centres for $\FrI\FrS$ and $\Leib_{\sFrS}$.  
Maximally uniform states serve furthermore as shape space centres turns out to be a common occurrence 
and is of likely interest to the Cosmology, Foundations of Physics and Quantum Gravity communities.
Moreover, $\Leib_{\sFrS}$ having a centre which is less distinguished -- both as a centre and as a notion of uniformity -- 
than $\FrI\FrS$ is a fourth and final price to pay in taking Leibniz's Identity of Indiscernibles to its logical conclusion for the (4, 1) similarity problem.  

\m

\n E) For the (4, 1) model, the number of qualitative types has increased by up to 2 orders of magnitude as compared to the (3, 1) model, 
as is evidenced by comparing Fig I.40, with Figs \ref{(4, 1)-Q-Top}, \ref{(4, 1)-Q-Lag}, \ref{(4, 1)-Q-Uni} and \ref{(4, 1)-Q-Jac}.    

\m

\n F) Consult Figures \ref{Collision-Set}, \ref{Uniform-Structure}, \ref{21} and \ref{29} regards the non-co-realization of the merger, 
uniform, and symmetric in each of space, Lagrangian separation space and Jacobi separation space for (4, 1).  
(4, 1) is moreover a first instance of somewhat nontrivial shape submanifolds: 2-$d$ shape space supports 1-$d$ submanifolds, 
which are less trivial than points, but are still geometrically trivial in a number of ways. 
These geometrical nontrivialities require (5, 1) -- so as to support 2-$d$ submanifolds, or (6, 1) or Part IV's (4, 2) model: so as to support 3-submanifolds.
Figure \ref{Uniform-Structure} moreover involves {\sl collections of points and arcs}, including some cases of missing end-points and topological nontriviality; 
these are one instance of stratified manifolds \cite{Whitney46, Thom55, Whitney65, Thom69, Pflaum, Kreck, Sniatycki} arising in Shape Theory; 
see \cite{Kendall, GT09, PE16, KKH16, III, Affine-Shape-2, Project-2} for others.
This is the setting for some of the Differential Geometry advances in Shape Theory announced in the current treatise.

\subsection{Future research directions stemming from Part II}

Aside from the aforementioned conceptual and theoretical applications to Cosmology, Foundations of Physics and Background-Independent Quantum Gravity, 
and the (4, 1) model's extension to Part IV's (4, 2) `quadrilateralland' model, some further future research directions of note stemming from the current treatise are as follows. 

\m

\n I) Increase in topological shape space graph size and complexity, occurrence of cusps and catastrophes in $\Leib_{\sFrS}$ and $\FrI\FrS$, 
and large increase in number of metric-level 
qualitative types are moreover permanent features as point number $N$ further increases in dimension 1.  
As we shall see in Papers III and IV, however, the first of these features goes away if the dimension $d$ increases away from 1.  
For this feature is rooted on points splitting space, which only holds if space is 1-$d$.  
One may expect the number of notions of shape receiving distinct realizations will increase similarly each time $N$ goes up by 1, 
by which (5, 1)'s increase in complexity relative to (4, 1) will be similar to the current treatise's increase in complexity relative to Part I's (3, 1) model. 
This is moreover of quite some relevance since various shape theories have $N = 5$ or $6$ as a minimal requirement \cite{AMech, Affine-Shape-1, Affine-Shape-2}. 
One can hope that the current treatise' analysis of shapes and shape spaces both contains a partial cover of the notions $N = 5$ and $6$ require, 
alongside its {\sl way of thinking} being extendible to produce whatever further notions at least the (5, 1), (6, 1), (5, 2) and (6, 2) models require.
We also note that (4, 1) is the first reasonably representative 1-$d$ similarity shape space in many ways.  

\m

\n II) A further application of the combinatorial shape measures introduced in the current treatise is in comparison between different $(N, d)$ shape models.
This is of particular relevance to the variable particle number extension of relational particle models, which, as laid out in Epilogues II.C and III.C of 
\cite{ABook}, is a model arena for General Relativity including spatial topology change \cite{GH92}, i.e\ Sec I.12.2's Big Superspace.
\n Moreover, Papers III and IV consider specifically relative-angle counterparts of some of these combinatorial shape measures.  

\m

\n III) The current treatise's consideration of globally-valid Killing vectors over $\FrS(4, 1)$, $\FrI\FrS(4, 1)$ and $\Leib_{\sFrS}(4, 1)$ 
is a crucial step for all of establishing Classical Dynamics conserved quantities on these, finding and optimally formulating probability distributions over these, 
and quantizing over these \cite{QLS, Quantum-Triangles}.   

\m

\n{\bf Acknowledgments} I thank Chris Isham and Don Page for discussions about configuration space topology, geometry, quantization and background independence. 
I also thank Jeremy Butterfield and Christopher Small for encouragement, and Reza Tavakol for reading the manuscript.  
I thank Don, Jeremy, Reza, Enrique Alvarez and Malcolm MacCallum for support with my career. 
I also thank Angela Lahee and the typesetting editors for my book manuscript this past summer for prompt and excellent work without which the current treatise could not 
have been completed by this date. 
I finally thank a very kind generous and patient person who has helped me a great deal.

\begin{appendices}

\section{Coincidence, merger, uniformity, symmetry and clumping quantifiers}\label{Appendix-II}

\subsection{A simple type of counting measure}\label{Counting-Measure-Sec}

{\bf Structure 1} Consider $m$ numbers $v_a$, $a$ = 1 to $m$, whose values which can be repeats. 
These numbers realize a partition, with each actualized value $v_b$ corresponding to a box with $n_b$ occupants, $b = 1$ to $k$.  
A suitable and widely known counting measure is then 
\be
{\cal C}(\bv) = \mbox{ln}
\left(
\prod_{b = 1}^k n_b!
\right)                   \m . \label{Counting-Measure}
\ee
Some advantages of using this sort of counting measure are as follows.  

\m

\n 1) Single-occupancy partitions -- numbers not participating in coincidences -- contribute nothing, by 
\be
\mbox{ln} \, 1 = 0   \m .   
\ee
\n 2) It is additive as regards unions of independent sets of boxes, by  
\be
{\cal C}(\bv_1) + {\cal C}(\bv_2) \es  \mbox{ln}  \left(  \prod_{b_1 = 1}^{k_1} n_{b_1}!  \prod_{b_2 = 1}^{k_2} n_{b_2}!  \right) 
                                  \es  \mbox{ln}  \left(  \prod_{b = 1}^{k_1 + k_2} n_{b}!                                \right)  \m .
\label{Indep}
\ee
{\bf Remark 1} Formula (\ref{Counting-Measure}) gives a simple notion of entropy for the partition; 
with a minus sign factor, it is a notion of negentropy alias information.  
Applying such a counting measure to shape theory is not however {\sl motivated} by entropic or informational considerations;  
rather both modelling these and modelling shape structures have a {\sl common need} for factorial-and-logarithm mathematics.  

\m

\n{\bf Structure 2} Our shape-theoretic considerations are, rather, more directly in the form of considering 
\be
Sym(\mbox{O-A}) \m : 
\ee 
a finite symmetry group acting on aspect A of object O.  
If moreover   
\be
Sym(\mbox{O-A})   \es  \bigtimes_{b = 1}^{k} S_b
\ee
-- a direct product of permutation groups, as occurs in our shape-theoretic applications -- then 
\be
|Sym(\mbox{O-A})| \es  \prod_{b = 1}^{k} n_b!       \m . 
\ee
Thus a counting measure of the aforementioned type, 
\be
{\cal AM}(\mO)    \:=  \mbox{ln}|Sym(\mbox{O-A})|  \m , 
\ee
arises.  

\m

\n There is moreover a relative version of ${\cal C}(\bv)$ -- i.e.\ the difference of two quantities of this type, which is thus of the computational form 
\be
\mbox{ logarithm of a ratio of products of factorials } .   
\ee
{\bf Definition 1} One context in which to use this is the removal of a common background contribution: 
\be
\mbox{distinctiveness}(\bv) := {\cal C}(\bv) - {\cal C}(\bv(\mbox{background}))  \m . 
\ee
{\bf Definition 2} Another is the comparison of an entity $\mv$ with its constituent parts $\mv_c$, $c = 1$ to $p$, 
\be
\mbox{excess} := {\cal C}(\bv) - \sum_{c = 1}^{p}{\cal C}(\bv_c)  \m . 
\ee
If the parts are independent, this is zero by (\ref{Indep}).
The whole can however be more than its parts: {\it positive excess}, corresponding to the counting measure exhibiting {\it superadditivity} at some point.
The whole can also be less than its parts: {\it negative excess}, corresponding to the counting measure exhibiting {\it subadditivity} at some point.  

\m

\n{\bf Remark 2} In the current treatise, 
        a cause of superadditivity is additional {\it correlation} between the parts, 
whereas a cause of subadditivity   is            {\it overcrowding}.

\subsection{Coincidence-or-collision measure}\label{Coincidence-Measure}

\n A quantifier for strength of point coincidences or particle collisions along such lines is as follows.  

\m

\n{\bf Definition 1} For shape S containing the coincidence-or-collision structure $\Co$ on which $Sym$ acts, the {\it point-or-particle measure} 
\be
{\cal P}{\cal M}(\mS) \:= \mbox{ln}|Sym(\mbox{S-${\cal P}$})| 
                      \es   
\mbox{ln}
\left(
\prod_{b = 1}^k n_b!
\right)               \m ,
\ee
where $n_b$ is the number of points-or-particles coincident at each position.   

\m

\n{\bf Examples} We give (2, 1), (3, 1) and (4, 1) examples of this, and all of this Appendix's other Lagrangian quantifiers, in Fig \ref{Measures}.  

\m

\n{\bf Remark 1} ${\cal PM}$ is such that a given coincidence-or-collision type's value is independent of which ($N$, $d$) problem one is considering.
This feature is useful as regards saving on computation and permitting ready comparison between models with different ($N$, $d$).   
It reflects that ${\cal P}{\cal M}$ relies solely on topological shape input, and, indeed (a fortiori for $d = 1$) solely on partition input.  
The other side of the coin is that this notion cannot probe metric clumping content (or, in 1-$d$, ordering of the partition along the line).  

\m

\n{\bf Remark 2} The identification of topological shapes as partitions points to a number of more sophisticated measures being available (see e.g.\ \cite{G02}).
This is of particular relevance once $N$ is large enough that ${\cal P}{\cal M}$ no longer discerns between distinct types of partition.

\subsection{Relative separation measure}\label{Separation-Measure}

\n{\bf Definition 1} For a shape S containing the relative separation structure ${\cal SM}$ on which $Sym$ acts, the {\it relative separation measure} 
\be
{\cal S}{\cal M}(\mS) \:= \mbox{ln}|Sym(\mbox{S-${\cal S}$})| 
                      \es   
\mbox{ln}
\left(
\prod_{b = 1}^k n_b!
\right)                                                           \m , 
\ee
where $n_b$ is the number of separations of each equal size.   

\m 

\n{\bf Remark 1} Both coincidences-or-collisions and (Lagrangian) uniformities contribute to this measure; see Fig \ref{Measures} for examples of this.   
Thus this measure is sensitive to both maximal homogeneity and maximal inhomogeneity extremes, and so is an unsigned quantifier of atypically large or small inhomogeneity.

\subsection{Uniformity strength measures}\label{Uniformity-Measures}

\n A first branch of natural sequels is, to restrict our measures so as to be able to detect uniformity separately from coincidence; 
this case's `naturality' follows from that of (Lagrangian) uniformity.  

\m

\n{\bf Definition 1} The {\it adjacent uniformity measure} of the shape S is
\be
{\cal AUM}(\mS) = \mbox{ln}|Sym(\mbox{S-(adjacent separations subject to restriction i)} )| 
                = \mbox{ln}
\left( 
\prod_{b = 1}^k n_b!
\right)                    \m , 
\ee
for $n_b$ now the number of equal adjacent relative separation values subject to Excision i) of Sec I.6.4. 

\m

\n{\bf Example 1} For (4, 1), this succeeds in giving a unique maximum value $\mbox{ln} \, 6  \approx 1.792$ for U, 
but returns just the same value $\mbox{ln}( 2! ) \approx 0.693$.  
This motivates defining the following more precise measure of uniformity strength; see Fig \ref{Measures} for its success.      

\m

\n{\bf Definition 2} The {\it total uniformity measure} 
\be
{\cal TUM}(\mS) \es  \mbox{ln}|Sym(\mbox{S-(${\cal S}$ subject to restrictions a) and b)  )}  ) | 
                \es  \mbox{ln}\left(\prod_{b = 1}^k n_b!\right)                                     \m ,
\ee
for $n_b$ now the number of equal relative separation values subject to Sec I.6.4's Excisions i) and ii).     

\m

\n{\bf Definition 3)} The {\it coincidence-or-collision-induced separation measure} ${\cal CSM}(\mS)$ of the shape S is the complementary count 
of separation coincidences which are induced by the collision structure, i.e.\ now solely the aforementioned restricted types i) and ii).

\subsection{Lagrangian-level examples}\label{Lagrangian-Examples}
%
{            \begin{figure}[!ht]
\centering
\includegraphics[width=1.0\textwidth]{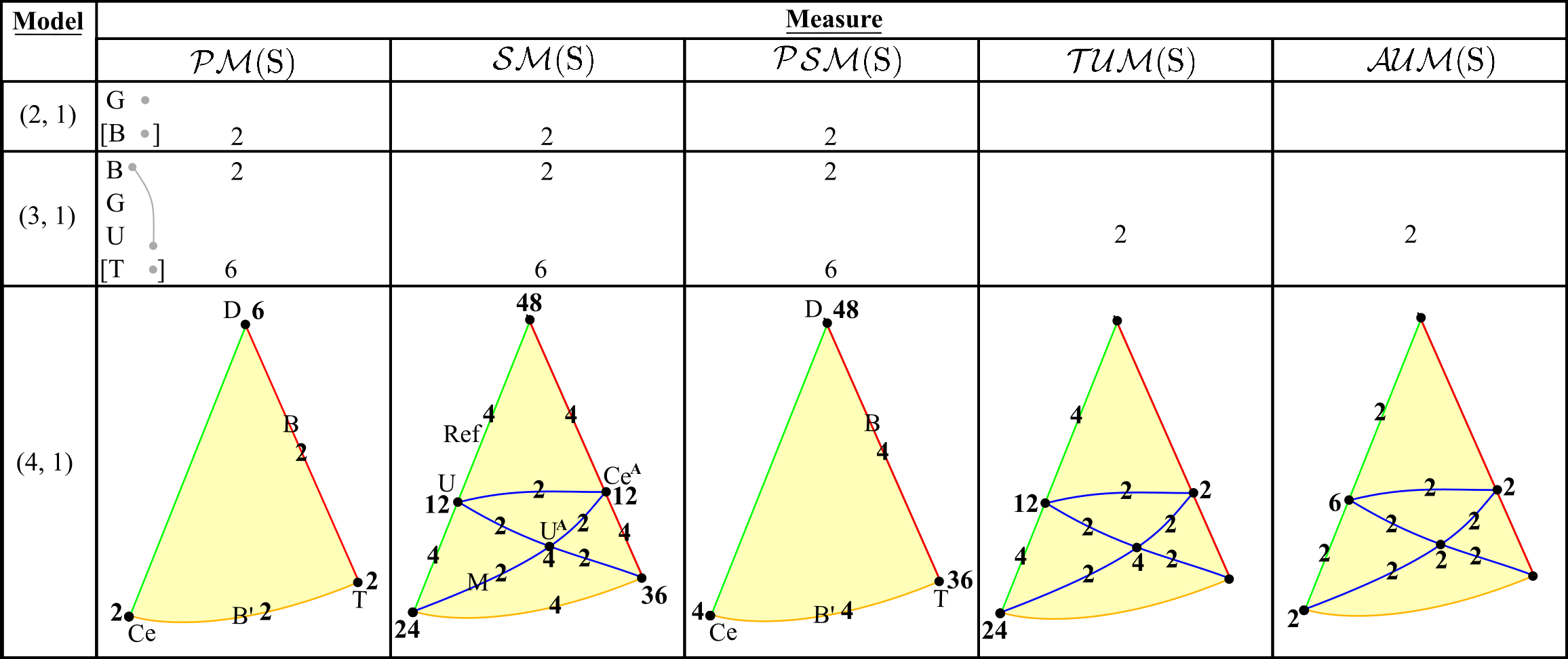}
\caption[Text der im Bilderverzeichnis auftaucht]{        \footnotesize{Examples 1, 2 and 3) of coincidence, separation and uniformity Lagrangian measures.
For convenience of presentation, the given value are prior to taking the logarithm.  
All points, arcs and regions not labelled by a number implicity return the trivial value 1, so $\mbox{ln} \, 1 = 0$} }
\label{Measures} \end{figure}          }

\n{Example 4)} The Ce and Ce$^{\sA}$ shapes -- the only intersections of the coincidence-or-collision structure and the uniformity structure -- provide a vivid example of excess. 
Reading off Fig \ref{Measures},
\be
\mbox{Excess-${\cal SM}$}(\mbox{Ce})         :=    \{{\cal SM} - {\cal CSM} - {\cal TUM}\}(\mbox{Ce}) 
                                              =    \mbox{ln} (24/4 \times 2) 
											  =    \mbox{ln}  \, 3 
										  \approx  1.099 
\m ,
\ee
\be
\mbox{Excess-${\cal SM}$}(\mbox{Ce$^{\sA}$})   :=     \{{\cal SM} - {\cal CSM} - {\cal TUM}\}(\mbox{Ce$^{\sA}$}) 
                                                =     \mbox{ln} (12/4 \times 2) 
											    =     \mbox{ln}  \, 3/2 
                                             \approx  0.406                                                       \m , 
\ee
whereas 
\be
\mbox{Excess-${\cal SM}$}(\mS) = 0 \mbox{ for all other (4, 1) shapes S } .  
\ee

\subsection{Hierarchical merger strength measure}\label{Merger-Measures}

\n Another branch of natural sequels is to generalize our measures so as to be able to detect the whole of the general merger hierarchy; 
this is now `natural' in a Jacobian sense.  

\m

\n{\bf Remark 1} A first approach is to detect each level of hierarchy's self-coincidences separately, 
and also each distinct pair of levels of hierarchy's mutual coincidences separately.   
This is done from passing from considering the Lagrangian inter-particle separations 
                            -- the ${\cal P} \times {\cal P}$ subcase of Jacobian inter-particle cluster separations -- 
to considering whichever choice of ${\cal A} \times {\cal B}$ Jacobian inter-particle cluster separations, 
where ${\cal A}$, ${\cal B}$ are each a single selection from the sequence of sets ${\cal P}$, ${\cal X}$, ${\cal T}$ ...

\m

\n{\bf Definition 1} The ${\cal A}$-{\it level self-merger strength measure} of the shape S is 
\be
{\cal S}({\cal A}){\cal M}{\cal M}(\mS) :=   \mbox{ln}|Sym(\mbox{S-(${\cal A}$ positions)})| 
                                        \es   
\mbox{ln}
\left(
\prod_{b = 1}^k n_b!
\right)                                                      \m , 
\ee
where $n_b$ is the number of separations of each equal size between ${\cal A}$-type centres of mass. 

\m

\n{\bf Definition 2} The $({\cal A}$, ${\cal B})$-{\it levels mutual-merger strength measure} of the shape S is   
\be
{\cal M}({\cal A}, {\cal B}){\cal M}{\cal M}(\mS) :=    \mbox{ln}|Sym(\mbox{S-((${\cal A}$, ${\cal B}$}) \mbox{ positions})| 
                                                  \es    
\mbox{ln}
\left(
\prod_{b = 1}^k n_b!
\right)                                                      \m , 
\ee
where $n_b$ is the number of separations of each equal size between ${\cal A}$-type centres of mass and ${\cal B}$ ones.  

\m

\n{\bf Remark 2} Clearly Definition 1 is a subcase of Definition 2: 
\be
{\cal M}({\cal A}, {\cal A}){\cal M}{\cal M} = {\cal S}({\cal A}){\cal M}{\cal M}                \m ;  
\ee
also ${\cal S}({\cal P}){\cal M}{\cal M}$ returns the Lagrangian particle concept ${\cal C}{\cal M}$.

\m

\n{Remark 3} A second approach is to consider the general merger hierarchy all at once, by the following sum construct (or product construct, once inside the logarithm).

\m

\n{\bf Definition 3}  The {\it general merger strength measure} of the shape S is 
$$
{\cal G}{\cal M}{\cal M}(\mS) \:=                     \sum_{{\cal A} \geq {\cal B}}            {\cal M}({\cal A}, {\cal B}){\cal M}{\cal M}(\mS)  
                              \es                     \sum_{{\cal A} \geq {\cal B}}   \mbox{ln}|Sym(\mbox{S-($({\cal A}$, ${\cal B}$) positions})|
$$
\be
							  \es  \mbox{ln}  \left(  \prod_{{\cal A} \geq {\cal B}}           |Sym(\mbox{S-($({\cal A}$, ${\cal B}$) positions})|  \right)
							  \es  \mbox{ln}  \left(  \prod_{{\cal A} \geq {\cal B}}  \prod_{b({\cal A}, {\cal B})}  n_{b({\cal A}, {\cal B})}!  \right)      \m .
\ee
Here $n_{b({\cal A}, {\cal B})}$ is a necessarily more explicit notation for 
the number of separations of each equal size between ${\cal A}$-type centres of mass and ${\cal B}$ ones, 
with reference also to ${\cal P}$, ${\cal X}$, ${\cal T}$... as the order involved in the sums and some of the products.   				   

\m

\n{\bf Definition 4}  The {\it acategorical merger strength measure} 
of the shape S draws its values, rather, from the full set of centres of mass without regard for their hierarchical type.   

\m

\n{\bf Example 1} Use of the above measures indeed succeed in picking out the M$^{\sfQ}$ arc, 
and in identifying its intersection points with other general merger arcs as more substantially special. 

\m

\n{\bf Remark 4} Merger strength measures can be used to compare a special point with the arcs it is an intersection of.  
Early on in the study, special points with excess merger strength may indicate that some of the merger arcs intersecting there have so far been overlooked.
On the other hand, at the end of the study, 
this approach can be used to quantify the extent to which excess merger strength is a final feature rather than just a symptom of having hitherto overlooked arcs.    

\m

\n{\bf Definition 5} 
\be
\mbox{(excess merger strength of intersection point $\mX$ of arcs $\mY_{c}$)} = {\cal G}{\cal M}{\cal M}(\mX) - \sum_{c = 1}^p{\cal G}{\cal M}{\cal M}(\mY_c)  \m .  
\ee
\n Then early on in the study, we may have (excess merger strength of intersection point $\mX$ of {\sl known} arcs $\mY_{c}$), 
which splits into (excess merger strength of intersection point $\mX$ of {\sl subsequently discovered} arcs $\mY_{d}$) and (final value of the excess).  

\m

\n{\bf Example 2} In D, Ref acts to `double' B's coincidences, but these already make use of two pairs of $\fX$'s (Fig \ref{No-Excess}).
Thus two of these have to be used twice in Ref's `doubling' action, by which overcrowding ensures that Ref's action is in practise less than a doubling.  
This illustrates Shape Theory's negative excess due to overcrowding, corresponding to being forced to re-use some centres of mass in establishing 
underlying hierarchical coincidences due to there being no more centres of mass of that order available.  

\m

\n{\bf Remark 5} One can also readily formulate hierarchical separation measures, 
applying the current subappendix's Jacobian generalization to CoM separation coincidences rather than CoM coincidences.  

\m

\n{\bf Remark 6} A problem of note which appears in this case is that some CoM separation coincidences already imply others.  
Thus it makes sense to subtract off a common background term which counts the always-implied coincidences.  

\m

\n{\bf Definition 6} Let us coin the term {\it CoM separation rigidity} for CoM separation coincidences that are universal among ($N$, $d$) shapes.

\m

\n {\bf Example 3)} The coincidence of the central particle with a 2-particle centre of mass in the (3, 1) M configuration 
            also implies coincidence with T and thus alignment of various $\fP\fX$ and $\fP\fT$ separations.  
This is not however universal.
That the three centres of mass form the same figure as the particles, but back-to-front and half-sized {\sl is} universal for (3, 1), 
but does not contribute any separation coincidences. 
For (4, 1), however, the centre of mass rigidity of Fig \ref{Pre-Varignon} is both universal and contributes CoM separation coincidences.  
Finally, once we consider 2-$d$ shapes in Papers III and IV, we will be able to pin well-known geometrical names on these particular 3- and 4-particle examples of CoM rigidities.  
%
{            \begin{figure}[!ht]
\centering
\includegraphics[width=0.7\textwidth]{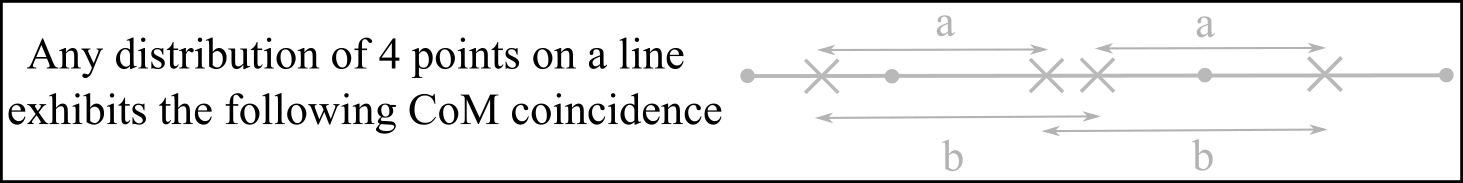}
\caption[Text der im Bilderverzeichnis auftaucht]{        \footnotesize{Centre of mass position and separation rigidity for (4, 1).} }
\label{Pre-Varignon} \end{figure}          }

\m

\n{\bf Definition 7} 
\be
(\mbox{CoM separation distinction}) = (\mbox{CoM separation measure}) - (\mbox{CoM separation rigidity})
\ee 
is a more conceptually desirable final form of CoM separation measure.

\subsection{Quantifying non-extremal configurations}\label{Non-Extremal}

\n{\bf Observation 1} A remaining problem with the quantifiers so far is that they are picking out {\sl just} D, T, Ce, U 
rather than configurations very near to D, T, Ce, U as well.

\m

\n{\bf Approach 1} One way of handling this involves feeding in counts of how many elements of the $\epsilon$-tolerant coincidence, 
merger, Lagrangian-uniform and Jacobian-uniform equations into the preceding subappendices' counting devices.  

\m

\n{\bf Approach 2} An alternative method is to tessellate space into equal-sized cells, which are to be treated as data bins for point-or-particle positions.
This discretization approach is along the lines of Roach's study \cite{Roach} for Statistics on $\mathbb{S}^1$.  

\m

\n{\bf End-Comment} Part IV includes the counterpart of this Appendix as regards quantifying the 2-$d$ angular counterparts of clumping and uniformity 
introduced in Part III: collinearity, perpendicularity and angle-uniformity, in both exact and approximate terms.

\section{Graph-Theoretic Analysis of Number of Qualitative Types}

\subsection{Preliminary definitions and lemmas}

\n{\bf Lemma 1 (Computational Formula for number of exact qualitative types)}  
\be 
Q(\mG)  =   2 \, e(\mG) + \chi(\mG) 
       \es \sum_{v \in \sG} d(v) + \chi(\mG)  \m .  
\ee 
\Proof Subtract $\chi(\mG) = V - E + F $ from the definition $Q(\mG) = V + E + F$. 
Then apply the first equation of Appendix I.A to obtain the second equality. $\Box$ 

\m

\n{\bf Definition 1} The {\it perimeter graph} $\mP(\mG)$ of a geometrically embedded graph $\mG$ consists of the external vertices and external edges.  
We use $\mbox{Int}(\mG)$ (`{\it interior part}' of $\mG$) to denote those parts of $\mG$ not in $\mP(\mG)$.  

\m

\n{\bf Remark 1} If this is not branching or self-intersecting (as for all shape space cases in the current treatise), 
\be
\mP(\mG) = \mC_k \m \mbox{ -- a cyclic graph --}
\ee
for some $k$. 
Under these circumstances, the following Lemma is thus immediate. 

\m

\n{\bf Lemma 2}
\be
|\mP(\mG)| =  e(\mP(\mG)) 
           =: p(\mG) \mbox{ : the {\it perimeter number} } . 
\ee
\n{\bf Remark 2} 
\be
\mbox{For compact without boundary shape spaces} \mma V(\mP(\mG)) = \emptyset \m , \m \mbox{so} \m \m p(\mG) = 0 \m . 
\ee
\n{\bf Definition 2} The {\it topological adjacency graph} $\mbox{Adj}(\mG)$ of a 2-$d$ geometrically embedded graph $\mG$ has vertex set 
\be
\{V(\mG), \m E(\mG), \m F(\mG)\}
\ee 
and edge set consisting of edges precisely whenever the two $\mbox{Adj}(\mG)$ vertex objects in question are topologically adjacent.  

\m

\n{\bf Remark 3} These conditions apply to all shape spaces in Papers II and III.

\m

\n{\bf Remark 4} It is immediately clear that 
\be
|\mbox{Adj}(\mG)| = V + E + F = Q              \m . 
\ee
Thus from Lemma 1, 
\be
|\mbox{Adj}(\mG)| = 2 \, e(\mG) + \chi(\mG)    \m . 
\ee
On the other hand, evaluating $e(\mP(\mG))$ requires further work.  

\m

\n{\bf Definition 3} Moreover 
\be
Q_{\sa\sp\sp\sr\so\sx}(\mG) := e(\mbox{Adj}(\mG)) 
\label{approx-def}
\ee
is itself one of the useful quantifiers of number of qualitative types (the `{\it first-order approximate qualitative types}').

\m

\n{\bf Remark 5} This provides a first justification for introducing Adj(G) in this analysis of qualitative types.  

\m

\n{\bf Definition 4} 
\be
Q_{\st\so\st\sa\sll}(\mG) := Q(\mG) + Q_{\sa\sp\sp\sr\so\sx}(\mG)
\label{Q-total}
\ee
is the {\it total number of exact and (first-order) approximate qualitative types}.

\m

\n{\bf Definition 5} $Q_{\sA\sp\sp\sr\so\sx}$ includes furthermore the second order terms around each point of interest 
which correspond to approaching this point approximately along one of the arcs of interest which intersect there. 

\m

\n{\bf Remark 6} This incorporates +1 qualitative type per external edge and +2 qualitative types per internal edge.    
Moreover, the approx to Approx distinction requires shape space dimension 2 to be nontrivial (Part I's 1-$d$ shape spaces support just one notion); 
this observation clearly also applies to total and the below definition of Total.  

\m

\n{\bf Definition 6} 
\be
Q_{\sT\so\st\sa\sll}(\mG) := Q(\mG) + Q_{\sA\sp\sp\sr\so\sx}(\mG)
\label{Q-Total}
\ee
is the corresponding new {\it grand total number of exact and approximate qualitative types}.

\subsection{Counting qualitative types}

\n{\bf Proposition 1} i) 
\be
Q_{\sa\sp\sp\sr\so\sx}(\mG)               =       2      \{  3 \, e(\mG) - p(\mG)\}  
                                         \es        \left\{ 3 \, \sum_{v \in \sG} - \sum_{v \in \sP(\sG)} \right\} \md(v)            \m . 
\label{1-i}
\ee
ii) 
\be
Q_{\st\so\st\sa\sll}(\mG) - \chi(\mG)     =       2      \{  4 \, e(\mG) - p(\mG) \}  
                                         \es        \left\{  4 \, \sum_{v \in \sG} - \sum_{v \in \sP(\sG)} \right\}  \md(v)          \m . 
\label{1-ii}
\ee
\Proof i) We can count out 
\be
e(\mbox{Adj}(\mG)) \es  \sum_{v \in \sG} w(v) + \sum_{f \in \sG} u(f) 
\label{First-Sum}
\ee
by attaching vertex and face weights to (an extended notion version of) the original graph $\mG$.  
\be 
w(v) = 2 \, n + \left\{     \s{    \mbox{3 \m $v \in \mP(\mG)$         }    }
                                     {    \mbox{4 \m $v \in \mbox{Int}(\mG)$  }    } \right. \mbox{ of degree $n$ + 2 }
\ee
can be trivially established by induction.
Thus 
\be
\sum_{v \in \sG} w(v) \es       \sum_{v \in \sI\sn\st(\sG)} \{2 \, n + 4 \} +  \sum_{v \in \sP(\sG)} \{2 \, n + 3\}
                      \es  2 \, \sum_{v \in \sG}            \{     n + 2 \} -  \sum_{v \in \sP(\sG)}        1 
					  \es  2 \, \sum_{v \in \sG}  \md(v)                    -             |\mP(\mG)|
                      \es  4 \, e(\mG)                                      -              \mp(\mG)                              \m .	
\label{1-I}					  
\ee
Here we have summed over the previous equation in the first equality, and  
expressed everything in terms of the whole original graph $\mG$ and the perimeter graph $\mP(\mG)$ in the second.
Then in the third equality, we have used the declared parametrization of the degree to rewrite the first sum as a sum of degrees 
and evaluated the second sum by the the definition of graph order.
We finally use the first equation of Appendix I.A in the fourth equality, alongside the Lemma 2 to bring in $p(\mG)$.

\m

\n On the other hand, 
\be 
u(f) = (\mbox{number of sides bounding $f$}) \m , 
\ee 
so 
\be
\sum_{f \in \sG} u(f)  \es  2 \, \sum_{e \in \sI\sn\st(\sG)} 1 + \sum_{e \in \sP(\sG)} 1 
                       \es  2 \, \sum_{e \in \sG}            1 - \sum_{e \in \sP(\sG)} 1
                       \es  2 \, e(\mG)                        - e(\mP(\mG))
                        =   2 \, e(\mG)                        -   p(\mG)                 \m .    
\label{1-II}						   
\ee
Here the face count uses internal edges twice and perimeter edges once in the first equality, 
and the second expresses everything in terms of the whole original graph $\mG$ and the perimeter graph $\mP(\mG)$. 
Then the third equality uses the definition of the size of a graph twice, and the fourth uses Lemma 2 to bring in $p(\mG)$.  

\m

\n Then using (\ref{approx-def}), (\ref{First-Sum}), (\ref{1-I}) and (\ref{1-II}) in succession, 
\be
Q_{\sa\sp\sp\sr\so\sx}(\mG) := e(\mbox{Adj}(\mG)) 
                             = 4 \, e(\mG) - p(\mG) + 2 \, e(\mG) - p(\mG) 
							 = 2 \{ 3 \, e(\mG) - p(\mG) \}                  \m , 
\ee
and we finish by using the first equation of Appendix I.A on each of $\mG$ and $\mP(\mG)$ for the last equality of (\ref{1-i}).  

\m

\n ii) Use Lemma 1 and the first equality of (\ref{1-i}) in (\ref{Q-total}) to get 
\be
Q_{\st\so\st\sa\sll}(\mG) = 2 \, e(\mG) + \chi(\mG) + 2 \{ 3 \, e(\mG) - p(\mG) \} 
                          = 2 \{ 4 \, e(\mG) - p(\mG) \} +  \chi(\mG)               \m ,
\ee
and we finish once again by using the first equation of Appendix I.A on each of $\mG$ and $\mP(\mG)$ for the last equality of (\ref{1-ii}).  $\Box$

\m

\n{\bf Proposition 2} i) 
\be
Q_{\sA\sp\sp\sr\so\sx}(\mG)               = 2      \{  5 \, e(\mG) - p(\mG)  \}  
                                         \es     \left\{  5 \, \sum_{v \in \sG} - 2 \, \sum_{v \in \sP(\sG)}  \right\}  \md(v)   \m . 
\label{2-i}
\ee
ii) 
\be
Q_{\sT\so\st\sa\sll}(\mG) - \chi(\mG)     =   4      \{ 3 \, e(\mG) - p(\mG)\} 
                                         \es  2 \left\{ 3 \, \sum_{v \in \sG} - \sum_{v \in \sP(\sG)}  \right\}  \md(v)   \m . 
\label{2-ii}
\ee
\Proof i) Once again, we can count out 
\be
e(\mbox{Adj}(\mG)) = \sum_{v \in \sG} W(v) + \sum_{f \in \sG} u(f) 
\label{Second-Sum}
\ee
by attaching vertex and face weights to (an extended notion version of) the original graph $\mG$.  
These are the same face weights as before, but different face weights as follows.  
\be 
W(v) = 4 \, n + \left\{   \s{    \mbox{5  \m $v \in \mP(\mG)$ }    }
                                   {    \mbox{8  \m $v \in \mbox{Int}(\mG)$    }    }     \right.     \mbox{ of degree $n$ + 2 }   \m , 
\label{W(v)} 								   
\ee
as can be trivially established by induction.
Thus 
\be
\sum_{v \in \sG} w(v) =      \sum_{v \in \sI\sn\st(\sG)} \{4 \, n + 8 \} +       \sum_{v \in \sP(\sG)} \{4 \, n + 5\}
                      = 4 \, \sum_{v \in \sG}            \{     n + 2 \} -  3 \, \sum_{v \in \sP(\sG)}        1 
					  = 4 \, \sum_{v \in \sG}  \md(v)                    -  3 \,           |\mP(\mG)|
                      = 8 \, e(\mG)                                      -  3 \,            \mp(\mG)                       \m .	
\label{2-I}					  
\ee
This manipulation uses the same series of moves as its counterpart in the proof of Proposition 1.

\m

\n So by (\ref{2-I}) and (\ref{1-II}) in (\ref{Second-Sum}),   
\be
Q_{\sA\sp\sp\sr\so\sx}(\mG)  = 8 \, e(\mG) - 3 \, p(\mG) + 2 \, e(\mG) - p(\mG) 
							 = 2 \{ 5 \, e(\mG) - 2 \, p(\mG) \}                 \m , 
\ee
and we finish yet again by using the first equation of Appendix I.A on each of $\mG$ and $\mP(\mG)$ for the last equality of (\ref{2-i}).  

\m

\n ii) Use Lemma 1 and the first equality of (\ref{1-i}) in (\ref{Q-total}) to get 
\be
Q_{\sT\so\st\sa\sll}(\mG) = 2 \, e(\mG) + \chi(\mG)      + 2 \{ 5 \, e(\mG) - 2 \, p(\mG) \} 
                          = 4 \{ 3 \, e(\mG) - p(\mG) \} +  \chi(\mG)                                \m ,
\ee
and we finish one last time by using the first equation of Appendix I.A on each of $\mG$ and $\mP(\mG)$ for the last equality of (\ref{2-ii}).  $\Box$

\m

\n{\bf Corollary 1} i) 
\be
Q_{\sA\sp\sp\sr\so\sx}(\mG) = 2  \{  Q_{\sa\sp\sp\sr\so\sx}(\mG) - e(\mG) \}     \m .
\ee
ii) 
\be
Q_{\sT\so\st\sa\sll}(\mG)   = 2 \, Q_{\sa\sp\sp\sr\so\sx}(\mG)     + \chi(\mG)   \m .
\ee
\Proof This is a simple matter of algebraic elimination that the reader can check.  $\Box$

\m

\n{\bf Remark 7} By ii), the adjacency graph edge number almost immediately gives the grand total of qualitative types, 
further justifying the adjacency graph's inclusion in this treatment of qualitative types.

\m

\n{\bf Corollary 2} For compact without boundary shape spaces, the following simplifications occur. 
 
\m

\n i) $Q_{\sa\sp\sp\sr\so\sx}(\mG)   = 6 \, e(\mG)   \m ,$

\m

\n ii) $Q_{\st\so\st\sa\sll}(\mG)    = 8 \, e(\mG) + \chi(\mG)   \m ,$

\m

\n iii) $Q_{\sA\sp\sp\sr\so\sx}(\mG) = 10 \, e(\mG)   \m ,$

\m

\n iv) $Q_{\sT\so\st\sa\sll}(\mG)    = 12 \, e(\mG) + \chi(\mG)   \m .$ 

\m

\Proof For compact without boundary shape spaces, $p(\mG) = 0$, which simplifies various formulae as shown.  $\Box$

\m

\n{\bf Remark 8} $e(\mG)$ and $\chi$ are now sufficient data for computing all types of $Q$ considered; 
in the more general setting, $p(\mG)$'s `boundary datum' is also required.

\subsection{Counting qualitative types of uniformity}

{\bf Remark 1} The uniformity structures in Papers II and III are connected sets of arcs and vertices.
The uniformity `graph' $\mG_{\sU}$ can contain edges with no vertex at one end.
In order to have an actual graph, we introduce the following. 

\m

\n{\bf Definition 1} The {\it completion} $\mG^{\scc}_{\sU}$ of the uniformity `graph' is the result of adding in the missing vertices. 

\m

\n{\bf Remark 1} Clearly 
\be
e(\mG^{\scc}_{\sU}) = e(\mG_{\sU})   \m .  
\label{GcU-GU}
\ee
\n{\bf Definition 2} 
\be
Q(\mU) = |\mG_{\sU}| + e(\mG_{\sU})   \m .  
\ee
The uniform structure for Papers II and III being 1-dimensional, I define $Q_{\sa\sp\sp\sr\so\sx}(\mU)$ and $Q_{\st\so\st\sa\sll}(\mU)$ 
in the same way as the approximate and total numbers of qualitative types of Part I's (3, 1) model. 

\m

\n{\bf Proposition 1} i) 
\be
Q_{\sa\sp\sp\sr\so\sx}(\mU) = 2 \, e(\mG_{\sU})   \m . 
\ee
ii) 
\be
Q_{\st\so\st\sa\sll}(\mU) = |\mG_{\sU}| + 3 \, e(\mG_{\sU})   \m . 
\ee
\Proof i)
\be
Q_{\sa\sp\sp\sr\so\sx}(\mU) = 2 \, e(G^{\sc}_{\sU}) 
                            = 2 \, e(G_{\sU})                 \m , 
\ee
since the absent vertices still have approximate regimes. 
Then use the first equation of Appendix I.A followed by (\ref{GcU-GU}). 

\m

\n ii) 
\be
Q_{\st\so\st\sa\sll}(\mU) = Q(\mU) + Q_{\sa\sp\sp\sr\so\sx}(\mU) 
                          = |\mG_{\sU}| + e(\mG_{\sU}) + 2 \, e(\mG_{\sU}) 
                          = |\mG_{\sU}| + 3 \, e(\mG_{\sU}) \m . 
\ee
by use of Definition 2 and i).

\end{appendices}


\end{document}